\documentclass[pra,twocolumn,10pt,superscriptaddress,nofootinbib,showpacs]{revtex4-1}

\usepackage{hyperref}
\usepackage{amsmath, amsfonts, amssymb, graphicx,cancel}

\newcommand{\nn}{\nonumber\\}
\newcommand{\hb}{\hat{b}}
\newcommand{\eq}{&=&}
\newcommand{\ceq}[1]{Eq.~(\ref{#1})}
\newcommand{\bea}{\begin{eqnarray}}
\newcommand{\eea}{\end{eqnarray}}

\renewcommand{\l}{\left(}
\renewcommand{\r}{\right)}

\usepackage{color}

\begin{document}
\title{Radio frequency spectroscopy of polarons in ultracold Bose gases}

\date{\today}

\begin{abstract}
Dynamical impurity in BEC
\end{abstract}

\author{Aditya Shashi}
\affiliation{Department of Physics, Harvard University, Cambridge, Massachusetts 02138, USA.}
\affiliation{Department of Physics and Astronomy, Rice University, Houston, Texas 77005, USA.}

\author{Fabian Grusdt}
\affiliation{Department of Physics, Harvard University, Cambridge, Massachusetts 02138, USA.}
\affiliation{Department of Physics and research center OPTIMAS, University of Kaiserslautern, Germany.}
\affiliation{Graduate School Materials Science in Mainz, Gottlieb-Daimler-Strasse 47, 67663 Kaiserslautern, Germany.}

\author{Dmitry A. Abanin}
\affiliation{Department of Physics, Harvard University, Cambridge, Massachusetts 02138, USA.}
\affiliation{Perimeter Institute for Theoretical Physics, Waterloo, Ontario N2L 6B9, Canada.}
\affiliation{Institute for Quantum Computing, Waterloo, Ontario N2L 3G1, Canada,}

\author{Eugene Demler}
\affiliation{Department of Physics, Harvard University, Cambridge, Massachusetts 02138, USA.}

\date{\today}

\begin{abstract}

Recent experimental advances enabled the realization of mobile impurities immersed in a Bose-Einstein condensate (BEC) of ultra-cold atoms. Here we consider impurities with two or more internal hyperfine states, and study their radio-frequency (RF) absorption spectra, which correspond to transitions between two different hyperfine states.

We calculate RF spectra for the case when one of the hyperfine states involved interacts with the BEC, while the other state is non-interacting, by performing a non-perturbative resummation of the probabilities of exciting different numbers of phonon modes. In the presence of interactions the impurity gets dressed by Bogoliubov excitations of the BEC, and forms a polaron. The RF signal contains a delta-function peak centered at the energy of the polaron measured relative to the bare impurity transition frequency with a weight equal to the amount of bare impurity character in the polaron state. The RF spectrum also has a broad incoherent part arising from the background excitations of the BEC, with a characteristic power-law tail that appears as a consequence of the universal physics of contact interactions.

We discuss both the direct RF measurement, in which the impurity is initially in an interacting state, and the inverse RF measurement, in which the impurity is initially in a non-interacting state. In the latter case, in order to calculate the RF spectrum, we solve the problem of polaron formation: a mobile impurity is suddenly introduced in a BEC, and dynamically gets dressed by Bogoliubov phonons. Our solution is based on a time-dependent variational ansatz of coherent states of Bogoliubov phonons, which becomes exact when the impurity is localized. Moreover we show that such an ansatz compares well with a semi-classical estimate of the propagation amplitude of a mobile impurity in the BEC. 
Our technique can be extended to cases when both initial and final impurity states are interacting with the BEC.

 \end{abstract}
\pacs{67.85.-d,78.40.-q,72.10.Di,47.70.Nd}

\maketitle

\section{Introduction}

The polaron problem\cite{landauorig,landaupekar46,frohlich1954electrons,feynman1955slow,devreese1996polaron} concerns the modification of the physical properties of an impurity by the quantum fluctuations of its environment. This ubiquitous problem naturally arises in a wide variety of physical situations including: electron-phonon interactions~\cite{frohlich1954electrons}, the propagation of muons in a solid~\cite{storchak1998quantum}, transport in organic transistors~\cite{hulea2006tunable}, the physics of giant magnetoresistance materials~\cite{von1993giant}, and high T$_{C}$ cuprates~\cite{salje2005polarons}. Recently in Refs.~ \cite{prokof2008fermi,punk2009polaron,Schirotzek2009,chevy2010ultra,schmidt2011excitation,Koschorreck2012,kohstall2012metastability,Zhang2012,massignan2012polarons,catani2012quantum,astrakharchik2004motion,cucchietti2006strong,sacha2006self,kalas2006interaction,Bruderer2008,Bruderer2008a,Bruderer2007,schmid2010dynamics,privitera2010polaronic,Casteels2011a,Casteels2012,Casteels2011,Tempere2009,blinova2013single,rath2013polaron} the polaron problem was considered in the context of quantum impurities in ultracold atomic gases.

The unprecedented control over interatomic interactions, external trapping potentials, and internal states of ultracold atoms, allows the realization of systems previously unattainable in condensed matter. Examples relevant to our study include Bose-Bose and Bose-Fermi mixtures with varying mass ratios. Moreover, specialized experimental probes like radio frequency (RF) spectroscopy~\cite{Gupta2003} and Ramsey interference~\cite{shin2004atom} enable detailed characterizations of these systems, including their coherent real-time dynamics. Furthermore these systems are very well characterized and can be theoretically described using simple models with just a few parameters. Such universality arises in cold atoms because they are well isolated from their environment, have simple dispersion relations, and the two particle scattering amplitudes have a universal form fully characterized by the scattering length (except in cases of narrow Feshbach resonances \cite{Chin.paper}, which we will not discuss here), while higher order scattering processes can be neglected due to diluteness. This is in contrast to generic condensed-matter systems where universal physics is manifested only at very low energies, while coherent dynamics is usually difficult to probe~\cite{bloch2008many}.

 \begin{figure}[h!]
\includegraphics[scale=.31]{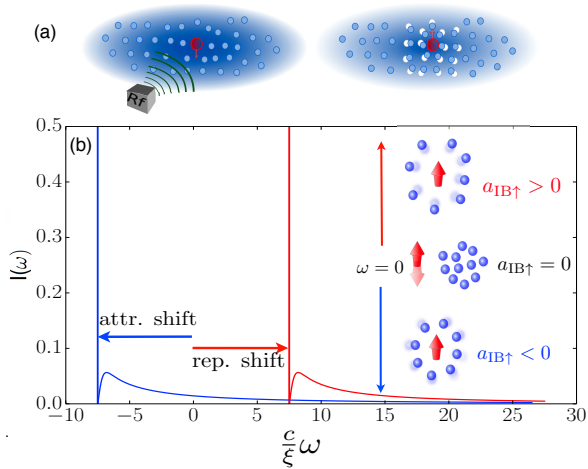}
\caption{\label{fig:RF_schematic} (a) Schematic representation of the system comprising a BEC (blue) with a small concentration of free impurities (red) that have two internal levels, $|\downarrow\rangle, |\uparrow\rangle$, in which the impurity-boson interaction characterized by the s-wave scattering length $a_{\rm IB,\sigma}$ is different. An RF pulse transfers impurities from $|\downarrow\rangle$ to $|\uparrow\rangle$, and probes the emergent polaronic state of the impurity, and the dynamics of its formation. (b) Typical RF signal shown when impurity-Bose scattering length $a_{\rm IB,\uparrow} >0$ (right), $a_{\rm IB,\uparrow} < 0$ (left). The RF signal contains a coherent peak centered at the energy of the polaron measured relative to the bare impurity transition frequency, and with weight corresponding to the impurity quasiparticle residue $Z$ (see Sec.~\ref{sec:qp_res}). There is an additional incoherent part capturing background excitations of the BEC, with a characteristic power-law tail (see Sec.~\ref{sec:universal_features}). (Inset) Attractive (repulsive) polarons corresponding to $a_{\rm\uparrow,\sigma}<0 (>0)$, with negative (positive) energy relative to the atomic transition frequency $\omega=0$.   
} \end{figure}

In the present article we consider dynamic impurities in a Bose-Einstein condensate (BEC), and demonstrate how the spectral and dynamical properties of the Fr\"{o}hlich polaron~\cite{frohlich1954electrons} can be probed using RF spectroscopy in dilute mixtures of ultracold atoms. For these systems we predict the essential spectroscopic features of an RF measurement, some of which are constrained by exact relations, and discuss the corresponding impurity dynamics.

RF spectroscopy~\cite{Zwierlein2003,Gupta2003,Chin2004,Bartenstein2005,Shin2007,Sommer2012}, along with its momentum resolved variant \cite{Stewart2008,Feld2011}, has emerged as an important experimental tool to study many-body physics in cold atoms. Pertinently, RF spectroscopy can directly probe the spectral properties of quantum impurities~\cite{Schirotzek2008,C.H.SchunckY.ShinA.SchirotzekM.W.Zwierlein2008,Schirotzek2009,kohstall2012metastability,Koschorreck2012}, and has prompted theoretical investigations of impurity spectral functions~\cite{veillette2008radio,Knap2012,rath2013polaron}.

 The subject of impurities in BECs has received some attention recently, but the focus has mainly been on the near-equilibrium-properties of these systems. The effect of impurities on BECs was studied using the Gross-Pitaevskii~\cite{stringarii.book} quantum hydrodynamic description of the coherent condensate wavefunction in Refs.~\cite{astrakharchik2004motion,cucchietti2006strong,sacha2006self,kalas2006interaction,Bruderer2008,blinova2013single}. Such an approach is restricted to weak impurity-boson couplings, and downplays the effects of quantum dynamics of the impurity, which appears only as a classical potential acting on the collective field of the bosons. The authors of Refs.~\cite{Casteels2011a,Casteels2012,Casteels2011,Tempere2009} took quantum impurities into account within a many-body treatment of the Bogoliubov excitations of the BEC, but considered equilibrium properties in the regime of weak impurity-boson interactions.

In contrast to these earlier works, we calculate the {\em full spectral response} of dilute quantum impurities in a BEC, and study their {\em non-equilibrium dynamics} which arise when applying an RF signal to the system. As shown in Fig~\ref{fig:RF_schematic}(a), we consider a BEC with a small concentration of free impurities with two internal states. The impurities are taken to initially be in the $|\downarrow\rangle$ internal state, and we consider the effect of an RF pulse which transfers them to the final $|\uparrow\rangle$ state. In Fig~\ref{fig:RF_schematic}(a) we present the so called ``inverse'' RF protocol~\cite{kohstall2012metastability}, in which impurities in the $|\downarrow\rangle$ state are non-interacting with the bosons, while in the $|\uparrow\rangle$ state they interact. Then the initially free impurities propagate as emergent quasiparticles, called polarons, that are dressed by a cloud of background BEC excitations. Correspondingly, the resulting RF absorption signal, shown in Fig~\ref{fig:RF_schematic}(b), contains a coherent peak centered at a frequency corresponding to the energy of the polaron measured 
from the transition frequency between the states $\downarrow$ and $\uparrow$ of the bare impurity. The peak has an exponentially suppressed weight that quantifies the amount of bare impurity character in the polaron state. Additionally the RF signal contains an incoherent part corresponding to the excitations of the background BEC, which displays a characteristic high frequency power-law 
tail. The latter is a manifestation of the universal two-body ``contact" physics studied by Tan~\cite{tan2008large,tan2008energetics}, and is a recurring feature of RF studies in ultracold atoms~\cite{punk2007theory,haussmann2009spectral,schneider2010universal,braaten2011universal,langmack2012clock}. We also discuss the ``direct'' RF protocol~\cite{Koschorreck2012}, in which  impurities initially in the $|\downarrow\rangle$ state interact with the bosons, and are transferred to a non-interacting $|\uparrow\rangle$ state. Our discussion can also be extended to the case where both internal states of the impurity are interacting, but with different interaction strengths. 

We calculate impurity RF spectra by resumming an infinite number of emitted Bogoliubov excitations, and thus capture the nonequilbrium dynamics of polaron formation. Moreover our treatment is mathematically exact for completely localized impurities.

The article is organized as follows: In Sec II we introduce an effective model describing the impurity-BEC system and discuss the time-dependent overlap required to calculate impurity RF spectra. In Section III we analyze the ground state properties of the system, and define the quantities we use to analyze the more complicated dynamical problem of RF spectra. In Section IV we present the main results concerning impurity RF spectra in three parts: first we demonstrate that the coherent and incoherent parts of the RF signal are both constrained by exact relations. Next we present the microscopic calculation of two types of RF measurements, so called ``direct'' and ``inverse'' RF spectroscopy, and lastly discuss non-equilibrium dynamics of the impurity which arise in the course of the inverse RF measurement. Finally, in Section V we summarize our results, point out connections to existing experiments, and highlight future directions of study.

\section{Microscopic Model}

We assume that the concentration of impurity atoms
is low, so we can neglect interactions between them, and discuss individual impurity atoms. 
Thus we consider a single impurity of mass $M$, which has two internal (e.g. hyperfine) states $|\uparrow\rangle, |\downarrow\rangle$, immersed in a BEC of a different type of atom of mass $m$. The Hamiltonian of the system is given by 
\begin{eqnarray}
\label{eq:schematic_model}
{\cal H} = {\cal H}_b +{\cal H}_I + |\uparrow\rangle \otimes \langle \uparrow| {\cal H}_{\rm int \uparrow}+ |\downarrow\rangle \otimes \langle \downarrow| {\cal H}_{\rm int \downarrow},  
\end{eqnarray}
where ${\cal H}_b$ is the BEC Hamiltonian, ${\cal H}_I = \frac{\hat{p}^2}{2M}$ is the Hamiltonian of the impurity atom with momentum $\hat{p}$,
and ${\cal H}_{\rm int \sigma}$ describes a density-density interaction of the bosons with impurity in state $\sigma$ at position ${\bf \hat{x}}$:
\bea
\label{eq:impurity_boson_int}
{\cal H}_{\rm int \sigma}=g_{\rm IB,\sigma} \rho_{\rm BEC}({\bf \hat{x}}),
\eea
where $g_{\rm IB,\sigma}$ models the microscopic short-range interaction between the atoms. Since we treat systems of ultracold atoms for which the effective range of interactions between atoms (on the order of the van der Waals length) is the smallest length scale, inter-atomic interactions can be modeled as having zero range\cite{Huang.book,bloch2008many}, and the microscopic host-impurity interaction can be described using the s-wave scattering length $a_{\rm IB,\sigma}$ of the impurity in state $\sigma$ with the surrounding BEC (see also Appendix A).

We will restrict our discussion to weakly-interacting Bose gases, well described by the Bogoliubov approximation \cite{stringarii.book}, in which the condensed ground state of the Bose gas is treated as a static ``mean field", and excitations are modeled as a bath of free phonons.
\begin{eqnarray}
\label{eq:BEC_hamiltonian}
{\cal H}_b \eq \sum_{\bf k\neq0}\omega_{\bf k}b^\dagger_{\bf k}b_{\bf k}, \ \ \omega_{\bf k} = ck\sqrt{1+\frac{(k\xi)^2}{2}},
\end{eqnarray}
 where $\xi = 1/(\sqrt{2}mc)$ is the healing length, $c$ the speed of sound in the BEC, $k = |{\bf k}|$, and where we took $\hbar=1$. In this framework the interaction (\ref{eq:impurity_boson_int}) between impurity and bosons can be rewritten as a sum of two terms. The first captures the ``mean-field'' interaction of the BEC ground state with the impurity, and the second encodes the impurity interactions with the Bogoliubov excitations. The density of the excitations can be expressed as a linear combination of phonon creation and annihilation operators, and leads to the following explicit form of the interaction Hamiltonian:
\begin{eqnarray}
{\cal H}_{\rm int \sigma} \eq \frac{2\pi a_{\rm IB,\sigma}}{\mu}n_0 + \sum_{\bf k} V_{{\bf k}\sigma} e^{i {\bf k.\hat{x}}} (\hb_{\bf k} +\hb_{-\bf k}^\dagger),
\label{eq:Frohlich_model1}
\end{eqnarray}
with \cite{Tempere2009}
\begin{eqnarray}
V_{{\bf k}\sigma} \eq \frac{2\pi a_{IB\sigma}\sqrt{N_0}}{\mu}\left(\frac{\xi k}{\sqrt{2+(\xi k)^2}}\right)^{1/2}.
\label{eq:Frohlich_model2}
\end{eqnarray} 
 Here $N_0$ is the number of atoms in the condensate, with the corresponding density $n_0$, and $\mu =(m^{-1} +M^{-1})^{-1}$ is the reduced mass of the impurity.

 The above approximations hold so long as the impurity-boson interaction does not significantly deplete the condensate, leading to the condition~\cite{astrakharchik2004motion,Bruderer2008}
\begin{eqnarray}
 |a_{\rm IB,\sigma}|\xi^{-1} \ll1. \label{eq:approximation_criterion}
\end{eqnarray}

Our treatment of the impurity-BEC system ignores the phenomenology of strong-coupling physics e.g., near a Feshbach resonance~\cite{rath2013polaron}, which lies beyond the parameter range (\ref{eq:approximation_criterion}). The model (\ref{eq:BEC_hamiltonian}), (\ref{eq:Frohlich_model1}), with parameters (\ref{eq:Frohlich_model2}), in its regime of validity, constitutes a generalized Fr\"{o}hlich model of polarons in ultracold BECs~\cite{Casteels2011a,Casteels2012,Casteels2011,Tempere2009}. 

\subsection{ RF spectroscopy as dynamical problem}\label{sec:RF_intro}

An RF pulse changes the internal state of the impurity atom without modifying its momentum. Thus for a $\downarrow$-impurity-BEC initial state with momentum $p$, energy $E_{i\downarrow}$, denoted $|i_{\downarrow p} \rangle$, the RF absorption cross section can be computed within Fermi's Golden Rule from
\begin{eqnarray}
I (p,\omega)=  \sum_{n} | \langle n_{\uparrow p} | \hat{V}_{\rm RF} | i_{\downarrow p} \rangle |^2\delta (\omega - ( E_{n\uparrow} -E_{i\downarrow})),\label{eq:RF_golden_rule}
\end{eqnarray}
where all states $|n_{\uparrow p}\rangle$ of $\uparrow$-impurity-BEC system with total momentum $p$ are summed over. The RF transition operator $\hat{V}_{\rm RF} \sim |\uparrow\rangle\langle\downarrow|$ instantaneously changes the internal state of the impurity, but the quantum mechanical state of the impurity-BEC system is otherwise unmodified by it, i.e. the initial state of the system $|i_{\downarrow p} \rangle$ is {\em quenched}. Using standard manipulations (see e.g. \cite{mahan2000many,Knap2012}) the last expression can be rewritten
as
\begin{eqnarray}
\label{eq:RF_response}
I(p,\omega)\eq{\rm Re}\frac{1}{\pi} \int_0^\infty dt e^{i\omega t} A_p(t)
\\
A_p(t) \eq e^{i E_{i\downarrow} t} \langle i_{\uparrow p} |
e^{ -i ( {\cal H}_b +{\cal H}_I + H_{\rm int \uparrow})t}
| i_{\uparrow p} \rangle,
\label{eq:overlap}
\end{eqnarray}
 where frequency $\omega$ is measured relative to the atomic transition frequency between states $|\downarrow\rangle$ and $|\uparrow\rangle$ of the bare impurity, and where we denoted $|i_{\uparrow p}\rangle=\hat{V}_{\rm RF}|i_{\downarrow p}\rangle$.  
 
Let us emphasize again: due to the instantaneous nature of the RF spin-flip, the state $|i_{\uparrow p}\rangle$ is identical to the initial state of the $\downarrow$-impurity BEC system in all respects, {\em except the internal state of the impurity}. Consequently, $|i_{\uparrow p}\rangle$ is different from, and therefore higher in energy than, the $\uparrow$-impurity-BEC ground state at momentum $p, |0_{\uparrow p}\rangle$. Thus it is more convenient to formulate the physical problem underlying the RF response as a dynamical one, rather than a traditional calculation of a ground state observable. Indeed, expression (\ref{eq:overlap}) has the form of the quantum propagation amplitude, related to the Loschmidt echo\cite{Silva2008}), 
where an eigenstate of the Hamiltonian
${\cal H}_b +{\cal H}_I +{\cal H}_{\rm int \downarrow}$ needs to be time evolved with ${\cal H}_b +{\cal H}_I +H_{\rm int \uparrow}$. $A_p(t)$ can also be measured directly in the time domain using the Ramsey sequence discussed in Ref.~\cite{Knap2012}. Analysis of (\ref{eq:overlap}) serves the central goal of this paper: the calculation of impurity RF spectra. 

\subsection{Direct and inverse RF: momentum resolved spectra}

Two varieties of RF spectroscopy are commonly used to probe impurity physics in cold atoms: direct and inverse RF. In the present context, direct RF involves preparing the system with the impurity initially in an interacting state, i.e. in Eq.(\ref{eq:RF_golden_rule}), the state $|i_{\downarrow p}\rangle = |0_{\downarrow p}\rangle$ will correspond to the interacting impurity-BEC state: a polaron with momentum $p$. The RF pulse then flips the impurities to a final state in which they are non-interacting, i.e., $a_{{\rm IB},\uparrow} \approx 0$. For the inverse RF measurement, the scenario above is reversed, and the impurity is initially in a non-interacting state, i.e. $|i_{\downarrow p}\rangle = |{\bf p}\rangle_{\downarrow}\otimes|0\rangle$ will correspond to the decoupled momentum ${\bf p}$ bare impurity-BEC ground state, with $|0\rangle$ the vacuum of Bogoliubov phonons, and the RF pulse flips the impurities to an interacting final state, i.e. $a_{{\rm IB},\uparrow} \neq 0$. 

Typically one is interested in performing a momentum resolved RF measurement. In the case of direct RF, a time-of-flight measurement following the RF pulse will directly yield the polaron momentum distribution since, after the impurity atoms are transferred to the $\uparrow$ state, they propagate ballistically without being scattered by the host BEC atoms. The combined time-of-flight and RF absorption measurements can be interpreted as momentum resolved RF spectroscopy~\cite{Stewart2008,Feld2011,Koschorreck2012}. Offsetting this advantage, the finite lifetime of the polaron~\cite{rath2013polaron}~\footnote{For positive scattering length, the pair-wise impurity-boson interaction potential admits a bound state, leading to an impurity-BEC ground state formed out of bound bosons, that is much lower in energy than the repulsive polaron which is formed out of scattered bosons. Consequently the repulsive polaron is a metastable state with a finite lifetime after which it will decay into the molecular state.} may pose a challenge to the initial adiabatic preparation of the system required for this measurement. On the other hand, for the inverse RF measurement, in which interactions are absent for the initial state of the impurity, the problem of finite polaron lifetime can be circumvented~\cite{kohstall2012metastability} but momentum resolution is more challenging to obtain. 

We propose the following momentum-resolved {\em inverse} RF measurement.  An external force that acts selectively on impurity atoms (e.g. through a magnetic field gradient) can be used to impart a finite initial momentum
\bea
p_0 = -\nabla V_{\rm ext,\downarrow}\Delta T,
\eea
where $p_0$, the center of the momentum distribution of $\downarrow$-impurities is the momentum transferred by applying a state-selective external potential gradient $\nabla V_{\rm ext, \downarrow}$ for a time $\Delta T$ to the impurities. 
An RF pulse would then transfer the initially weakly interacting impurities to an interacting final state. The known transferred momentum $p_0$, combined with the absorption of RF, would yield a momentum resolved RF spectrum. Since the experiment is done at a finite concentration of impurity atoms to obtain the total absorption cross section $I (p,\omega)$ would need to be averaged over the impurity momentum distrubtion (see e.g., the Supplementary materials of Ref.~\cite{Schirotzek2009}), with width given by the thermal de Broglie wavelength, or by the inverse of the distance between impurity atoms (if they are fermionic and obey the Pauli exclusion principle). Typically the width is expected to be small due to the low temperature and diluteness of the impurities. The advantage of such a measurement is its insensitivity to the polaron lifetime as it requires no adiabatic preparation \cite{kohstall2012metastability}, while also allowing a momentum resolved measurement, but at the cost of repeated measurements to resolve a finite momentum range.

\section{Polaron ground state in BEC }\label{sec:GS_prop}

In order to characterize polaronic phenomena manifested in RF spectra, it is useful to review the ground state properties of polarons in BECs. 

It is possible to tune interactions between ultracold atoms to be effectively attractive or repulsive using Feshbach resonances~\cite{Chin.paper}. Correspondingly, the Bose polaron comes in two varieties associated with effective attraction ($a_{\rm IB,\sigma} < 0$) and repulsion ($a_{\rm IB,\sigma} > 0$) between the impurity and the BEC. Moreover at strong coupling there is an additional transition of the attractive polaron into a bound molecular state \cite{rath2013polaron}. We will only discuss the regime of weak impurity-Bose interactions which satisfy the condition (\ref{eq:approximation_criterion}) and are captured by our Fr\"{o}hlich model  (\ref{eq:BEC_hamiltonian}), (\ref{eq:Frohlich_model1}), with parameters (\ref{eq:Frohlich_model2}).

We note that the authors of Ref.~\cite{rath2013polaron} also considered the spectral properties of impurities in a BEC, but considered the regime of strong impurity-bose coupling which occurs in the vicinity of the Feshbach resonance. Their approach, inspired in part by Chevy's variational wavefunction description of fermionic polarons \cite{Chevy2006,Combescot2007}, separates the spectral contributions of the bound molecules and the repulsive polarons on the repulsive side of the Feshbach resonance ($a_{\rm IB, \sigma}>0$). However their selective resummation scheme does not reduce to the exact solution in the case of a heavy impurity, and consequently misses the physics of the orthogonality catastrophe~\cite{Anderson.OC} in low dimensions. Thus it does not accurately describe the precise lineshape of the incoherent part of RF spectra.

Although the analysis of the ground state of the polaron model has been carried out previously in Refs.~\cite{alexandrov1995polarons, bei2009polaron}, we present it here to motivate our later study of dynamics as a generalization of the approach to the ground state.

\subsection{ Lee-Low-Pines transformation} 
\label{sec:LLP}

There exists a canonical transformation introduced by Lee, Low, and Pines\cite{lee1953motion} (LLP), that singles out the conserved total momentum of the system:
\begin{eqnarray}
\tilde{{\cal H}} \eq e^{iS}{\cal H}e^{-iS}, \text{with} \  S = {\bf \hat{\bf x}}.\sum_{\bf k }{\bf k} \hat{b}^\dagger_{\bf k} \hat{b}_{\bf k},\label{eq:S_LLP}\\
e^{iS} \hat{b}_{\bf k} e^{-iS} \eq \hat{b}_{\bf k} e^{-i{\bf k}.{\bf x}}, e^{iS}\hat{\bf p}e^{-iS} = {\bf \hat{p}} -\sum_{\bf k} {\bf k} \hat{b}^\dagger_{\bf k} \hat{b}_{\bf k}.
\end{eqnarray}
We may write the transformed Hamiltonian as
\begin{eqnarray}
\tilde{\cal H} \eq \frac{1}{2M}\left({\bf p} - \sum_{\bf k} {\bf k} \hb^\dagger_{\bf k}\hb_{\bf k}\right)^2+ \sum_{\bf k} V_{\bf k}(\hb^\dagger_{\bf k} + \hb_{-\bf k})\nn
&&\hspace{2 cm} + \sum_{\bf k}\omega_{\bf k} \hb^\dagger_{\bf k} \hb_{\bf k},\label{eq:transformed_Hamiltonian}
\end{eqnarray}
where without loss of generality we projected the full Hamiltonian onto the sector $\sigma = \uparrow$; the same can be done in the other sector. 

 The LLP transformation eliminates the impurity degree of freedom by isolating the conserved total momentum ${\bf p}$ of the system  which becomes a parameter of the effective Hamiltonian (\ref{eq:transformed_Hamiltonian}). The simplification comes at the cost of an induced interaction between the Bogoliubov excitations, which enocodes the quantum dynamics of the impurity, and vanishes in the $M\to\infty$ limit of a static localized impurity.

It was argued in Refs.~\cite{gerlach1987proof,gerlach1991analytical} that the existence of a finite momentum ground state implies symmetry breaking, and consequently, a phase transition corresponding to the ``self-localization'' transition of Landau and Pekar \cite{landaupekar46}. Although we will discuss states of the Hamiltonian (\ref{eq:transformed_Hamiltonian}) with arbitrary total momentum $p$, it was established rigorously in Ref.~\cite{spohn1986roughening} that a large class of Fr\"{o}hlich type models with gapless phonons, including the present one, can only admit a ground state with $p=0$. 

We will consider eigenstates of Hamiltonian (\ref{eq:transformed_Hamiltonian}) with finite total momentum $p$, which are not ``true" global ground states in the above sense, but are nonetheless required to calculate momentum resolved RF spectra using the time dependent overlap (\ref{eq:overlap}). The symmetry breaking in the present context is not spontaneous, but rather due to the injection of an impurity with finite momentum into the BEC. We will use the term ``polaron ground state'' to refer to the lowest-energy eigenstate of Hamiltonian (\ref{eq:transformed_Hamiltonian}) with a given total momentum $p$. We approximate such states using a mean-field treatment.

\subsection{Mean-field polaron solution}

For a localized impurity $M\to\infty$, Hamiltonian (\ref{eq:transformed_Hamiltonian}) decouples into a sum of independent harmonic oscillators, each of which has a coherent state as its ground state \cite{Glauber.book}. Consequently the many-body ground state in this limit is a decoupled product of coherent states:
\begin{eqnarray}
| 0_{M\rightarrow \infty}\rangle =  \prod_{\bf k}e^{\beta_{\bf k}\hat{b}^\dagger_{\bf k} - \beta_{\bf k}^*\hat{b}_{\bf k}}|0\rangle,\ \ \beta_{\bf k} = -\frac{V_{\bf k}}{\omega_{\bf k}}.
\end{eqnarray}

Moroever we expect by continuity that for an impurity with a large, finite mass $M$, we can approximate the true ground state by an optimally chosen product of coherent states:
\bea
|0_{\downarrow p}\rangle = \prod_{\bf k}e^{\alpha^{\rm MF}_{\bf k}\hat{b}^\dagger_{\bf k} - (\alpha^{\rm MF}_{\bf k})^*\hat{b}_{\bf k}}|0\rangle,\label{eq:MF_state}
\eea
with $\alpha_{\bf k}^{\rm MF}$ determined by minimizing the total energy of the system $E(\{\alpha_{\bf k}\}) = \langle  0_{\downarrow p}|\tilde{\cal H}|0_{\downarrow p}\rangle$, which can be cast as a mean field self-consistency condition
\bea
 \alpha^{\rm MF}_{\bf k} \eq - \frac{V_{\bf k}}{ \omega_{\bf k} + \frac{k^2}{2 M} -\frac{k_{\parallel}}{M} \l p-\Xi[\alpha^{\rm MF}_{\bf k}] \r}, \label{eq:MFpolaron0}\\
 \Xi[\alpha_{\bf k}] &\equiv& \sum_{\bf k} k_{\parallel} |\alpha_{\bf k}|^2.\nonumber
\eea
where we denote the total phonon momentum projected in the direction $k_{\parallel} \equiv \frac{\bf p}{|\bf p|}$ by the parameter $\Xi$. The set of self-consistency conditions (\ref{eq:MFpolaron}) can then be reformulated as a {\em single scalar equation} for $\Xi$:
\bea
 \Xi = \sum_{\bf k} \frac{k_{\parallel}V_{\bf k}^2}{ \left( \omega_{\bf k} + \frac{k^2}{2 M} -\frac{k_{\parallel}}{M} \l p-\Xi \r \right)^2}. \label{eq:MFpolaron}
\eea

Having approximated the polaron ground state wavefunction using \ceq{eq:MFpolaron}, we can calculate the polaron binding energy, effective mass, and the  overlap with the bare impurity. 

%which in three spatial dimensions (3D) has the explicit form:
%\begin{widetext}
%\bea
% \Xi = \frac{M^2}{(p-\Xi)^2} \int_0^\infty \frac{dk}{(2\pi)^2} ~ \frac{\xi k^2}{(2+(k\xi)^2)^{1/2}}\left[ \frac{2 \l \omega_k + \frac{k^2}{2M} \r \frac{k}{M}\l p-\Xi \r}{\l \omega_k + \frac{k^2}{2M} \r^2 - \l \frac{k}{M} (p-\Xi)\r^2} + \log \l \frac{\omega_k + \frac{k^2}{2M}-\frac{k}{M} \l p-\Xi \r}{\omega_k + \frac{k^2}{2M}+\frac{k}{M} \l p-\Xi \r} \r  \right],
% \label{eq:XiselfConsLong}\\
% \alpha_{\bf k_{\parallel}} = -\frac{V_{k}}{\omega_k + \frac{k^2}{2M}+\frac{k}{M} \l p-\Xi \r}.
%\eea
%\end{widetext}
%We took the continuum limit $\sum_{\bf k} \to \int \frac{d^D{\bf k}}{(2\pi)^D}$ and used the cylindrical symmetry of the system to perform the angular integration in 3D. The procedure can be straightforwardly generalized to $D<3$. 

\subsection{Binding energy of the polaron}\label{sec:binding_energy}
The binding energy is defined as the difference between the ground state energy of the polaron at zero momentum and the energy of a BEC with a non-interacting impurity atom:
\bea
E_B \eq \langle 0_{\uparrow p=0}|{\cal H}| 0_{\uparrow p=0}\rangle  -\langle 0|\otimes\langle p = 0|{\cal H}_b + {\cal H}_I| 0\rangle\otimes|p = 0\rangle \nn
\eq \langle 0_{\uparrow p=0}|{\cal H}| 0_{\uparrow p=0}\rangle  - 0 \nn
\eq \sum_{\bf k}\frac{ \left[\l1 + \frac{k_{\parallel}}{M}\Xi \l \omega_k +\frac{k^2}{2M} \r^{-1}\r^{-1}-2 \right]}{\l \omega_k +\frac{k^2}{2M} + \frac{k_{\parallel}}{M}\Xi \r} V_k^2 + \frac{\Xi^2}{2M},\nn
\label{eq:binding_energy}
\eea
where we took an expectation value using the state (\ref{eq:MF_state}) optimized according to \ceq{eq:MFpolaron}. Note that we did not include the mean field energy of the interactions between condensend bosons and the impurity $E_{\rm MF} = \frac{2\pi a_{\rm IB,\sigma}n_0}{\mu}$, in the binding energy.

The binding energy is a well defined physical observable, which must moreover be expressible in terms of the $s$-wave scattering length, by virtue of the universality of interactions in cold atoms (see Appendix A). However, a naive evaluation of the sum in \ceq{eq:binding_energy} leads to an ultraviolet (UV) divergence. The appearance of UV divergences in physical observables is a direct consequence of poorly approximating the fundamentally different physics at atomic length scales. Indeed, our zero-range model \ceq{eq:impurity_boson_int} pathologically couples microscopic degrees of freedom to the physically relevant long distance degrees of freedom. However, in order to describe universal properties which are insensitive to microscopic physics, we require a means of safely and justifiably decoupling microscopic and macroscopic scales.

To this end we found it most convenient to evalute \ceq{eq:binding_energy} using dimensional regularization~\cite{Zinn-Justin.book}, which is equivalent to the regularization scheme based on a momentum cutoff used in Refs.~\cite{astrakharchik2004motion,Tempere2009,rath2013polaron}. The regularization amounts to the subtraction of the leading divergence in the binding energy which takes the form
\bea
E_{\rm B}^{\rm div}\to-\left(\frac{2\pi a_{\rm IB,\sigma}}{\mu}\right)^2 n_0\sum_{\bf k}\frac{2\mu}{k^2}.\label{eq:subtracted_infinity}
\eea
Physically such a subtraction can be justified by considering the total interaction energy of the BEC and impurity:
\bea
E_{\rm int} = E_{\rm B} + E_{\rm MF}.
\eea 
and expressing the mean field interaction energy of the condensate in terms of the ``bare'' coupling to the impurity $g_{\rm IB,\sigma}$ from Eq.~(\ref{eq:impurity_boson_int}):
\bea
E_{\rm MF} = g_{\rm IB,\sigma}n_0.
\eea
The bare coupling can be related to the physical impurity-boson s-wave scattering length using the Lippman-Schwinger equation
\bea
\frac{1}{g_{\rm IB,\sigma}} = \frac{\mu}{2\pi a_{\rm IB,\sigma}} - \sum_{\bf k}\frac{2\mu}{k^2},
\eea
which yields the following expression for the mean-field energy, accurate to second order in $a_{\rm IB,\sigma}$:
\bea
E_{\rm MF} = \frac{2\pi a_{\rm IB,\sigma}n_0}{\mu} + \left(\frac{2\pi a_{\rm IB,\sigma}}{\mu}\right)^2 n_0 \sum_{\bf k}\frac{2\mu}{k^2}.
\eea
Indeed, the second term on the right hand side is precisely the ``subtracted infinity'' required to eliminate the diveregence (\ref{eq:subtracted_infinity}). Thus we obtain a well-behaved binding energy which can be expressed in closed form for a localized impurity with $M\to\infty$
\bea
E_{B, \rm reg.}^{\rm M \to \infty}= - \frac{2\sqrt{2}\pi a_{\rm IB,\sigma}^2n_0}{\mu\xi} < 0,\label{eq:loc_bind_en}
\eea
and must be evaluated numerically for finite mass impurities. The details of the regularization procedure used to obtain \ceq{eq:loc_bind_en} are presented in Appendix A. 

We will later need the generalized binding energy for a finite momentum polaron, i.e. \ceq{eq:binding_energy} with $p\neq 0$. As shown in Sec.~ \ref{sec:direct_RF}, the latter quantity will contribute a shift of the RF signal relative to the atomic transition rate between $\uparrow$ and $\downarrow$ of the bare impurity. 

\subsection{Effective mass of the polaron}
In the absence of interactions, the bare impurity propagates as a free particle with a quadratic dispersion $\varepsilon_{\rm I} = \frac{p^2}{2M}$. It is useful to conceptualize the polaron also as a propagating object -- a wave packet --  composed of an impurity dragging a cloud of bosonic excitations. Such a dressing of the impurity will naturally imply propagation with an effectively heavier mass. We can identify the effective mass of the polaron from its group velocity by requiring the polaron dispersion to take the form $\varepsilon_{\rm polaron} = \frac{p^2}{2M^*}$. Then from the definition of the polaron group velocity we find
\bea
 v_{\rm polaron} &\equiv& \frac{\partial}{\partial p} \varepsilon_{\rm polaron} = \frac{p}{M^*}\nn
 \eq \partial_p \left(\langle 0_{\uparrow p}|\tilde{\cal H}| 0_{\uparrow p}\rangle - \langle 0_{\uparrow p=0}|\tilde{\cal H}| 0_{\uparrow p=0}\rangle\right)\nn
 \frac{p}{M^*}\eq \frac{p}{M} - \langle 0_{\uparrow p}|\sum_{\bf k}k \hat{b}^\dagger_{\bf k}\hat{b}_{\bf k}|0_{\uparrow p}\rangle,\label{eq:v_grp_polaron}
\eea
where in the second line we expressed the polaron dispersion as the energy difference between the system at finite momentum $p$ and zero momentum. We can express \ceq{eq:v_grp_polaron} in terms of the mean field solution to find
\bea
\frac{M}{M^*} = 1-\frac{ \Xi}{p}, \label{eq:eff_mass_def}
\eea
with the parameter $\Xi$, the total momentum of the bosons, obtained by solving \ceq{eq:MFpolaron}.

Here we note an interesting feature of the mean-field treatment above. One finds that for a certain parameter regime, no mean-field solution can be found due to a singularity in the self-consistency \ceq{eq:MFpolaron}. The singularity arises when the denominator of the right hand side of \ceq{eq:MFpolaron} admits a zero for small $k$:
$$
0= \omega_{\bf k} + \frac{k^2}{2 M} -\frac{k_{\parallel}}{M} \l p-\Xi[\alpha^{\rm MF}_{\bf k}]\r \xrightarrow{k\ll 1/\xi}  c k - \frac{pk_{\parallel}}{M^*},
$$
where we used \ceq{eq:eff_mass_def} to obtain the right hand side.

Thus we find that the mean-field treatment breaks down when
\bea
v^* = \frac{p}{M^*} > c. \label{eq:mod_Landau_crit}
\eea

The criterion (\ref{eq:mod_Landau_crit}) is reminiscent of Landau's criterion for dissipationless transport through a superfluid \cite{astrakharchik2004motion}, with one important difference. The usual criterion is a purely kinematic bound obtained by weighing the relative advantage for an impurity to emit excitations, and does not include the effects of interactions. The remarkable feature of \ceq{eq:mod_Landau_crit} is the role of interactions: it is not the {\em bare impurity} velocity that is compared to the sound speed, but rather the {\em effective polaron} velocity. Due to the strong dependence of the effective mass on interactions, one finds that for a large enough interaction the {\em polaron} is subsonic, although the corresponding bare impurity in the absence of interactions would be supersonic. 

\begin{figure}[htpb]
\includegraphics[scale=.3]{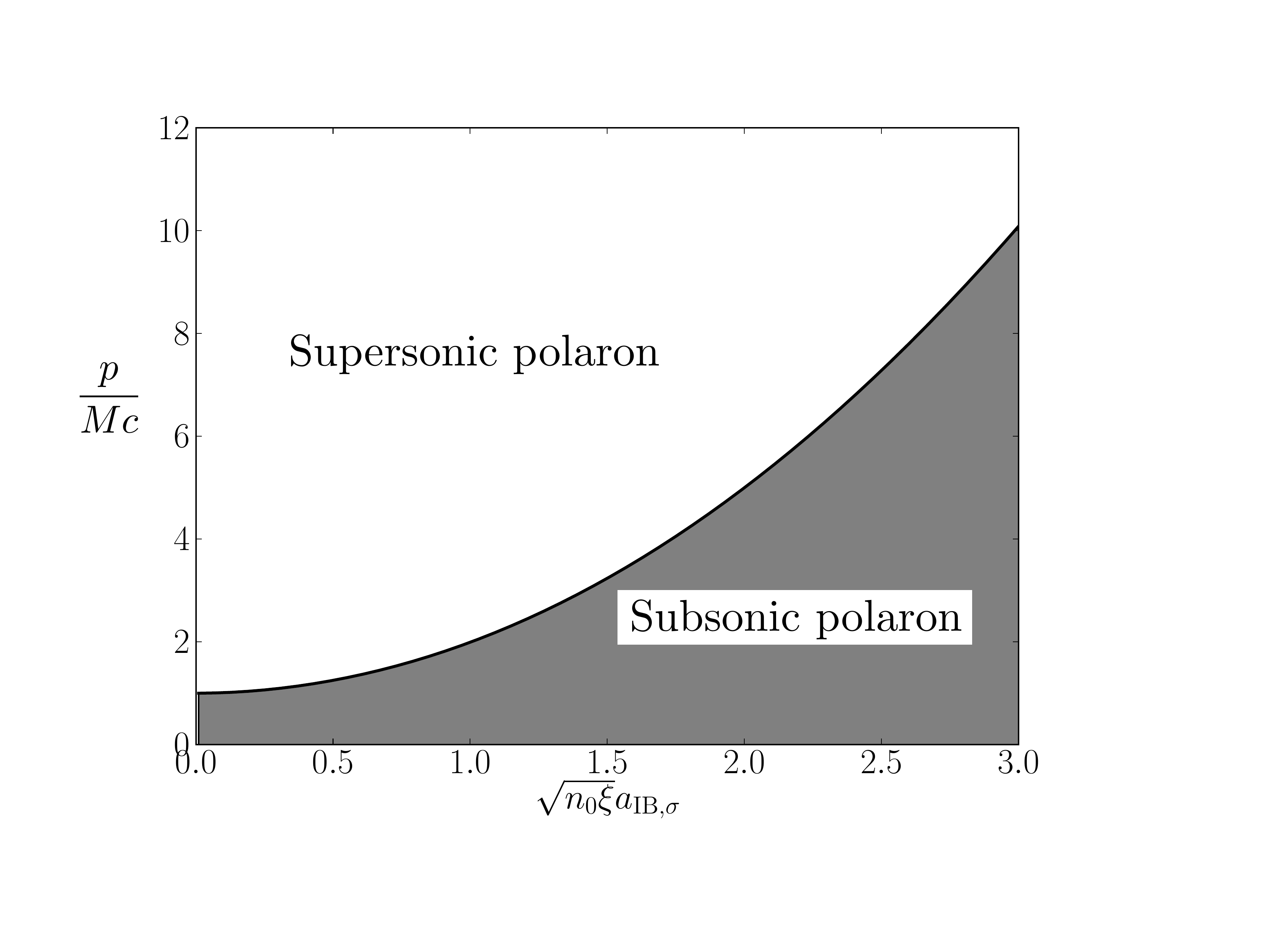}
\caption{\label{fig:subsonic_sol}  Mean-field solutions are obtained in the shaded region, while in the upper unshaded region no solutions can be found within our ansatz. The line separating the regions corresponds to the condition (\ref{eq:mod_Landau_crit}) reminiscent of the Landau criterion. In the absence of interactions the separation occurs at the usual subsonic to supersonic transition point $p/M = c$.} \end{figure}

 In Fig.~\ref{fig:subsonic_sol} we plot the critical strength of interactions for which we find polaronic solutions. We interpret the lack of solutions in the unshaded region of the figure as a break down of our ansatz; presumably the true ground state for supersonic polarons involves correlations between phonon excitations, and goes beyond the mean field description presented here. 

\subsection{Quasiparticle residue}\label{sec:qp_res}

The quasiparticle residue directly quantifies the component of the bare impurity that remains in the interacting ground state. Although it is usually extracted from the residue of the pole of the impurity Green's function~\cite{abrikosov1965quantum}, it may also be obtained as the overlap between the free and dressed impurity wavefunction. Since the impurity degrees of freedom drop out of the problem due to the Lee-Low-Pines transformation, we obtain the quasiparticle weight from the overlap of the phonon vacuum $|0\rangle$ and the interacting phonon ground state $|0_{\downarrow,p}\rangle$: 
\bea
Z \eq |\langle 0|0_{\uparrow p}\rangle|^2 \nn
\eq {\rm exp}\left[ - \sum_{\bf k} \frac{V_{\bf k}^2}{\left( \omega_{\bf k} + \frac{k^2}{2 M} -\frac{k_{\parallel}}{M} \l p-\Xi[\alpha^{\rm MF}_{\bf k}] \r\right)^2}\right]\nn
\eq {\rm exp}\left[ - \sum_{\bf k} \frac{V_{\bf k}^2}{ \left(\omega_{\bf k} + \frac{k^2}{2 M} -\frac{p k_{\parallel}}{M^*} \right)^2}\right],\label{eq:qp_residue}\eea
where we used \ceq{eq:eff_mass_def} in the last line to relate the quasiparticle weight and the effective mass.

\begin{figure}[htpb]
\includegraphics[scale=.32]{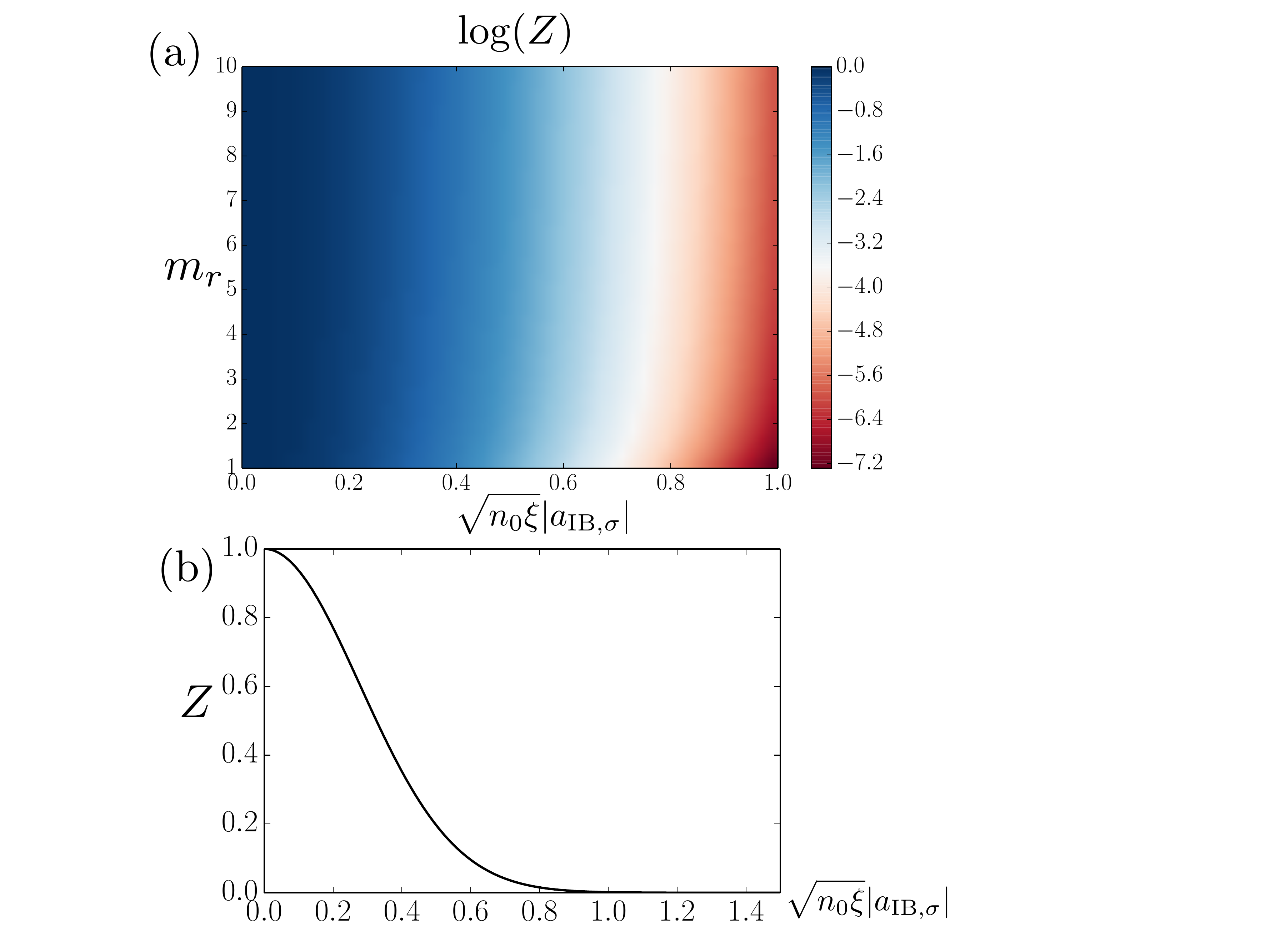}
\caption{\label{fig:qp_residue} (a)Log plot of the quasiparticle weight (which is exponentially small) as a function of interaction strength, represented by the dimensionless quantity  $a_{\rm IB,\sigma}\sqrt{n_0\xi}$, i.e. ratio between the mean free path of the impurity and the length scale over which bosons are localized), and mass ratio between impurity and bosons $m_r = M/m$.  For any moderate interaction strength, the quasiparticle weight is almost negligible, corresponding to an extremely strong renormalization of the impurity. (b) Quasiparticle weight $Z$ plotted as a function of interaction strength $a_{\rm IB,\sigma}\sqrt{n_0\xi}$ for a fixed mass ratio of $m_r = 2.5$} \end{figure}

In Fig.~\ref{fig:qp_residue}, we plotted the quasiparticle residue on a logarithmic scale, in the 3-$D$ case as a function of the impurity-BEC mass ratio, and interaction strength; strong interactions as well as small mass ratio quickly suppresses $Z$. One finds that in spatial dimensions $D=2,3$, a quantum impurity in a weakly-interacting BEC always forms a quasiparticle, although with exponentially suppressed weight for growing interaction strength. Moreover, at a given impurity-BEC interaction strength, quasiparticle residue is {\em larger} for {\em heavier} impurities, and retains a finite value even in the $M\to\infty$ limit. This should be contrasted to impurities in a Fermi gas with quasiparticle residue that has the opposite dependence on mass. In particular due to Anderson's Orthogonality Catastrophe (OC)~\cite{Anderson.OC} the quasiparticle residue $Z=0$ for localized impurities with $M\to\infty$ in a Fermi sea in 1-,2-,and 3-$D$. Interestingly, for $D=1$, the expression (\ref{eq:qp_residue}) contains an infrared divergence which again leads to $Z=0$, and signals OC even for localized impurities in 1-$D$ {\em Bose} gases. The mechanism of the OC, namely the catastrophic emission of excitations in response to an impurity, occurs independently of the exchange statistics of the many-body environment and is mainly due to the kinematic confinement of 1-$D$ systems~\cite{Tsvelik.book}.

We will in Sec.~\ref{sec:RF} show that the quasiparticle residue $Z$ is directly measurable via RF spectroscopy, and manifests as the weight of the coherent part of the signal.

\section{Analysis of RF spectra}\label{sec:RF} 

In Sec.~\ref{sec:RF_intro} we showed that in order to obtain RF spectra, the relevant quantity is the time-dependent overlap (\ref{eq:overlap}), i.e. the propagation amplitude of the initial $\downarrow$-impurity-BEC state by the Hamiltonian associated with the $\uparrow$-impurity-BEC system:
\bea
A_p(t) \eq e^{i E_{i\downarrow} t} \langle i_{\uparrow p} |
e^{ -i \tilde{\cal H}t}
| i_{\uparrow p} \rangle,\label{eq:overlap_LLP}
\eea 
where we used $|i_{\uparrow p}\rangle=\hat{V}_{\rm RF}|i_{\downarrow p}\rangle$, with $|i_{\downarrow p}\rangle$ the initial state of the $\downarrow$-impurity-BEC system at momentum $p$ energy $E_{i\downarrow}$, and $\hat{V}_{\rm RF} = |\uparrow\rangle\langle \downarrow|$. Note that in order to use the LLP transformed $\uparrow$-impurity-BEC Hamiltonian we must consider the effect of the transformation on $|i_{\uparrow p}\rangle$, however in the cases of interest to us $|i_{\uparrow p}\rangle$ involves the phonon vacuum, which is invariant under LLP. 

The RF spectral response of the impurity is simply obtained as the Fourier transform of \ceq{eq:overlap_LLP}.
 First, in Sec.~\ref{sec:RF} A we discuss general features of the time dependent overlap (\ref{eq:overlap_LLP}). In Sec~\ref{sec:RF} B,C, we explicitly calculate the overlap and corresponding RF spectra for direct and inverse RF protocols.

\subsection{Generic features of the RF response}\label{sec:universal_features}

Starting from a straightforward Lehmann expansion \cite{abrikosov1965quantum} of the RF response, and resolving the identity in terms of eigenstates $|m_{\uparrow p}\rangle$ of the time-evolving Hamiltonian, with energy $E_{m\uparrow}$, we obtain:
\bea
I(p,\omega)
\eq{\rm Re}\frac{1}{\pi} \int_0^\infty dt e^{i(\omega + E_{i\downarrow}) t} \langle i_{\uparrow p} |e^{ -i \tilde{\cal H}_\uparrow t}|i_{\uparrow p} \rangle\nn
\eq \sum_{m}{\rm Re}\frac{1}{\pi} \int_0^\infty dt e^{i(\omega + E_{i\downarrow}  - E_{m\uparrow})t} |\langle  m_{\uparrow p}|i_{\uparrow p} \rangle|^2\nn
\eq{\rm Re}\frac{1}{\pi} \int_0^\infty dt e^{i(\omega - \Delta_0)t} Z_{\uparrow\downarrow}\times\nn
&&\hspace{1.5 cm}\left(1+\sum_{ m \neq 0}e^{i\Delta_m t}\frac{ |\langle  m_{\uparrow p}|i_{\uparrow p} \rangle|^2 }{Z_{\uparrow\downarrow }}\right),\label{eq:Lehmann-analysis}
\eea
with
\bea
Z_{\uparrow\downarrow }\eq |\langle 0_{\uparrow p}|i_{\uparrow p}\rangle|^2,\Delta_m =  E_{m \uparrow} - E_{i\downarrow},\label{eq:qp_delta}
\eea
where $|0_{\uparrow p}\rangle$ is the ground state of the $\uparrow$-impurity-BEC Hamiltonian (\ref{eq:transformed_Hamiltonian}). 

We expect the low energy contribution to $I(p,\omega)$ to be dominated by the long time limit of the integrand for which, due to dephasing, we find:
\bea
I\left(p,\omega \ll \frac{c}{\xi}\right) \eq \lim_{t\to\infty} Z_{\uparrow\downarrow}\left(1+\sum_{ m \neq 0}e^{i\Delta_m t}\frac{ |\langle m_{\uparrow p}|i_{\uparrow p} \rangle|^2 }{Z_{\uparrow\downarrow }}\right)\nn
&\to& Z_{\uparrow\downarrow}.\nn 
\eea
This dephasing mechanism separates a coherent and incoherent contribution which constitute the total RF signal:
\bea
I(p,\omega) = I_{\rm coh}(p,\omega) + I_{\rm incoh}(p,\omega),
\eea
with the coherent part given by
\bea
I_{\rm coh}(p,\omega-\Delta_0)= Z_{\uparrow\downarrow}\delta(\omega - \Delta_0).\label{eq:gen_coherent_response}
\eea

From \ceq{eq:qp_delta} we find that the weight of the coherent peak of the impurity RF response is determined by the overlap between the initial state of the $\downarrow$-impurity-BEC system, and the {\em ground state} of the final $\uparrow$-impurity-BEC system (the RF operator $\hat{V}_{\rm RF}$ abruptly changes the impurity internal state, but otherwise leaves the impurity-BEC state unmodified, i.e.  $|i_{p\downarrow}\rangle \to |i_{p\uparrow}\rangle$ must be thought of as a sudden quench). The center of the peak occurs at the energy difference between the initial and final states $E_{0,\uparrow}-E_{i\downarrow}$ measured with respect to the bare atomic transition rate of the impurity between its internal states.

In the case of the direct and inverse RF protocols considered here, the weight of the coherent peak is in fact the quasiparticle weight $Z$ defined in \ceq{eq:qp_residue}. Indeed, for the direct RF protocol the impurity is initially in the polaronic ground state $|i_{\downarrow p}\rangle = |0_{\downarrow p}\rangle$, while the ground state of the non-interacting $\uparrow-$impurity-BEC system is decoupled, i.e. in this case $|0_{\uparrow p}\rangle = |{\bf p}\rangle_{\uparrow}\otimes|0\rangle$, thus
\bea
\label{eq:Z_direct}
Z^{\rm direct RF}_{\uparrow\downarrow} = |\langle 0|0_{\uparrow p}\rangle|^2.
\eea
For the inverse RF protocol the $\downarrow-$impurity is initially non-interacting with the bosons, and after the RF spin-flip, $|i_{\uparrow p}\rangle = |{\bf p}\rangle_{\uparrow}\otimes|0\rangle$, while the ground state of the interacting $\uparrow-$impurity-BEC system is the polaronic ground state $|0_{\uparrow p}\rangle$, leading to 
\bea
\label{eq:Z_inverse}
Z^{\rm inverse RF}_{\uparrow\downarrow}= |\langle 0_{\uparrow p}|0\rangle|^2.
\eea
Since the impurity degrees of freedom drop out of the problem due to the LLP transformation, in both Eqs.~(\ref{eq:Z_direct}) and (\ref{eq:Z_inverse}), the overlap between initial state and final ground state defined in \ceq{eq:qp_delta} reduces to the overlap of the {\em phonon vacuum} $|0\rangle$ and the interacting {\em phonon} ground state $|0_{\uparrow p}\rangle$ (see also Sec.~\ref{sec:qp_res}).

Although the Lehmann analysis (\ref{eq:Lehmann-analysis}) demonstrates the existence of an incoherent contribution to the RF signal, it does not specify its structure without additional knowledge about the many body eigenstates of the system. Interestingly, again for the particular case where one of the two internal states of the impurity is non-interacting with the BEC, the asymptotic behavior of the incoherent part of the RF is also constrained by exact relations. 

This fact was demonstrated e.g. by the authors of Refs.~\cite{schneider2010universal,langmack2012clock}, by relating the high-frequency impurity RF response to the momentum distribution of the many-body system $n(k)$. Fermi's golden rule for the RF transition rate of impurity atoms between non-interacting and interacting internal states can be expressed as the convolution\cite{langmack2012clock} of the free propagator of the impurity in the non-interacting state, and its spectral function $A(k,\omega) = -2{\rm Im}G(k,\omega)$ in the interacting state, where $G$ is the interacting Green's function:
\bea
I(\omega)
\eq \sum_{\bf k} \int d\Omega A(k,\Omega)n(\Omega)\delta(\Omega - \omega - \varepsilon_k ).\label{eq:convolution}
\eea
Here $n(\Omega)$ is the distribution function of the many-body environment at energy $\Omega$. To isolate the high frequency contribution, one can integrate the expression 
\ceq{eq:convolution} by parts, and use the sum rule $\int d\Omega A({\bf k},\Omega)n(\Omega) = n({\bf k})$~\cite{abrikosov1965quantum} to obtain
\bea
I(\omega \to \infty) \approx \sum_{\bf k} n({\bf k} \to \infty) \delta(\omega - \varepsilon_{\bf k}),
\eea
where $n(k)$ is the momentum distribution of the many-body environment of the impurity. The authors of Refs.~\cite{schneider2010universal,langmack2012clock} considered RF spectroscopy of fermions, but in the expression above, exchange statistics only enter through $n(k)$. Interestingly the large momenta structure of $n(k)$, which determines the high frequency RF response, is {\em insensitive} to exchange statistics~\cite{combescot2009particle,braaten2011universal} and allows us to directly generalize the argument for bosons. In particular, for large momenta $n(k)$ displays a universal power-law tail~\cite{tan2008large,tan2008energetics,braaten2008exact,braaten2011universal}:  
\bea
n({\bf k}\to\infty) \to C/k^4.\label{eq:contact_tail}
\eea
This form was discovered by Tan~\cite{tan2008large,tan2008energetics} who identified the ``contact'' $C$ as the density of pairs of atoms, whose binary collisions are responsible for the emergence of this universal feature. The asymptotic behavior (\ref{eq:contact_tail}) of the momentum distribution in turn constrains the asymptotic behavior of the RF response:
\bea\label{eq:incoh_gen}
I(\omega\to\infty) \propto   \begin{cases}
   C\omega^{-3/2} & \text{in 3-$D$},\\
   C\omega^{-2}  & \text{in 2-$D$},\\
  \end{cases}
\eea
leading to universal high-frequency RF tails that have been noted in various contexts for systems of interacting bosons and fermions\cite{punk2007theory,haussmann2009spectral,schneider2010universal,braaten2011universal,langmack2012clock}. 

Dimensionality of the system plays a crucial role in determining the precise form of the RF singal. For the high frequency incoherent part of the RF discussed above, different power law tails emerged in 2-$D$ and 3-$D$, due to the dimensional dependence of the many-body density of states. Moreover, as discussed in Sec.~\ref{sec:qp_res} the quasiparticle weight, $Z$, which controls the coherent part of the RF signal, attains a finite albeit exponentially small value in 2-$D$, and 3-$D$, while it displays a characteristic infrared divergence in 1-$D$. The latter phenomenon signals the orthogonality catastrophe intrinsic to the kinematically constrained phase space of 1-$D$ systems. Here, the spectrum is dominated by a power-law decay (the 1$D$ generalization of the incoherent part adds a subleading $1/\omega$ correction to the leading log-divergence):
\bea
I(\omega-\Delta) \approx C|\omega-\Delta|^{-\alpha},
\eea
where the exponent $\alpha(a_{IB})$ depends on the phase shift induced by scattering of the impurity \cite{Knap2012} and within our formalism is given by the first order Born result $\alpha \sim n_0^2 a_{\rm IB,\uparrow}^2.$

\begin{figure}[htpb]
\includegraphics[scale=.35]{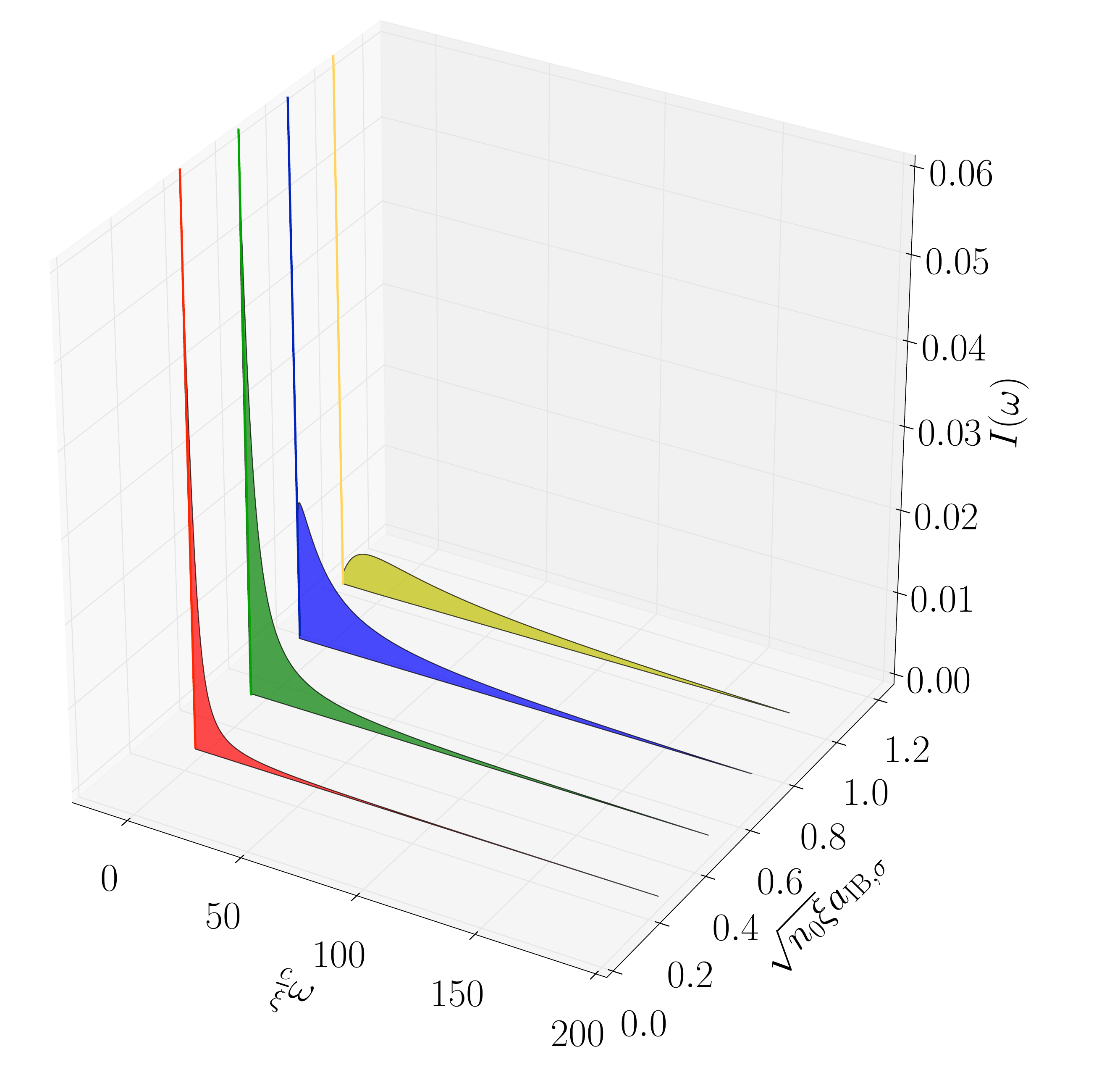}
\caption{\label{fig:fig3-spectraaib} RF spectra for different initial impurity interaction strengths. The quantity $a_{\rm IB,\sigma}\sqrt{n_0\xi}$ is a dimensionless ratio between the mean free path of the impurity and the length scale over which bosons are localized (a non-interacting BEC has completely delocalized bosons). We observe that the spectral weight starts almost entirely in the coherent part of the spectrum, corresponding to a nearly free impurity, and gradually shifts to higher energies as more excitations of the BEC are generated by increasing impurity-bose interactions. The spectra presented above were obtained for an experimentally relevant mass ratio $M/m$ of 2.5; there is a weak dependence of the spectra on mass ratio, and is not observable on the scale shown here. } \end{figure}

With this general phenomenology of the RF response in mind, we performed a detailed microscopic calculation of the time dependent overlap (\ref{eq:overlap}) by generalizing the mean-field approach to polaron ground states of Sec.~\ref{sec:GS_prop} to the problem of impurity dynamics.

\subsection{Direct RF: Transition from interacting to non-interacting state}\label{sec:direct_RF}

In the direct RF measurement, the system is first adiabatically prepared in the polaronic ground state, i.e. $|i_{\downarrow p}\rangle =  |0_{\downarrow p}\rangle$. Since the system is non-interacting in its final state, the time evolving Hamiltonian in this case is simply that of free Bogoliubov bosons, ${\cal H}_b = \sum_{\bf k}\omega_{\bf k}\hat{b}^\dagger_{\bf k}\hat{b}_{\bf k}$. 

We showed in Sec.\ref{sec:GS_prop} that the ground state can be approximated as a product of coherent states, see \ceq{eq:MF_state}, which moreover becomes exact in the case of an infinitely heavy impurity. Thus the problem of calculating the time-dependent overlap reduces to free evolution of product coherent states:
\bea
A_p(t)\eq \langle 0_{\uparrow p}|e^{-i{\cal H}_b t}|0_{\uparrow p}\rangle\nn
\eq \prod_{\bf k} \langle 0|e^{\alpha_{\bf k}^{\rm MF} \hat{b}^\dagger_{\bf k}e^{-i\omega_{\bf k}t} - (\alpha^{\rm MF}_{\bf k})^*\hat{b}_{\bf k}e^{i\omega_{\bf k} t}}|0\rangle,\label{eq:direct_td_overlap}
\eea
with $\alpha_{\bf k}^{\rm MF}$ obtained from solving \ceq{eq:MFpolaron}; in the limit of a localized impurity with $M\to\infty$, $\alpha_{\bf k} \to -\frac{V_{\bf k}}{\omega_{\bf k}}$, and there one obtains the exact solution to the time dependent overlap. 

We find that the overlap amplitude decays quickly from unity to an exponentially small limiting value with an oscillatory envelope:
\bea
\label{eq:overlap_limit}
&&A_p(t\to\infty)\to  Ze^{-i\Delta t},\ \ \Delta =\Delta_1 + \Delta_2,\nn
&&Z = \exp\left[-\sum_{\bf k} \frac{V_{\bf k}^2}{\left(\omega_{\bf k}+\frac{k^2}{2M} -\frac{\bf p.k}{M^*}\right)^2}\right], \nn
&&\Delta_1 = \sum_{\bf k} \frac{V_{\bf k}^2}{(\omega_{\bf k}+\frac{k^2}{2M} -\frac{\bf p.k}{M^*})^2} +   \frac{2\pi}{\mu}n_0a_{\rm IB,\sigma},\nn
&& \Delta_2 =  \frac{p^2}{2M}\left(1-\frac{M}{M^*}\right)
\eea
Here $Z$ is the quasiparticle residue defined in \ceq{eq:qp_residue}, and is in agreement with the general analysis of Sec.~\ref{sec:universal_features}. $\Delta$ denotes the energy difference between interacting and non-interacting ground states, and consists of two contributions: $\Delta_1$ includes the ``mean-field'' shift due to the interaction of impurity with the static BEC ground state, and the finite momentum generalization of the binding energy defined in \ceq{eq:binding_energy}, and $\Delta_2$ which accounts for the change in effective mass of the impurity. As in the ground state case, the (generalized) binding energy was regularized as described in Appendix A. 

The RF absorption spectrum can be simply obtained by Fourier transforming \ceq{eq:direct_td_overlap}. We present a few sample spectra in Fig. \ref{fig:fig3-spectraaib}. The RF absorption spectrum of the impurity contains a coherent and incoherent contribution as expected from the general analysis presented in Sec.~\ref{sec:universal_features}
$$
I(p,\omega) = I_{\rm coh}(p,\omega) + I_{\rm incoh}(p,\omega).
$$
The coherent peak is determined entirely by the long time limit of \ceq{eq:direct_td_overlap} which is the quasiparticle residue defined in \ceq{eq:qp_residue}. 
\bea
\label{eq:RF_coh}
I_{\rm coh}(p,\omega - \Delta) = Z\delta(\omega - \Delta),
\eea
with $\Delta$ defined in \ceq{eq:overlap_limit}

The spectrum contains additionally a broad incoherent part corresponding to the short time dynamics of polaron destruction due to excitations generated when the impurity-BEC interactions are removed in the course of the direct RF:
\bea
\label{eq:RF_incoh}
I_{\rm incoh}(p,\omega - \Delta) = \frac{\rm Re}{\pi} \int\limits_0^\infty dt'e^{i(\omega - \Delta)t}\left(A_p(t)e^{i\Delta t} - Z\right).
\eea

For concreteness, we present the leading high and low frequency behavior of the RF spectrum in the exactly solvable case of a localized impurity; it is straighforward but tedious to obtain identical results for mobile impurities. By expanding the exponential in \ceq{eq:RF_incoh} to leading order, we can approximate \ceq{eq:RF_incoh} using 
\bea
I_{\rm incoh}(\omega - \Delta) &\approx&  {\rm Re}\frac{Z}{\pi} \int_0^\infty dt e^{i(\omega - \Delta)t}\sum_{\bf k} \left|\frac{V_{\bf k}}{\omega_{\bf k}}\right|^2e^{-i\omega_{\bf k} t}\nn
\eq Z \sum_{\bf k} \left|\frac{V_{\bf k}}{\omega_{\bf k}}\right|^2\delta(\omega - \Delta - \omega_{\bf k})\nn
\eq Z\int \frac{d \Omega}{2\pi^2} \frac{(\sqrt{2 \Omega^2 + 1} - 1)^{d/2}}{\Omega^2 \sqrt{2\Omega^2+1}}\delta(\omega - \Delta - \Omega)\nn
\eq\frac{Z}{2\pi^2} \frac{(\sqrt{2 (\omega-\Delta)^2 + 1} - 1)^{d/2}}{(\omega-\Delta)^2 \sqrt{2(\omega - \Delta)^2+1}}.\eea
Thus we find the following limiting behaviors of the incoherent RF response:
\bea
\label{eq:RF_asymptotics}
I_{\rm incoh}\left(\omega - \Delta \gg \frac{c}{\xi}\right) &\propto&
  \begin{cases}
   (\omega-\Delta)^{-3/2},  & \text{in 3-$D$}.\\
   (\omega-\Delta)^{-2}  & \text{in 2-$D$},\\
  \end{cases}\\
I_{\rm incoh}\left(\omega - \Delta \ll \frac{c}{\xi}\right) &\propto&
  \begin{cases}
   (\omega-\Delta),  & \text{in 3-$D$}.\\
   C_1 + C_2(\omega-\Delta)^{2}  & \text{in 2-$D$}.\\
  \end{cases}
\eea

We see that the high frequency tails of the RF spectra in Eqs.~(\ref{eq:RF_incoh})-(\ref{eq:RF_asymptotics}) are in agreement with the general functional form required by Eq.~(\ref{eq:incoh_gen}). This provides a non-trivial consistency check to our microscopic approach. We now generalize our approach to consider the more complicated dynamics involved in the inverse RF measurement.

\subsection{Inverse RF: Transition from noninteracting to interacting state}\label{sec:inv_RF}

In the inverse RF measurement impurities are transferred from an initially non-interacting state to an interacting state, with $a_{\rm IB,\uparrow}$ finite and $a_{\rm IB,\downarrow}\approx 0$. We again consider the time dependent overlap (\ref{eq:overlap}), but the associated dynamics cannot be reduced to free evolution as in the direct RF in Sec.~\ref{sec:direct_RF}. However, the case of the localized impurity is once again amenable to an exact solution, and inspires an approximate treatment of the mobile impurity.

\subsubsection{Dynamics of a localized impurity}
Like the ground state of the localized impurity-BEC system, the time evolving wavefunction of the system is also a product of coherent states, but with {\em time dependent} parameters. 

The initial free Hamiltonian ${\cal H}_b$ is modified after the switch on of interactions to ${\cal H}_b + {\cal H}_{\rm int}$. Crucially, the two Hamiltonians are related by a canonical transformation. We introduce the displacement operators
$D(\alpha)=e^{\sum_{\bf k}(\alpha_{\bf k} \hat{b}_{\bf k}^\dagger -\alpha^*_{\bf k} \hat{b}_{\bf k})}$
which shift the mode operators 
$$
D^{-1}(\alpha) \hat{b}_{\bf k} D(\alpha)=\hat{b}_{\bf k} + \alpha_{\bf k}.
$$
Then, for the appropriate choice of shift $\alpha_{\bf k} = \frac{V_{\bf k}}{\omega_{\bf k}},$ we find $D^{-1}({\cal H}_b + {\cal H}_{\rm int})D = {\cal H}_b + \Delta$, with $\Delta$ a constant number. Thus we can directly solve the time-evolution of the initial state using the displacement operators as follows:
\begin{eqnarray}
&&|\phi_{M\to\infty}(t)\rangle = e^{i({\cal H}_b + {\cal H}_{\rm int})t}|0 \rangle\nn
&&=e^{-i\Delta t} D^{-1}\left(\frac{V_{\bf k}}{\omega_{\bf k}}\right)e^{iH_bt}D\left(\frac{V_{\bf k}}{\omega_{\bf k}}\right)|0\rangle\nn
%&&=e^{-i\Delta t} \prod_{\bf k}e^{\frac{V_{\bf k}}{\omega_{\bf k}}(b_{\bf k} - b_{\bf k}^\dagger)}e^{i\omega_{\bf k}b^\dagger_{\bf k}b_{\bf k}t} e^{\frac{V_{\bf k}}{\omega_{\bf k}}(b_{\bf k}^\dagger-b_{\bf k})}e^{-i\omega_{\bf k}b^\dagger_{\bf k}b_{\bf k} t}|0\rangle\nn
&&= e^{-i\Delta t} \prod_{\bf k}e^{\frac{V_{\bf k}}{\omega_{\bf k}}(\hb_{\bf k} - \hb_{\bf k}^\dagger)} e^{\frac{V_{\bf k}}{\omega_{\bf k}}(\hb_{\bf k}^\dagger e^{-i\omega_{\bf k} t}-\hb_{\bf k}e^{i\omega_{\bf k} t})}|0\rangle,\nonumber
\eea
leading to an expression for the wavefunction of the form:
\bea
|\phi_{M\to\infty}(t)\rangle = e^{-\Psi(t) - i\Delta t}\prod_{\bf k} e^{\frac{V_{\bf k}}{\omega_{\bf k}}(e^{-i\omega_{\bf k}t}-1)b^\dagger_{\bf k}}|0\rangle,\label{eq:time_evo2}
\eea
with
\bea
\Psi(t) \equiv \sum_{\bf k}\left|\frac{V_{\bf k}}{\omega_{\bf k}}\right|^2(1-e^{-i\omega_{\bf k} t}), \ \Delta \equiv\sum_{\bf k} \frac{V_{\bf k}^2}{\omega_{\bf k}} + \frac{2\pi}{\mu} a_{\rm IB,\downarrow}n_0.\nonumber
\end{eqnarray}

\subsubsection{Dynamics of a finite mass impurity}

 Inspired by the exact time evolving wavefunction of the localized impurity-BEC system, a product of time dependent coherent states, we make an analogous ansatz for finite mass impurity-BEC system:
\begin{eqnarray}
|\phi (t)\rangle = e^{-i\chi (t)}e^{\sum_{\bf k}\alpha_{\bf k}(t)\hb^{\dagger}_{\bf k} - \frac{1}{2}|\alpha_{\bf k}(t)|^2}|0\rangle. \label{eq:DMF_state}
\end{eqnarray}
The variational wavefunction (\ref{eq:DMF_state}) represents a mean-field approach to dynamics: the wavefunction factorizes for individual phonons, so each phonon indexed by momentum ${\bf k}$ evolves in an effective time-dependent oscillator Hamiltonian, whose frequency $\omega_{\bf k}(t)$ is renormalized by the other phonon modes.

Projecting the Schr\"odinger equation onto the variational state (\ref{eq:MF_state}) (see e.g. \cite{altman2005superfluid,demler2011semiclassical}) we obtain equations of motion for the variational coherent state parameters:
\begin{eqnarray}
\label{eq:DMF_phonons_EOM}
\dot \chi(t) \eq \frac{p^2}{2M} - \sum_{\bf k,k'}\frac{\bf k.k'}{2M} |\alpha_{\bf k}|^2|\alpha_{\bf k'}|^2 + \frac{1}{2}\sum_{\bf k}V_{\bf k}(\alpha_{\bf k} + \alpha^*_{\bf k}),\nn
i\dot\alpha_{\bf k}(t) \eq \left(\Omega_{\bf k} - \frac{\bf p.k}{M}+ \frac{\bf k}{M} .\sum_{\bf k'} {\bf k'} |\alpha_{\bf k'}(t)|^2\right) \alpha_{\bf k}(t) + V_{\bf k},\nn
\end{eqnarray}
with $\Omega_{\bf k} = \omega_{\bf k} + \frac{k^2}{2M}$.

% While a formal analytic solution of these equations can be obtained
%\bea
%\alpha_{\bf k}(t) \eq iV_{\bf k} \int_0^t dt_1\cos\left[\int_{t_1}^t dt'\tilde\omega_{\bf k}(t')\right],\nn
%\chi(t) \eq  \sum_{\bf k} V_{\bf k}^2 \int\limits_0^tdt_1\chi_{\bf k}(t_1)+\frac{P^2}{2M}t+\int\limits_0^t dt'\frac{(P - P_I(t'))^2}{2M},\nn
%\chi_{\bf k}(t)\eq\int\limits_0^{t}dt_1 \sin\left[\int\limits_{t_1}^{t} dt'\tilde\omega_{\bf k}(t')\right],\nn
%\tilde\omega_{\bf k}(t) \eq\omega_{\bf k} + \frac{k^2}{2M} - \frac{\bf p.k}{M}+ \frac{\bf k}{M} .\sum_{\bf k'} {\bf k'} |\alpha_{\bf k'}(t)|^2,
%\eea
We solved the differential \ceq{eq:DMF_phonons_EOM} numerically using a standard computational package~\footnote{Solutions of Eqs.~(\ref{eq:DMF_phonons_EOM}) are naively UV divergent. Imposing a sharp cut-off gives rise to unphysical oscillations at the cut-of frequency. To avoid this problem we introduced a soft cut-off $V_k \rightarrow V_k e^{-k^2/2\Lambda^2}$, choosing $\Lambda$ large enough to obtain converged results for relevant observables.}. We found that the inverse RF spectrum is qualitatively quite similar to the direct RF spectrum calculated in the previous subsection. In light of the general phenomenology of RF responses presented in Sec.~\ref{sec:universal_features}, the similarity between the two RF spectra is not surprising, since both involve transitions between interacting and non-interacting impurity-BEC states, which constrains the high and low frequency parts of the RF response. 

\subsubsection{Dynamical ansatz as optimal estimate of time-dependent overlap}
Here we demonstrate that the time-dependent mean-field approach, which is tailored to solve the {\em general} dynamics of the interacting Hamiltonian, gives a good semiclassical approximation to the specific propagation amplitude in \ceq{eq:overlap}. Using the LLP transformation, this amplitude can be written as 
\bea
\label{eq:time-dep-preaction}
A_p(t)\eq \langle i_{\uparrow p}|e^{-i{\cal H}t}|i_{\uparrow p}\rangle =  \langle 0|e^{-i{\cal H}t}|0\rangle\\ 
\eq \langle 0|e^{-i\left[\frac{1}{2M}\left(p - \sum_{\bf k}{\bf k}\hat{b}^\dagger_{\bf k}\hat{b}_{\bf k}\right)^2 + \sum_{\bf k}(\omega_{\bf k}\hat{b}^\dagger_{\bf k}\hat{b}_{\bf k}+V_{\bf k}(\hat{b}^\dagger_{\bf k} + \hat{b}_{\bf k}))\right]t}|0\rangle, \nonumber
\eea
where the phonon vacuum $|0\rangle$, is time evolved by the Hamiltonian (\ref{eq:transformed_Hamiltonian}) for a given time $t$, and the overlap of the resulting state is measured with respect to the initial vacuum.

 As an alternative approach to calcuating such a propagation amplitude, we may formulate \ceq{eq:time-dep-preaction} as a path integral, i.e. a sum over configurations of the semi-classical velocity profile of the impurity, and compare the mean-field ansatz with the saddle point of such a path integral (see Appendix B for more details).
 
  We obtain the path integral formulation by introducing into the time-dependent overlap (\ref{eq:time-dep-preaction}) a classical field $\varphi(t)$, corresponding to the fluctuating impurity velocity. This is justified by the Hubbard-Stratonovich (HS) identity, which is typically used in equilbrium quantum field theory to decouple interacting systems by using a random variable to mimic fluctuations of the system. In a similar spirit, we use $\varphi(t)$ to decouple the interaction between bosons in \ceq{eq:time-dep-preaction} and introduce a corresponding path integral to sum over all configurations of $\varphi(t)$:   
\bea
\label{eq:time-dep-pathint}
&& A_p(t) = \int{\cal D}[\varphi(t)] e^{i\int_0^t dt' \frac{M}{2}\varphi(t')^2}\\
&\times& \langle 0| e^{-i\int_0^t dt'\left[\varphi(t').\left(p - \sum_{\bf k}{\bf k}\hat{b}^\dagger_{\bf k}\hat{b}_{\bf k}\right)- \sum_{\bf k}(\omega_{\bf k}\hat{b}^\dagger_{\bf k}\hat{b}_{\bf k}+V_{\bf k}(\hat{b}^\dagger_{\bf k} + \hat{b}_{\bf k}))\right]}|0\rangle. \nonumber 
\eea
As seen above, the HS decoupling reduces the originally interacting bosonic Hamiltonian to a quadratic form, allowing us to integrate out the bosons exactly. We may then approximate the resulting path integral, now over $\varphi(t)$ alone, by a saddle point treatment:
\bea
\label{eq:saddle_pt_eq}
\varphi_s(t') \eq \frac{{\bf p}}{M} + \sum_{\bf k} \frac{V_{\bf k}^2{\bf k}}{M}\int_{t'}^t dt_1 \int_0^{t'} dt_2\nn
&&\hspace{1 cm}\times e^{-i\int_{t_2}^{t_1} dt'' (\omega_{\bf k} - {\bf k}.\varphi_s(t''))}.
\eea

\begin{figure}[h!]
 \includegraphics[scale=.32]{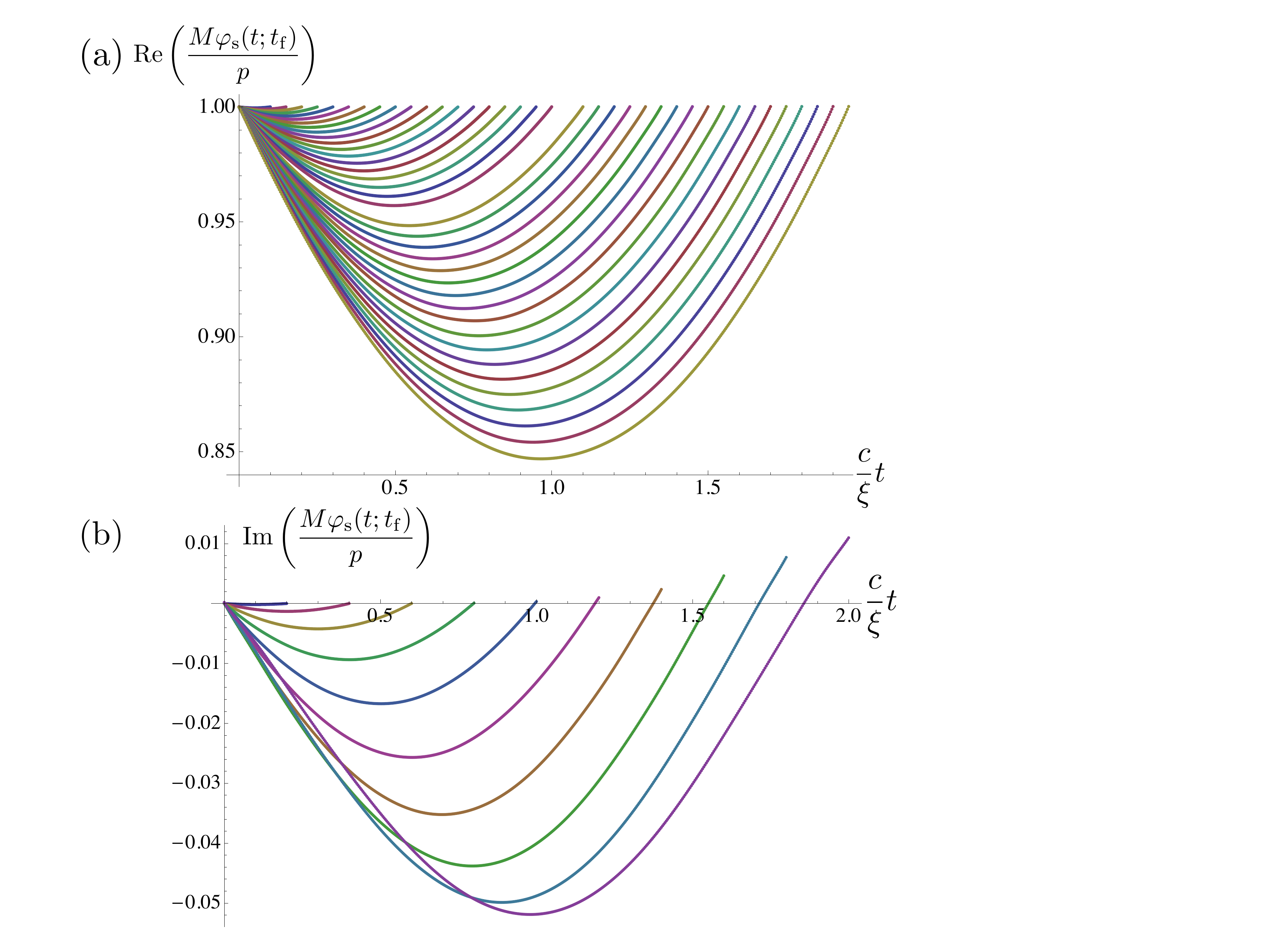}
 \caption{\label{fig:sp_traj}(a) Real part of the (rescaled) solution of saddle point Eq.~(\ref{eq:saddle_pt_eq}) plotted for $m_r = 75.0, \sqrt{n_0\xi}a_{\rm IB,\sigma} = .25, p/Mc = .6$; we obtained a family of trajectories parameterized by $t_{\rm f}$, the total propagation time for which the amplitude \ceq{eq:time-dep-preaction} was required. Each individual trajectory is a time-evolving function of $t < t_{\rm f},$ and can be interpreted (after rescaling) as the time-dependent momentum of the impurity. Note the symmetry of the saddle point trajectories around $t = t_{\rm f}/2$, which arises because they optimize \ceq{eq:time-dep-preaction}, the amplitude for a time-evolving state to return to its initial value. This is in contrast to the time-dependent mean-field solution which simply propagates forward to the steady state at time $t_{\rm f}$ (c.f. Fig.~\ref{fig:posc}). (b) Imaginary part of the saddle point trajectories are shown for the same parameters. The imaginary part shares the symmetry property of the real part, but is typically smaller in magnitude. While it does not lend itself to direct interpretation as the physical momentum of the impurity, it is necessary to properly optimize the propagation amplitude when expressed as a path integral \ceq{eq:time-dep-pathint}.} \end{figure}

 The details of our derivation of \ceq{eq:time-dep-pathint} and its saddle point Eq.~(\ref{eq:saddle_pt_eq}) are provided in Appendix~B. Our saddle point approximation yields an optimal $\varphi_s(t)$, shown in Fig.~\ref{fig:sp_traj}, which we can then use to evaluate the time-dependent overlap \ceq{eq:time-dep-preaction}. We checked that this approach is in agreement with the results of the time-dependent mean-field analysis, but at significantly greater numerical effort.
 
 Thus we conclude that the mean-field ansatz for the dynamics of the impurity, optimally estimates the RF response. In the remainder we present the main features of the dynamical mean-field solution.  

\begin{figure}[htpb]
\includegraphics[scale=.28]{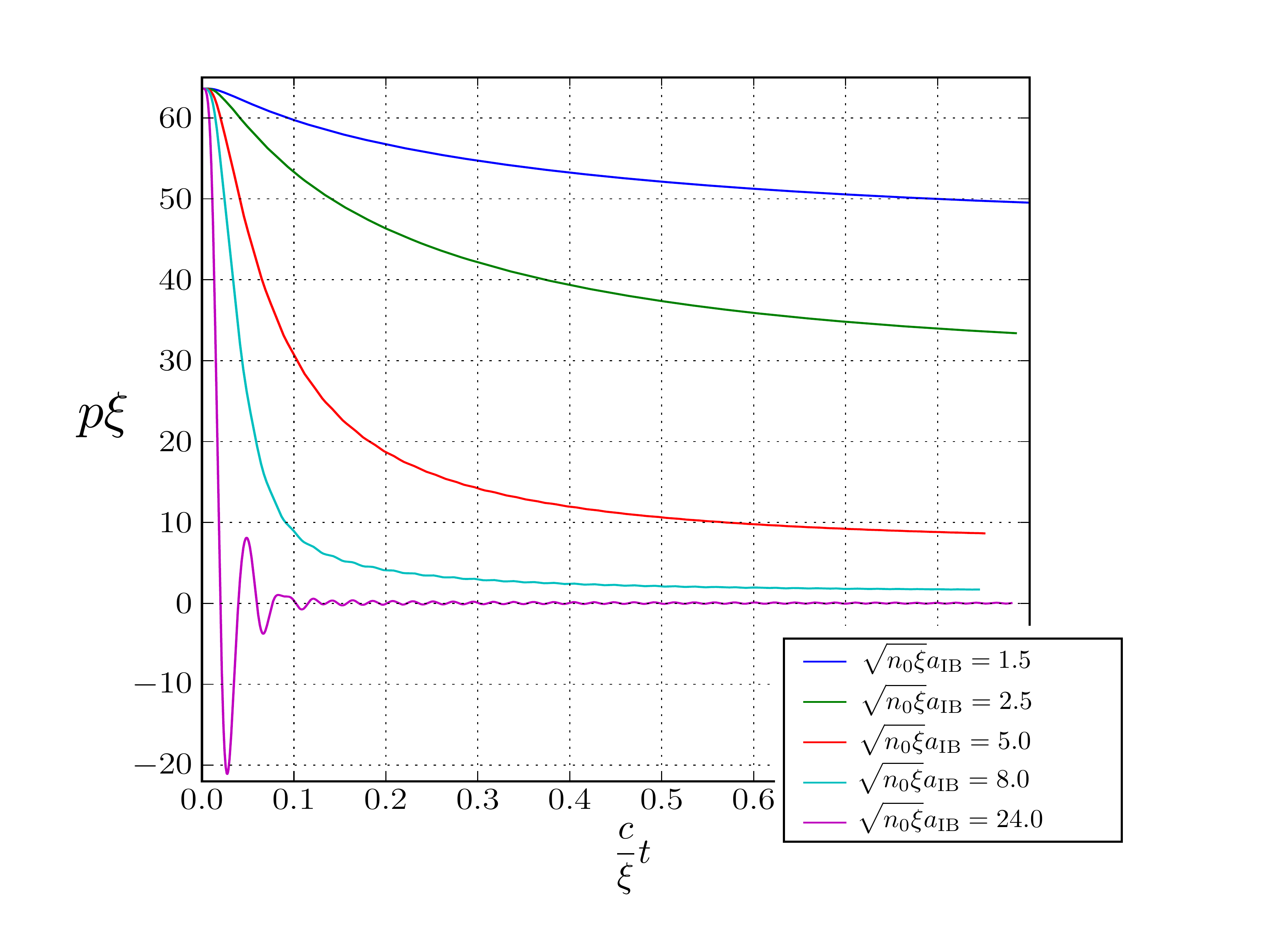}
\caption{\label{fig:posc}Impurity momentum as a function of $t$ after switching on interactions. Strong interactions lead to small asymptotic impurity momentum (corresponding to heavy effective mass). Additionally the momentum develops decaying oscillations associated with internal mode of the polaron.} \end{figure}

\begin{figure}[htpb]
\includegraphics[scale=.28]{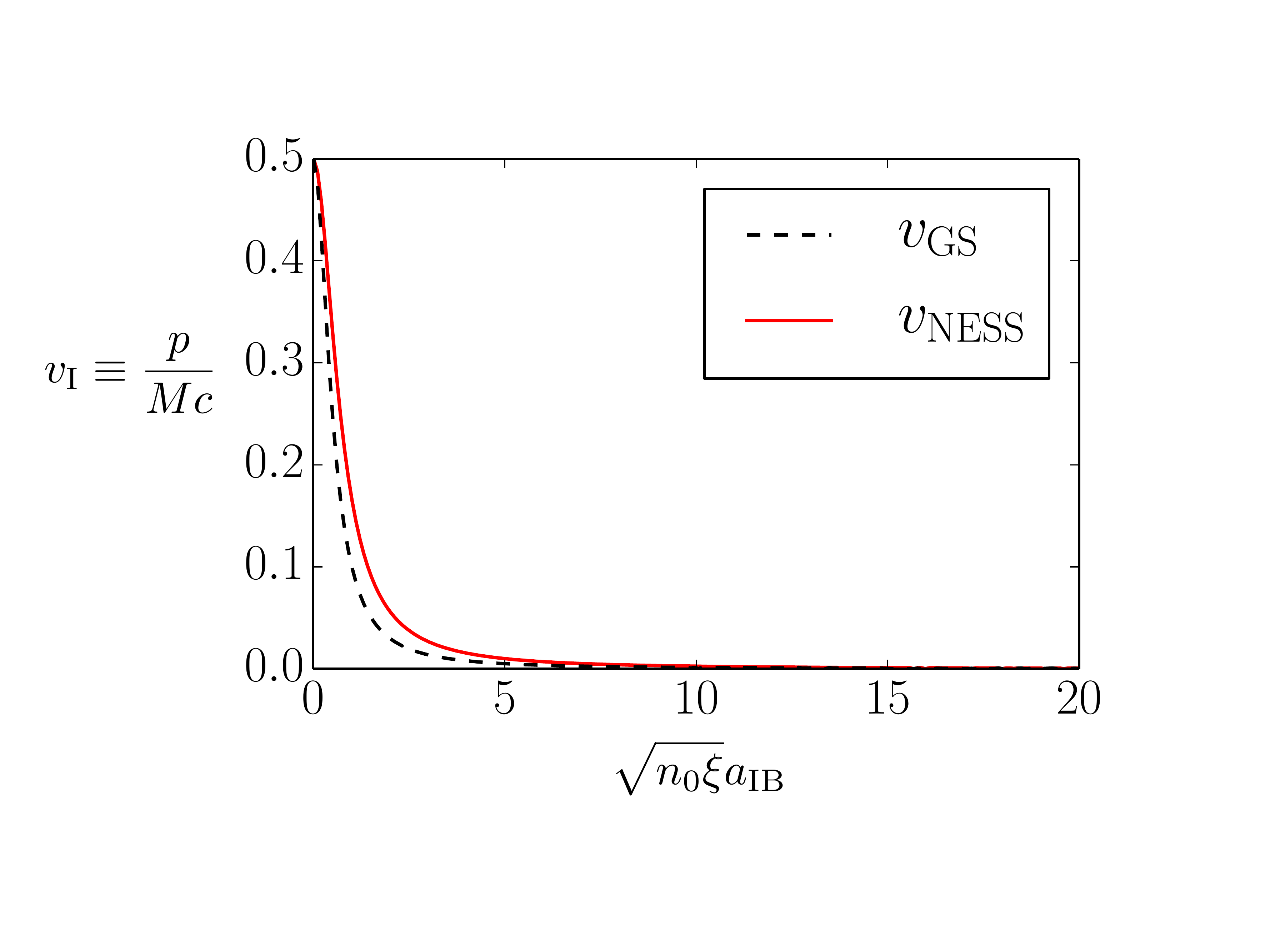}
\caption{\label{fig:vstar} The asymptotic velocity attained by the impurity as a function of impurity-BEC interaction $a_{\rm IB}$, for a given initial momentum in the non-equilibrium steady-state (NESS, solid red) and the ground state (GS, dashed red).} \end{figure}

\subsubsection{Inverse RF and non-equilbrium dynamics}

Although the prominent features of the RF spectrum appear identical for the direct and inverse RF there are differences in the details: both measurements involve Hamiltonian evolution of a non-eigenstate (see \ceq{eq:overlap}), however the inverse measurement involves more complicated dynamics compared to the direct RF; the dynamics of the latter are trivially determined by a non-interacting Hamiltonian (see Sec.~\ref{sec:direct_RF}). However, due to the strong impurity renormalization by BEC interactions, the complicated non-equilibrium dynamics of the impurity do not manifest in spectra, which are enveloped by the exponentially small spectral weight $Z$ (see Eq.~(\ref{eq:RF_incoh})). 

Fortunately our dynamical mean field solution \ceq{eq:DMF_phonons_EOM} approximates the full time dependence of the system and can be used to study observables beyond the RF spectrum. 

We studied the time-evolution of the momentum of the impurity, following the abrupt switch on of interactions. The results plotted in Fig.~\ref{fig:posc} show how the impurity relaxes to a steady state at long times. For weak interactions, the impurity loses a small portion of its momentum to the bosonic bath, corresponding to a minimally dressed polaron with large quasiparticle weight. The steady state momentum of the impurity decreases rapidly with interactions which we interpret as the onset of strong dressing and a reduction in quasi-particle weight. We also point out a surprising feature emerging at strong interactions -- decaying oscillations in the impurity momentum. We conjecture that quenching the impurity interaction to large values excites a long lived internal excitation of the emergent polaron; unfortunately no signature of this phenomenon manifests in the RF spectrum due to exponential suppression of weight for strong interactions, but it would be interesting to study this behavior in an experiment directly probing the non-equilibrium dynamics of the impurity, e.g. exciting the internal structure of the polaron by resonantly driving it in a trap. 

We emphasize that although the coherent peak of the RF spectrum is characterized by the ground state of the interacting impurity-BEC system (see Sec.~\ref{sec:universal_features}), the steady-state reached by the impurity following a sudden switch on is {\em different} from the interacting ground state. This can be seen formally by taking the long-time limit of the expectation value of an arbitrary observable $\hat{O}$. Performing a spectral decomposition of this quantity highlights the appropriate ensemble description of the steady state of the system:
\bea
\lim_{\rm t\to\infty} \langle i_{\uparrow p}|\hat{O}(t)|i_{\uparrow p}\rangle \to \sum_{n} |\langle i_{\uparrow p}|n_{\uparrow p}\rangle|^2\langle n_{\uparrow p}|\hat{O}| n_{\uparrow p}\rangle, \label{eq:diagonal_ensemble}
\eea 
 The right hand side expressed in terms of $|n_{\uparrow p}\rangle$, the time-independent eigenstates of the final Hamiltonian, represents the Diagonal Ensemble which characterizes the long time behavior of a generic closed quantum system \cite{rigol2008thermalization}. Clearly the steady state of the system is different from its ground state and is in fact an ensemble which includes the ground state, but also contains additional excitations. 
 
 Within our formalism we approximate the dynamics of the system using a time dependent product of coherent states. We expect that such an approximation can also capture the long-time steady state expectation value of operators, i.e. the long time limit of the coherent state product approaches \ceq{eq:diagonal_ensemble}. We found strong evidence of this fact; we plotted in Fig.~\ref{fig:vstar} the steady state (SS) and ground state (GS) group velocity of the impurity defined as:
\bea
v_{\rm SS, GS} = \frac{p_{\rm SS, GS}}{M},
\eea
where the steady-state value of the impurity velocity was calculated using the long time limit of our coherent state product \ceq{eq:DMF_state}, while the ground state value was calculated using \ceq{eq:MF_state}. We observe a quantitative difference between the two quantities. The quasiparticle residue Z (see \ceq{eq:qp_residue}) on the other hand is approximately equal (difference typically less than 1 part in $10^6$ for many different parameters) when calculated using the two states. This supports the picture of the impurity steady-state we put forward in \ceq{eq:diagonal_ensemble}, and is also consistent with the general argument about the coherent peak of the RF response presented in Sec.~\ref{sec:universal_features}

\section{Conclusions and Outlook}
We studied the fate of quantum impurities in BECs, and discussed the manifestation of polaron physics in RF spectroscopy. Population imbalanced dilute mixtures of degenerate ultracold atoms, either Bose-Fermi\cite{truscott2001observation,stan2004observation,gunter2006bose,inouye2004observation,fukuhara2009all,wu2012ultracold} or Bose-Bose\cite{catani2008degenerate,Shin2008,pilch2009observation,wernsdorfer2010lattice,mccarron2011dual,spethmann2012dynamics} mixtures, in which the role of the majority many-body environment is played by bosons, are the ideal settings in which to explore this rich physics. We require sufficiently low temperatures for which the bosonic environment will condense and can be modelled as a weakly interacting BEC. Crucially the atoms playing the role of quantum impurities should have hyperfine structure which can typically be addressed by RF pulses, and we require control over the interactions between impurity in different hyperfine levels and BEC. Ideally one of the hyperfine levels should be weakly interacting with the BEC, which will allow the faithful realization of the predictions in our article. The modest requirements discussed above are attainable using currently available experimental systems and techniques, thus we expect that our predictions can be tested in the near future. We consider a few particularly relevant experiments below.

\subsection{ Relation to experimental systems} \label{sec:experiments}

 Bose-Bose mixtures of Rb$^{87}$-K$^{41}$ \cite{Shin2008,wernsdorfer2010lattice} and Rb$^{87}$-Cs$^{133}$ \cite{pilch2009observation,spethmann2012dynamics}, as well as the Bose-Fermi mixture of Na$^{23}$-K$^{40}$\cite{wu2012ultracold}, are promising candidates in which to realize the polaronic physics of heavy impurities in BECs. In the three systems considered the heavy impurities, respectively Rb$^{87}$, Cs$^{133}$, K$^{40}$, have intrinsic mass ratio $M/m\approx 2$with respect to the BEC atoms, which can be further enhanced by a state-selective optical lattice. Moreover all of the experimental systems satisfy the criteria outlined previously: low temperatures sufficient to achieve BEC are routinely attained, atoms can be reliably trapped, inter-atom interactions can be tuned via carefully mapped out Feshbach resonances, and impurity atoms have hyperfine levels which can be addressed using RF. To quantify the impurity-BEC interactions which can be attained in these systems, we define a dimensionless ratio, $g_{\rm eff} = \xi n_0 a_{IB}^2$, between the average correlation length of the BEC $\sim \xi$ to the mean-free path of the impurity $\sim 1/(n_0 a_{IB}^2)$. We find that for the systems considered, intermediate interactions up to $g_{\rm eff} \approx 2-3$ can be attained using resonant tuning of scattering lengths, while ensuring the condition \ceq{eq:approximation_criterion} for the validity of our theoretical approach is satisfied.

\subsection{Related problems}

Our treatment in the present article missed aspects of strong coupling physics near a Feshbach resonance which are experimentally accessible, and theoretically rich. Given the possibility to form bound molecules for large positive impurity-boson interactions, it is quite possible that the system admits a polaron to molecule phase transition -- this is especially pertinent, given the impossibility of quantum phase transitions in Fr\"ohlich type models, and thus will clearly involve physics beyond such a model. Moreover, as a more non-trivial probe of the rich phase diagram afforded by the impurity-BEC system, it would be interesting to study the decay of the attractive polaron into the ``true'' molecular ground state of the system. 

The dynamics of polaron formation, and internal excitation structure of polarons are relatively unexplored areas of research. Indeed, within our current framework we observed coherent oscillations in the course of the relaxation of the impurity into a polaronic state (see Fig.~\ref{fig:posc}), which we interpreted as signatures of the internal structure of the polaron. It would be worthwhile devising a more elaborate theoretical description of the internal structure of the polaron, which may be probed in an experiment by resonantly driving the impurity-BEC system, and could shed light on the dynamics of polaron formation. One can also consider other non-trivial probes of polaron dynamics, such as the effect of driving Bloch oscillations of lattice impurities~\cite{fabiprep}. Such a scenario is particularly exciting as it is experimentally feasible using optical lattices. 

\section{Acknowledgements}

We would like to thank Dr. Richard Schmidt for numerous invaluable discussions, and thoughtful comments. We also thank Dr. Sebastian Will for stimulating discussions from an experimentalist's perspective. 

The authors acknowledge support from Harvard-MIT CUA, DARPA OLE program, AFOSR MURI on Ultracold Molecules, the ARO-MURI on Atomtronics, and the ARO MURI Quism program. AS acknowledges support from the Alfred P. Sloan Foundation (BR-5123), the National Science Foundation (DMR-1049082), the Norman Hackerman Advnace Research Program (01889), and the Welch Foundation (C-1739). FG thanks the Graduate School MAINZ for financial support. Research at the Perimeter Institute is supported by the Government of Canada through Industry Canada and by the Province of Ontario through the Ministry of Research and Innovation. 
  
\appendix

 \section{UV regularization of polaron binding energy}

Here we describe the regularization of UV divergences which arise in our model of impurity-BEC interactions
\bea
H_{\rm int} = \int d{\bf x}d{\bf x'} g_{\rm IB,\sigma}\delta({\bf x - x'})\rho_{\rm BEC}({\bf x})\rho_{\rm I}({\bf x'}), \label{eq:zero_range_model}
\eea 
which assumes zero interaction range. Such a model is a reasonable treatment of interactions in dilute atoms~\cite{Huang.book,bloch2008many}, that occur predominantly via two-particle collisions. Moreover for low-energy collisions, the two-particle scattering amplitude attains a universal form given by
\bea
f_{\rm IB, \sigma}(k) = \frac{-1}{1/a_{\rm IB,\sigma}+ ik}, \label{eq:scattering_amplitude}
\eea
which depends only on the s-wave scattering length, $a_{\rm IB,\sigma}.$ Consequently the effect of interactions enters all physical observables {\em only through the measurable s-wave scattering length}, which completely encodes the physics of two-particle collisions, and leads to universality in ultracold atoms. 

However, for large enough energies the scattering amplitude (\ref{eq:scattering_amplitude}) is no longer universal, and is sensitive to the microscopic details of the true interatomic potential. The appearance of UV divergences in physical observables is a direct consequence of poorly approximating this fundamentally different atomic-scale physics. Indeed, the zero range model \ceq{eq:zero_range_model} pathologically couples short (atomic) distance to long distance degrees of freedom. On the other hand, if one is only interested in universal properties, which are insensitive to microscopic physics, then one requires a means of safely and justifiably decoupling microscopic and macroscopic degrees of freedom. The renormalization group provides the formal means of achieving such a decoupling~\cite{Zinn-Justin.book}, but in the present case we require only a very trivial example of renormalization, which amounts to ``the subtraction of an infinity''. We demonstrate this approach, called dimensional regularization, on the binding energy defined in \ceq{eq:binding_energy}.

Consider the limit of a localized impurity $M\to\infty$ where the binding energy simplifies to
\bea
E^{M\to\infty}_{B} = -\sum_{\bf k} \frac{V_{\bf k}^2}{\omega_{\bf k}}  \xrightarrow{k\gg 1/\xi} -n_0g_{\rm IB,\sigma}^2\sum_{\bf k} \frac{2\mu}{k^2}.
\eea
We wish to subtract the leading UV divergence on the right hand side, but this procedure is a priori unjustified. To construct a rigorous prescription we invoke analytic continuity: we take the continuum limit $\sum_{\bf k} \to \int \frac{d^{D}{\bf k}}{(2\pi)^D}$ letting spatial dimension $D$ temporarily be a complex valued parameter. We will restore it to integer dimension, e.g. $D=3$, at the end of the calculation.  Such a procedure leads to the important identity:
\bea
\label{eq:regularization_prescription}
\int \frac{ d^{D}{\bf k}}{(2\pi)^D} \frac{1}{{\bf k}^2} = 0, \ \ \ \ D \ \in  \ \mathbb{C}.
\eea

Identity (\ref{eq:regularization_prescription}) allows us to subtract the leading UV divergence from all quantities which require regularization, including the binding energy, since it amounts to the mathematically allowed subtraction of zero by analytically continuing to complex dimension $D$. Thus we find the following regularized finite expression for the energy
\bea
E_{B, \rm reg.}^{\rm M \to \infty} &=&- \lim_{D \to 3} \int \frac{d^D {\bf k}}{(2\pi)^D} \left(\frac{V_{\bf k}^2}{\omega_{\bf k}}  - \frac{2\mu n_0g_{\rm IB,\sigma}^2}{k^2}\right) \nn
\eq - \frac{2\sqrt{2}\pi a_{\rm IB,\sigma}^2n_0}{\mu\xi} < 0.\label{eq:explicit_reg}
\eea
 
Moreover, we can use the same prescription to obtain the binding energy of a polaron formed by a finite mass impurity. There, the subtracted quantity retains the form of identity \ceq{eq:regularization_prescription}, but has a different prefactor. The actual computation of the energy (\ref{eq:binding_energy}) needs to be performed numerically in this case.  

\section{Path integral formulation of time dependent overlap}

Here we consider the time-dependent overlap, 
\bea
\label{app:t-dep-overlap}
A(t)\eq \langle \psi_i|e^{-i{\cal H}t}|\psi_i\rangle,
\eea
which describes the return probability of a non-stationary initial state $|\psi_i\rangle$, following time-evolution by a Hamiltonian ${\cal H}$. It typically arises in the context of quantum quenches, where it plays the same role as the partition function in equilibrium statistical mechanics. To take this analogy further, we wish to formulate the time dependent overlap (\ref{app:t-dep-overlap}) as a path integral, which is a standard formulation of the usual partition function. 

In addition to being of general theoretical interest, in the present context it provides a practical means of calculating the response of an impurity in a BEC to an RF signal. Indeed, as described in Sec.~\ref{sec:RF_intro} of the main text, the impurity RF response is in fact the Fourier transform of \ceq{app:t-dep-overlap}. 

Specifically, we consider the return amplitude of an initially decoupled impurity-BEC state, after time evolution by an interacting Hamiltonian, leading to the expression 
\bea
\label{app:t-dep-ovlerap-start}
A_p(t)\eq \langle 0|e^{-i\int_0^t{\cal H}dt'}|0\rangle\\
\eq \langle 0|e^{-i\left[\frac{1}{2M}\left({\bf p} - \sum_{\bf k}{\bf k}\hat{b}^\dagger_{\bf k}\hat{b}_{\bf k}\right)^2 + \sum_{\bf k}(\omega_{\bf k}\hat{b}^\dagger_{\bf k}\hat{b}_{\bf k}+V_{\bf k}(\hat{b}^\dagger_{\bf k} + \hat{b}_{\bf k}))\right]t}|0\rangle. \nonumber
\eea
This quantity determines the ``inverse'' RF response, see Sec.~\ref{sec:inv_RF} of the main text. Note that by using the Lee-Low-Pines (LLP) transformation outlined in Sec.~\ref{sec:LLP}, we dispensed with the impurity degree of freedom in the Hamiltonian, and mapped the impurity dynamics onto an interaction between phonons. Additionally, the initial state $|0\rangle$ is simply the phonon vacuum, and is unaffected by the LLP. 

Using the Hubbard-Stratonovich (HS) identity
\bea
&& e^{-\frac{i}{2M}\left(p - \sum_{\bf k}{\bf k}\hat{b}^\dagger_{\bf k}\hat{b}_{\bf k}\right)^2}\nn
\eq \frac{\int_{-\infty}^\infty d\varphi(t') e^{i\left[\frac{M}{2}\phi(t')^2 - i\varphi(t').\left({\bf p} - \sum_{\bf k}{\bf k}\hat{b}^\dagger_{\bf k}\hat{b}_{\bf k}\right)\right]dt'}}{\int_{-\infty}^\infty d\varphi(t')e^{i\frac{M}{2}\varphi(t')^2}},
\eea
in each interval $dt'$ we introduce a time-dependent classical field $\varphi(t')$. This leads to the following path integral formulation of the time-dependent overlap (\ref{app:t-dep-ovlerap-start}): 
\bea
\label{app:time-dep-pathint}
&& A_p(t) = {\cal N}\int{\cal D}[\varphi(t)] e^{i\int_0^t dt' \frac{M}{2}\varphi(t')^2}\\
&\times& \langle 0| e^{-i\int_0^t dt'\left[\varphi(t').\left({\bf p} - \sum_{\bf k}{\bf k}\hat{b}^\dagger_{\bf k}\hat{b}_{\bf k}\right)- \sum_{\bf k}(\omega_{\bf k}\hat{b}^\dagger_{\bf k}\hat{b}_{\bf k}+V_{\bf k}(\hat{b}^\dagger_{\bf k} + \hat{b}_{\bf k}))\right]}|0\rangle, \nonumber 
\eea
normalized by ${\cal N} = \int {\cal D}[\varphi(t)] e^{i\int_0^t dt' \frac{M}{2}\varphi(t')^2}$.

The path integral notation is a compact representation of the measure
$$
\int {\cal D}[\varphi(t)] = \lim_{N\to\infty}\prod_{j=1}^N \int^\infty_{-\infty}d\varphi(t_j),
$$
which accounts for our discretization of the time interval $t$ into $N\to\infty$ infinitesimal windows of size $dt'$. Correspondingly, we also decomposed the bosonic Hamiltonian
\bea
\label{app:bose_path_int_h}
H[\hat{b}^\dagger_{\bf k},\hat{b}_{\bf k},\varphi] = \sum_{\bf k}\left[(\omega_{\bf k} - \varphi.{\bf k})\hat{b}^\dagger_{\bf k}\hat{b}_{\bf k}+V_{\bf k}(\hat{b}^\dagger_{\bf k} + \hat{b}_{\bf k})\right],
\eea
 into a sum of $N$ discrete terms which we rewrote as an integral, to precision $dt'$:
$$
e^{i\int_0^t H dt'} = \prod_{j=1}^N e^{iH[\hat{b}^\dagger_{\bf k},\hat{b}_{\bf k},\varphi(t_j)]} +{\cal O}(dt')= e^{i\int_0^t dt'H[\hat{b}^\dagger_{\bf k},\hat{b}_{\bf k},\varphi(t')]}.
$$

Hamiltonian (\ref{app:bose_path_int_h}) contains at most quadratic terms in bosons, enabling us to ``integrate them out''. We do so by noting that the dynamics of bosons due to such a quadratic Hamiltonian can be exactly described by a decoupled product of time-dependent coherent states (c.f. the discussion of localized impurities in Sec.~\ref{sec:inv_RF} of the main text). Thus we demand
\bea
e^{-i \int_0^t dt'H[\hat{b}^\dagger_{\bf k},\hat{b}_{\bf k},\varphi(t')]}|0\rangle = \prod_{\bf k}|\alpha_{\bf k}(t)\rangle,\label{app:coherent_state_evolution}
\eea   
with $|\alpha_{\bf k}(t)\rangle$ of the coherent state form:
\bea
|\alpha_{\bf k}(t)\rangle = e^{i\chi_{\bf k}(t)}e^{\alpha_{\bf k}(t)\hat{b}^\dagger_{\bf k} - \alpha^*_{\bf k}(t)\hat{b}_{\bf k}}|0\rangle.\label{app:coherent_state_explicit}
\eea
By taking the time derivative of the two sides of \ceq{app:coherent_state_evolution}, and using the explicit form (\ref{app:coherent_state_explicit}) to differentiate the right hand side, we obtain differential equations for the coherent state parameters:
\bea
\dot{\alpha}_{\bf k}(t) \eq - i[(\omega_{\bf k} - \varphi(t).{\bf k})\alpha_{\bf k}(t) + V_{\bf k}],\label{app:alpha_eq}\\
\dot{\chi}_{\bf k}(t) \eq -\frac{V_{\bf k}}{2}(\alpha_{\bf k}(t)+\alpha^*_{\bf k}(t)),
\eea
 which can be solved by recognizing that \ceq{app:alpha_eq} contains the total time-derivative of $\alpha_{\bf k}(t){\rm exp}\left[-i{\bf k}.\int_0^t \varphi(t')dt' + i\omega_{\bf k}t\right]$. 
 
 Thus we obtain:
 \bea\label{app:alpha_sol}
 \alpha_{\bf k}(t) \eq -iV_{\bf k}\int_0^t dt_1 e^{-i{\bf k}.\int_{t_1}^t dt'(\omega_{\bf k}- \varphi(t'))},\\
 \chi_{\bf k}(t) \eq V_{\bf k}^2 \int_0^t dt_1 \int_0^{t_1}dt_2 \sin\left[\int_{t_2}^{t_1}dt'(\omega_{\bf k} - \varphi(t').{\bf k})\right].\nn\label{app:chi_sol}
 \eea
 
 The expectation value
 \bea
 E[\varphi(t)] = \langle 0| e^{-i\int_0^t dt'H[\hat{b}^\dagger_{\bf k},\hat{b}_{\bf k},\varphi(t)] } |0\rangle,
 \eea
  appearing in \ceq{app:time-dep-pathint}, can be rewritten using \ceq{app:coherent_state_evolution} and the coherent state property
  $$
  \langle 0|e^{\alpha b^\dagger - \alpha b}|0\rangle = e^{-\frac{1}{2}|\alpha|^2},
  $$
  to yield
  \bea
  E[\varphi(t)] = e^{\sum_{\bf k}\left[i\chi_{\bf k}(t) - \frac{1}{2}|\alpha_{\bf k}(t)|^2\right]},\label{app:eval_coh}
  \eea
 which allows us to rewrite the time-dependent overlap (\ref{app:time-dep-pathint}) in the form:
\bea
A_p(t) \eq {\cal N}\int{\cal D}[\varphi(t)] e^{i\int_0^t dt' \left(\frac{M}{2}\varphi(t')^2 - {\bf p}.\varphi(t')\right)}E[\varphi(t)]\nn
\eq  {\cal N}\int{\cal D}[\varphi(t)] e^{i\int_0^t dt' \left(\frac{M}{2}\varphi(t')^2 - {\bf p}.\varphi(t')\right)}\nn
&&\hspace{1.5 cm}\times e^{\sum_{\bf k}\left[i\chi_{\bf k}(t) - \frac{1}{2}|\alpha_{\bf k}(t)|^2\right]}.\label{app:time-dep-pathint2}
\eea
 Eqs.~(\ref{app:alpha_sol}),(\ref{app:chi_sol}) can be substituted in \ceq{app:time-dep-pathint2}, leading to a path integral over $\varphi(t)$ alone:
 \bea
 \label{app:path_int_fin}
A_p(t) = {\cal N}\int{\cal D}[\varphi(t)] e^{i{\cal A}[\varphi(t)]},
\eea
with action given by
\bea
\label{app:gen_action}
&&{\cal A}[\varphi(t)] = \int_0^t dt'\left[\frac{M}{2} \varphi(t')^2 - \varphi(t).{\bf p}\right] + i\sum_{\bf k} V_{\bf k}^2\nn
&&\times\int_0^t dt_1 \int_0^{t_1}dt_2\exp\left[-i\int_{t_2}^{t_1}dt'(\omega_{\bf k} - \varphi(t').{\bf k})\right].\ \ \ \ 
\eea

Thus, using the HS identity and the exact solution of the bosonic Hamiltonian (\ref{app:bose_path_int_h}) in terms of decoupled coherent states, we showed that the path integral (\ref{app:path_int_fin}) with the action (\ref{app:gen_action}) is an {\em exact reformulation} of the time-dependent overlap (\ref{app:t-dep-ovlerap-start}). However further progress requires an approximation scheme to treat the non-Gaussian path integral, which involves a retarded self-interaction of the impurity velocity field $\varphi(t)$. To this end we estimate \ceq{app:path_int_fin} within a saddle point treatment, by extremizing action (\ref{app:gen_action}) with respect to $\varphi(t)$. Thus we obtain the following saddle point equation
\bea
\label{app:saddle_pt_eq}
\varphi_s(t') \eq \frac{{\bf p}}{M}\nn
&& + \sum_{\bf k} \frac{V_{\bf k}^2{\bf k}}{M}\int_{t'}^t dt_1 \int_0^{t'} dt_2 e^{-i\int_{t_2}^{t_1} dt'' (\omega_{\bf k} - {\bf k}.\varphi_s(t''))}.\nn
\eea
The solution of \ceq{app:saddle_pt_eq} represents a single trajectory that approximates the path integral form of the overlap (\ref{app:path_int_fin}) by identifying the most dominant contribution to it. The solution is a time-dependent velocity profile defined up to the propagation time $t$ at which the time-dependent overlap is evaluated. Moreover, as can be seen from \ceq{app:saddle_pt_eq} it is symmetric around $t/2$ and $\varphi_s(0) = \varphi_s(t) = \frac{\bf p}{M}$, the bare velocity of the impurity. This unique feature of the velocity profile is due to the requirement of the time-evolving state to return to its initial value, by construction of the quantum propagation amplitude (\ref{app:t-dep-overlap}).

We solved \ceq{app:saddle_pt_eq} iteratively, taking a lattice of time and momentum points. Moreover in the numerical procedure we dealt with the UV divergence inherent to the zero-range model (see Appendix. A) by introducing a soft cutoff for large momenta, into the interaction of the form $e^{-k^2/\Lambda^2}$, and choosing $\Lambda$ large enough to obtain converged results. The numerical effort required to solve \ceq{app:saddle_pt_eq} was significantly greater than the mean-field approach outlined in the main text, see Ref.~\ref{sec:inv_RF}. On the other hand the difference in value of the time-dependent overlap was negligible when computed using the two approaches. Thus we evaluated RF spectra using the time-dependent mean-field approach, confirming its validity based on this agreement.  
\bibliography{PolaronRF}

%merlin.mbs apsrev4-1.bst 2010-07-25 4.21a (PWD, AO, DPC) hacked
%Control: key (0)
%Control: author (72) initials jnrlst
%Control: editor formatted (1) identically to author
%Control: production of article title (-1) disabled
%Control: page (0) single
%Control: year (1) truncated
%Control: production of eprint (0) enabled
\begin{thebibliography}{91}%
\makeatletter
\providecommand \@ifxundefined [1]{%
 \@ifx{#1\undefined}
}%
\providecommand \@ifnum [1]{%
 \ifnum #1\expandafter \@firstoftwo
 \else \expandafter \@secondoftwo
 \fi
}%
\providecommand \@ifx [1]{%
 \ifx #1\expandafter \@firstoftwo
 \else \expandafter \@secondoftwo
 \fi
}%
\providecommand \natexlab [1]{#1}%
\providecommand \enquote  [1]{``#1''}%
\providecommand \bibnamefont  [1]{#1}%
\providecommand \bibfnamefont [1]{#1}%
\providecommand \citenamefont [1]{#1}%
\providecommand \href@noop [0]{\@secondoftwo}%
\providecommand \href [0]{\begingroup \@sanitize@url \@href}%
\providecommand \@href[1]{\@@startlink{#1}\@@href}%
\providecommand \@@href[1]{\endgroup#1\@@endlink}%
\providecommand \@sanitize@url [0]{\catcode `\\12\catcode `\$12\catcode
  `\&12\catcode `\#12\catcode `\^12\catcode `\_12\catcode `\%12\relax}%
\providecommand \@@startlink[1]{}%
\providecommand \@@endlink[0]{}%
\providecommand \url  [0]{\begingroup\@sanitize@url \@url }%
\providecommand \@url [1]{\endgroup\@href {#1}{\urlprefix }}%
\providecommand \urlprefix  [0]{URL }%
\providecommand \Eprint [0]{\href }%
\providecommand \doibase [0]{http://dx.doi.org/}%
\providecommand \selectlanguage [0]{\@gobble}%
\providecommand \bibinfo  [0]{\@secondoftwo}%
\providecommand \bibfield  [0]{\@secondoftwo}%
\providecommand \translation [1]{[#1]}%
\providecommand \BibitemOpen [0]{}%
\providecommand \bibitemStop [0]{}%
\providecommand \bibitemNoStop [0]{.\EOS\space}%
\providecommand \EOS [0]{\spacefactor3000\relax}%
\providecommand \BibitemShut  [1]{\csname bibitem#1\endcsname}%
\let\auto@bib@innerbib\@empty
%</preamble>
\bibitem [{\citenamefont {Landau}(1933)}]{landauorig}%
  \BibitemOpen
  \bibfield  {author} {\bibinfo {author} {\bibfnamefont {L.~D.}\ \bibnamefont
  {Landau}},\ }\href@noop {} {\bibfield  {journal} {\bibinfo  {journal} {Phys.
  Z. Sowj Un.}\ }\textbf {\bibinfo {volume} {3}},\ \bibinfo {pages} {664}
  (\bibinfo {year} {1933})}\BibitemShut {NoStop}%
\bibitem [{\citenamefont {Landau}\ and\ \citenamefont
  {Pekar}(1946)}]{landaupekar46}%
  \BibitemOpen
  \bibfield  {author} {\bibinfo {author} {\bibfnamefont {L.~D.}\ \bibnamefont
  {Landau}}\ and\ \bibinfo {author} {\bibfnamefont {S.~I.}\ \bibnamefont
  {Pekar}},\ }\href@noop {} {\bibfield  {journal} {\bibinfo  {journal} {Zh.
  Exp. Teor. Fiz.}\ }\textbf {\bibinfo {volume} {16}},\ \bibinfo {pages} {341}
  (\bibinfo {year} {1946})}\BibitemShut {NoStop}%
\bibitem [{\citenamefont {Fr{\"o}hlich}(1954)}]{frohlich1954electrons}%
  \BibitemOpen
  \bibfield  {author} {\bibinfo {author} {\bibfnamefont {H.}~\bibnamefont
  {Fr{\"o}hlich}},\ }\href@noop {} {\bibfield  {journal} {\bibinfo  {journal}
  {Adv. in Phys.}\ }\textbf {\bibinfo {volume} {3}},\ \bibinfo {pages} {325}
  (\bibinfo {year} {1954})}\BibitemShut {NoStop}%
\bibitem [{\citenamefont {Feynman}(1955)}]{feynman1955slow}%
  \BibitemOpen
  \bibfield  {author} {\bibinfo {author} {\bibfnamefont {R.~P.}\ \bibnamefont
  {Feynman}},\ }\href@noop {} {\bibfield  {journal} {\bibinfo  {journal} {Phys.
  Rev.}\ }\textbf {\bibinfo {volume} {97}},\ \bibinfo {pages} {660} (\bibinfo
  {year} {1955})}\BibitemShut {NoStop}%
\bibitem [{\citenamefont {Devreese}(2007)}]{devreese1996polaron}%
  \BibitemOpen
  \bibfield  {author} {\bibinfo {author} {\bibfnamefont {J.}~\bibnamefont
  {Devreese}},\ }\href@noop {} {\bibfield  {journal} {\bibinfo  {journal} {J.
  Phys.}\ }\textbf {\bibinfo {volume} {19}},\ \bibinfo {pages} {255201}
  (\bibinfo {year} {2007})}\BibitemShut {NoStop}%
\bibitem [{\citenamefont {Storchak}\ and\ \citenamefont
  {Prokof’ev}(1998)}]{storchak1998quantum}%
  \BibitemOpen
  \bibfield  {author} {\bibinfo {author} {\bibfnamefont {V.~G.}\ \bibnamefont
  {Storchak}}\ and\ \bibinfo {author} {\bibfnamefont {N.~V.}\ \bibnamefont
  {Prokof’ev}},\ }\href@noop {} {\bibfield  {journal} {\bibinfo  {journal}
  {Rev. Mod. Phys.}\ }\textbf {\bibinfo {volume} {70}},\ \bibinfo {pages} {929}
  (\bibinfo {year} {1998})}\BibitemShut {NoStop}%
\bibitem [{\citenamefont {Hulea}\ \emph {et~al.}(2006)\citenamefont {Hulea},
  \citenamefont {Fratini}, \citenamefont {Xie}, \citenamefont {Mulder},
  \citenamefont {Iossad}, \citenamefont {Rastelli}, \citenamefont {Ciuchi},\
  and\ \citenamefont {Morpurgo}}]{hulea2006tunable}%
  \BibitemOpen
  \bibfield  {author} {\bibinfo {author} {\bibfnamefont {I.}~\bibnamefont
  {Hulea}}, \bibinfo {author} {\bibfnamefont {S.}~\bibnamefont {Fratini}},
  \bibinfo {author} {\bibfnamefont {H.}~\bibnamefont {Xie}}, \bibinfo {author}
  {\bibfnamefont {C.}~\bibnamefont {Mulder}}, \bibinfo {author} {\bibfnamefont
  {N.}~\bibnamefont {Iossad}}, \bibinfo {author} {\bibfnamefont
  {G.}~\bibnamefont {Rastelli}}, \bibinfo {author} {\bibfnamefont
  {S.}~\bibnamefont {Ciuchi}}, \ and\ \bibinfo {author} {\bibfnamefont
  {A.}~\bibnamefont {Morpurgo}},\ }\href@noop {} {\bibfield  {journal}
  {\bibinfo  {journal} {Nature Mat.}\ }\textbf {\bibinfo {volume} {5}},\
  \bibinfo {pages} {982} (\bibinfo {year} {2006})}\BibitemShut {NoStop}%
\bibitem [{\citenamefont {von Helmolt}\ \emph {et~al.}(1993)\citenamefont {von
  Helmolt}, \citenamefont {Wecker}, \citenamefont {Holzapfel}, \citenamefont
  {Schultz},\ and\ \citenamefont {Samwer}}]{von1993giant}%
  \BibitemOpen
  \bibfield  {author} {\bibinfo {author} {\bibfnamefont {R.}~\bibnamefont {von
  Helmolt}}, \bibinfo {author} {\bibfnamefont {J.}~\bibnamefont {Wecker}},
  \bibinfo {author} {\bibfnamefont {B.}~\bibnamefont {Holzapfel}}, \bibinfo
  {author} {\bibfnamefont {L.}~\bibnamefont {Schultz}}, \ and\ \bibinfo
  {author} {\bibfnamefont {K.}~\bibnamefont {Samwer}},\ }\href@noop {}
  {\bibfield  {journal} {\bibinfo  {journal} {Phys. Rev. Lett.}\ }\textbf
  {\bibinfo {volume} {71}},\ \bibinfo {pages} {2331} (\bibinfo {year}
  {1993})}\BibitemShut {NoStop}%
\bibitem [{\citenamefont {Salje}\ \emph {et~al.}(2005)\citenamefont {Salje},
  \citenamefont {Alexandrov},\ and\ \citenamefont {Liang}}]{salje2005polarons}%
  \BibitemOpen
  \bibfield  {author} {\bibinfo {author} {\bibfnamefont {E.~K.}\ \bibnamefont
  {Salje}}, \bibinfo {author} {\bibfnamefont {A.}~\bibnamefont {Alexandrov}}, \
  and\ \bibinfo {author} {\bibfnamefont {W.}~\bibnamefont {Liang}},\
  }\href@noop {} {\emph {\bibinfo {title} {Polarons and bipolarons in high-Tc
  superconductors and related materials}}}\ (\bibinfo  {publisher} {Cambridge
  University Press},\ \bibinfo {year} {2005})\BibitemShut {NoStop}%
\bibitem [{\citenamefont {ProkofÕev}\ and\ \citenamefont
  {Svistunov}(2008)}]{prokof2008fermi}%
  \BibitemOpen
  \bibfield  {author} {\bibinfo {author} {\bibfnamefont {N.}~\bibnamefont
  {ProkofÕev}}\ and\ \bibinfo {author} {\bibfnamefont {B.}~\bibnamefont
  {Svistunov}},\ }\href@noop {} {\bibfield  {journal} {\bibinfo  {journal}
  {Phys. Rev. B}\ }\textbf {\bibinfo {volume} {77}},\ \bibinfo {pages} {020408}
  (\bibinfo {year} {2008})}\BibitemShut {NoStop}%
\bibitem [{\citenamefont {Punk}\ \emph {et~al.}(2009)\citenamefont {Punk},
  \citenamefont {Dumitrescu},\ and\ \citenamefont {Zwerger}}]{punk2009polaron}%
  \BibitemOpen
  \bibfield  {author} {\bibinfo {author} {\bibfnamefont {M.}~\bibnamefont
  {Punk}}, \bibinfo {author} {\bibfnamefont {P.~T.}\ \bibnamefont
  {Dumitrescu}}, \ and\ \bibinfo {author} {\bibfnamefont {W.}~\bibnamefont
  {Zwerger}},\ }\href@noop {} {\bibfield  {journal} {\bibinfo  {journal} {Phys.
  Rev. A}\ }\textbf {\bibinfo {volume} {80}},\ \bibinfo {pages} {053605}
  (\bibinfo {year} {2009})}\BibitemShut {NoStop}%
\bibitem [{\citenamefont {Schirotzek}\ \emph {et~al.}(2009)\citenamefont
  {Schirotzek}, \citenamefont {Wu}, \citenamefont {Sommer},\ and\ \citenamefont
  {Zwierlein}}]{Schirotzek2009}%
  \BibitemOpen
  \bibfield  {author} {\bibinfo {author} {\bibfnamefont {A.}~\bibnamefont
  {Schirotzek}}, \bibinfo {author} {\bibfnamefont {C.-H.}\ \bibnamefont {Wu}},
  \bibinfo {author} {\bibfnamefont {A.}~\bibnamefont {Sommer}}, \ and\ \bibinfo
  {author} {\bibfnamefont {M.~W.}\ \bibnamefont {Zwierlein}},\ }\href
  {http://link.aps.org/doi/10.1103/PhysRevLett.102.230402
  papers2://publication/doi/10.1103/PhysRevLett.102.230402} {\bibfield
  {journal} {\bibinfo  {journal} {Phys. Rev. Lett.}\ }\textbf {\bibinfo
  {volume} {102}},\ \bibinfo {pages} {230402} (\bibinfo {year}
  {2009})}\BibitemShut {NoStop}%
\bibitem [{\citenamefont {Chevy}\ and\ \citenamefont
  {Mora}(2010)}]{chevy2010ultra}%
  \BibitemOpen
  \bibfield  {author} {\bibinfo {author} {\bibfnamefont {F.}~\bibnamefont
  {Chevy}}\ and\ \bibinfo {author} {\bibfnamefont {C.}~\bibnamefont {Mora}},\
  }\href@noop {} {\bibfield  {journal} {\bibinfo  {journal} {Rep. Prog. Phys.}\
  }\textbf {\bibinfo {volume} {73}},\ \bibinfo {pages} {112401} (\bibinfo
  {year} {2010})}\BibitemShut {NoStop}%
\bibitem [{\citenamefont {Schmidt}\ and\ \citenamefont
  {Enss}(2011)}]{schmidt2011excitation}%
  \BibitemOpen
  \bibfield  {author} {\bibinfo {author} {\bibfnamefont {R.}~\bibnamefont
  {Schmidt}}\ and\ \bibinfo {author} {\bibfnamefont {T.}~\bibnamefont {Enss}},\
  }\href@noop {} {\bibfield  {journal} {\bibinfo  {journal} {Phys. Rev. A}\
  }\textbf {\bibinfo {volume} {83}},\ \bibinfo {pages} {063620} (\bibinfo
  {year} {2011})}\BibitemShut {NoStop}%
\bibitem [{\citenamefont {Koschorreck}\ \emph {et~al.}(2012)\citenamefont
  {Koschorreck}, \citenamefont {Pertot}, \citenamefont {Vogt}, \citenamefont
  {Fr\"{o}hlich}, \citenamefont {Feld},\ and\ \citenamefont
  {K\"{o}hl}}]{Koschorreck2012}%
  \BibitemOpen
  \bibfield  {author} {\bibinfo {author} {\bibfnamefont {M.}~\bibnamefont
  {Koschorreck}}, \bibinfo {author} {\bibfnamefont {D.}~\bibnamefont {Pertot}},
  \bibinfo {author} {\bibfnamefont {E.}~\bibnamefont {Vogt}}, \bibinfo {author}
  {\bibfnamefont {B.}~\bibnamefont {Fr\"{o}hlich}}, \bibinfo {author}
  {\bibfnamefont {M.}~\bibnamefont {Feld}}, \ and\ \bibinfo {author}
  {\bibfnamefont {M.}~\bibnamefont {K\"{o}hl}},\ }\href {\doibase
  10.1038/nature11151} {\bibfield  {journal} {\bibinfo  {journal} {Nature}\
  }\textbf {\bibinfo {volume} {485}},\ \bibinfo {pages} {619} (\bibinfo {year}
  {2012})}\BibitemShut {NoStop}%
\bibitem [{\citenamefont {Kohstall}\ \emph {et~al.}(2012)\citenamefont
  {Kohstall}, \citenamefont {Zaccanti}, \citenamefont {Jag}, \citenamefont
  {Trenkwalder}, \citenamefont {Massignan}, \citenamefont {Bruun},
  \citenamefont {Schreck},\ and\ \citenamefont
  {Grimm}}]{kohstall2012metastability}%
  \BibitemOpen
  \bibfield  {author} {\bibinfo {author} {\bibfnamefont {C.}~\bibnamefont
  {Kohstall}}, \bibinfo {author} {\bibfnamefont {M.}~\bibnamefont {Zaccanti}},
  \bibinfo {author} {\bibfnamefont {M.}~\bibnamefont {Jag}}, \bibinfo {author}
  {\bibfnamefont {A.}~\bibnamefont {Trenkwalder}}, \bibinfo {author}
  {\bibfnamefont {P.}~\bibnamefont {Massignan}}, \bibinfo {author}
  {\bibfnamefont {G.~M.}\ \bibnamefont {Bruun}}, \bibinfo {author}
  {\bibfnamefont {F.}~\bibnamefont {Schreck}}, \ and\ \bibinfo {author}
  {\bibfnamefont {R.}~\bibnamefont {Grimm}},\ }\href@noop {} {\bibfield
  {journal} {\bibinfo  {journal} {Nature}\ }\textbf {\bibinfo {volume} {485}},\
  \bibinfo {pages} {615} (\bibinfo {year} {2012})}\BibitemShut {NoStop}%
\bibitem [{\citenamefont {Zhang}\ \emph {et~al.}(2012)\citenamefont {Zhang},
  \citenamefont {Ong}, \citenamefont {Arakelyan},\ and\ \citenamefont
  {Thomas}}]{Zhang2012}%
  \BibitemOpen
  \bibfield  {author} {\bibinfo {author} {\bibfnamefont {Y.}~\bibnamefont
  {Zhang}}, \bibinfo {author} {\bibfnamefont {W.}~\bibnamefont {Ong}}, \bibinfo
  {author} {\bibfnamefont {I.}~\bibnamefont {Arakelyan}}, \ and\ \bibinfo
  {author} {\bibfnamefont {J.~E.}\ \bibnamefont {Thomas}},\ }\href {\doibase
  10.1103/PhysRevLett.108.235302} {\bibfield  {journal} {\bibinfo  {journal}
  {Phys. Rev. Lett.}\ }\textbf {\bibinfo {volume} {108}},\ \bibinfo {pages}
  {235302} (\bibinfo {year} {2012})}\BibitemShut {NoStop}%
\bibitem [{\citenamefont {Massignan}(2012)}]{massignan2012polarons}%
  \BibitemOpen
  \bibfield  {author} {\bibinfo {author} {\bibfnamefont {P.}~\bibnamefont
  {Massignan}},\ }\href@noop {} {\bibfield  {journal} {\bibinfo  {journal}
  {Eur. Phys. Lett.}\ }\textbf {\bibinfo {volume} {98}},\ \bibinfo {pages}
  {10012} (\bibinfo {year} {2012})}\BibitemShut {NoStop}%
\bibitem [{\citenamefont {Catani}\ \emph {et~al.}(2012)\citenamefont {Catani},
  \citenamefont {Lamporesi}, \citenamefont {Naik}, \citenamefont {Gring},
  \citenamefont {Inguscio}, \citenamefont {Minardi}, \citenamefont {Kantian},\
  and\ \citenamefont {Giamarchi}}]{catani2012quantum}%
  \BibitemOpen
  \bibfield  {author} {\bibinfo {author} {\bibfnamefont {J.}~\bibnamefont
  {Catani}}, \bibinfo {author} {\bibfnamefont {G.}~\bibnamefont {Lamporesi}},
  \bibinfo {author} {\bibfnamefont {D.}~\bibnamefont {Naik}}, \bibinfo {author}
  {\bibfnamefont {M.}~\bibnamefont {Gring}}, \bibinfo {author} {\bibfnamefont
  {M.}~\bibnamefont {Inguscio}}, \bibinfo {author} {\bibfnamefont
  {F.}~\bibnamefont {Minardi}}, \bibinfo {author} {\bibfnamefont
  {A.}~\bibnamefont {Kantian}}, \ and\ \bibinfo {author} {\bibfnamefont
  {T.}~\bibnamefont {Giamarchi}},\ }\href@noop {} {\bibfield  {journal}
  {\bibinfo  {journal} {Phys. Rev. A}\ }\textbf {\bibinfo {volume} {85}},\
  \bibinfo {pages} {023623} (\bibinfo {year} {2012})}\BibitemShut {NoStop}%
\bibitem [{\citenamefont {Astrakharchik}\ and\ \citenamefont
  {Pitaevskii}(2004)}]{astrakharchik2004motion}%
  \BibitemOpen
  \bibfield  {author} {\bibinfo {author} {\bibfnamefont {G.~E.}\ \bibnamefont
  {Astrakharchik}}\ and\ \bibinfo {author} {\bibfnamefont {L.~P.}\ \bibnamefont
  {Pitaevskii}},\ }\href@noop {} {\bibfield  {journal} {\bibinfo  {journal}
  {Phys. Rev. A}\ }\textbf {\bibinfo {volume} {70}},\ \bibinfo {pages} {013608}
  (\bibinfo {year} {2004})}\BibitemShut {NoStop}%
\bibitem [{\citenamefont {Cucchietti}\ and\ \citenamefont
  {Timmermans}(2006)}]{cucchietti2006strong}%
  \BibitemOpen
  \bibfield  {author} {\bibinfo {author} {\bibfnamefont {F.~M.}\ \bibnamefont
  {Cucchietti}}\ and\ \bibinfo {author} {\bibfnamefont {E.}~\bibnamefont
  {Timmermans}},\ }\href@noop {} {\bibfield  {journal} {\bibinfo  {journal}
  {Phys. Rev. Lett.}\ }\textbf {\bibinfo {volume} {96}},\ \bibinfo {pages}
  {210401} (\bibinfo {year} {2006})}\BibitemShut {NoStop}%
\bibitem [{\citenamefont {Sacha}\ and\ \citenamefont
  {Timmermans}(2006)}]{sacha2006self}%
  \BibitemOpen
  \bibfield  {author} {\bibinfo {author} {\bibfnamefont {K.}~\bibnamefont
  {Sacha}}\ and\ \bibinfo {author} {\bibfnamefont {E.}~\bibnamefont
  {Timmermans}},\ }\href@noop {} {\bibfield  {journal} {\bibinfo  {journal}
  {Phys. Rev. A}\ }\textbf {\bibinfo {volume} {73}},\ \bibinfo {pages} {063604}
  (\bibinfo {year} {2006})}\BibitemShut {NoStop}%
\bibitem [{\citenamefont {Kalas}\ and\ \citenamefont
  {Blume}(2006)}]{kalas2006interaction}%
  \BibitemOpen
  \bibfield  {author} {\bibinfo {author} {\bibfnamefont {R.~M.}\ \bibnamefont
  {Kalas}}\ and\ \bibinfo {author} {\bibfnamefont {D.}~\bibnamefont {Blume}},\
  }\href@noop {} {\bibfield  {journal} {\bibinfo  {journal} {Phys. Rev. A}\
  }\textbf {\bibinfo {volume} {73}},\ \bibinfo {pages} {043608} (\bibinfo
  {year} {2006})}\BibitemShut {NoStop}%
\bibitem [{\citenamefont {Bruderer}\ \emph
  {et~al.}(2008{\natexlab{a}})\citenamefont {Bruderer}, \citenamefont {Klein},
  \citenamefont {Clark},\ and\ \citenamefont {Jaksch}}]{Bruderer2008}%
  \BibitemOpen
  \bibfield  {author} {\bibinfo {author} {\bibfnamefont {M.}~\bibnamefont
  {Bruderer}}, \bibinfo {author} {\bibfnamefont {A.}~\bibnamefont {Klein}},
  \bibinfo {author} {\bibfnamefont {S.~R.}\ \bibnamefont {Clark}}, \ and\
  \bibinfo {author} {\bibfnamefont {D.}~\bibnamefont {Jaksch}},\ }\href
  {\doibase 10.1088/1367-2630/10/3/033015} {\bibfield  {journal} {\bibinfo
  {journal} {New J. Phys.}\ }\textbf {\bibinfo {volume} {10}},\ \bibinfo
  {pages} {033015} (\bibinfo {year} {2008}{\natexlab{a}})}\BibitemShut
  {NoStop}%
\bibitem [{\citenamefont {Bruderer}\ \emph
  {et~al.}(2008{\natexlab{b}})\citenamefont {Bruderer}, \citenamefont {Bao},\
  and\ \citenamefont {Jaksch}}]{Bruderer2008a}%
  \BibitemOpen
  \bibfield  {author} {\bibinfo {author} {\bibfnamefont {M.}~\bibnamefont
  {Bruderer}}, \bibinfo {author} {\bibfnamefont {W.}~\bibnamefont {Bao}}, \
  and\ \bibinfo {author} {\bibfnamefont {D.}~\bibnamefont {Jaksch}},\ }\href
  {\doibase 10.1209/0295-5075/82/30004} {\bibfield  {journal} {\bibinfo
  {journal} {Eur. Phys. Lett.}\ }\textbf {\bibinfo {volume} {82}},\ \bibinfo
  {pages} {30004} (\bibinfo {year} {2008}{\natexlab{b}})}\BibitemShut {NoStop}%
\bibitem [{\citenamefont {Bruderer}\ \emph {et~al.}(2007)\citenamefont
  {Bruderer}, \citenamefont {Klein}, \citenamefont {Clark},\ and\ \citenamefont
  {Jaksch}}]{Bruderer2007}%
  \BibitemOpen
  \bibfield  {author} {\bibinfo {author} {\bibfnamefont {M.}~\bibnamefont
  {Bruderer}}, \bibinfo {author} {\bibfnamefont {A.}~\bibnamefont {Klein}},
  \bibinfo {author} {\bibfnamefont {S.~R.}\ \bibnamefont {Clark}}, \ and\
  \bibinfo {author} {\bibfnamefont {D.}~\bibnamefont {Jaksch}},\ }\href
  {\doibase 10.1103/PhysRevA.76.011605} {\bibfield  {journal} {\bibinfo
  {journal} {Phys. Rev. A}\ }\textbf {\bibinfo {volume} {76}},\ \bibinfo
  {pages} {011605} (\bibinfo {year} {2007})}\BibitemShut {NoStop}%
\bibitem [{\citenamefont {Schmid}\ \emph {et~al.}(2010)\citenamefont {Schmid},
  \citenamefont {H{\"a}rter},\ and\ \citenamefont
  {Denschlag}}]{schmid2010dynamics}%
  \BibitemOpen
  \bibfield  {author} {\bibinfo {author} {\bibfnamefont {S.}~\bibnamefont
  {Schmid}}, \bibinfo {author} {\bibfnamefont {A.}~\bibnamefont {H{\"a}rter}},
  \ and\ \bibinfo {author} {\bibfnamefont {J.~H.}\ \bibnamefont {Denschlag}},\
  }\href@noop {} {\bibfield  {journal} {\bibinfo  {journal} {Phys. Rev. Lett.}\
  }\textbf {\bibinfo {volume} {105}},\ \bibinfo {pages} {133202} (\bibinfo
  {year} {2010})}\BibitemShut {NoStop}%
\bibitem [{\citenamefont {Privitera}\ and\ \citenamefont
  {Hofstetter}(2010)}]{privitera2010polaronic}%
  \BibitemOpen
  \bibfield  {author} {\bibinfo {author} {\bibfnamefont {A.}~\bibnamefont
  {Privitera}}\ and\ \bibinfo {author} {\bibfnamefont {W.}~\bibnamefont
  {Hofstetter}},\ }\href@noop {} {\bibfield  {journal} {\bibinfo  {journal}
  {Phys. Rev. A}\ }\textbf {\bibinfo {volume} {82}},\ \bibinfo {pages} {063614}
  (\bibinfo {year} {2010})}\BibitemShut {NoStop}%
\bibitem [{\citenamefont {Casteels}\ \emph
  {et~al.}(2011{\natexlab{a}})\citenamefont {Casteels}, \citenamefont
  {Tempere},\ and\ \citenamefont {Devreese}}]{Casteels2011a}%
  \BibitemOpen
  \bibfield  {author} {\bibinfo {author} {\bibfnamefont {W.}~\bibnamefont
  {Casteels}}, \bibinfo {author} {\bibfnamefont {J.}~\bibnamefont {Tempere}}, \
  and\ \bibinfo {author} {\bibfnamefont {J.~T.}\ \bibnamefont {Devreese}},\
  }\href {\doibase 10.1103/PhysRevA.83.033631} {\bibfield  {journal} {\bibinfo
  {journal} {Phys. Rev. A}\ }\textbf {\bibinfo {volume} {83}},\ \bibinfo
  {pages} {033631} (\bibinfo {year} {2011}{\natexlab{a}})}\BibitemShut
  {NoStop}%
\bibitem [{\citenamefont {Casteels}\ \emph {et~al.}(2012)\citenamefont
  {Casteels}, \citenamefont {Tempere},\ and\ \citenamefont
  {Devreese}}]{Casteels2012}%
  \BibitemOpen
  \bibfield  {author} {\bibinfo {author} {\bibfnamefont {W.}~\bibnamefont
  {Casteels}}, \bibinfo {author} {\bibfnamefont {J.}~\bibnamefont {Tempere}}, \
  and\ \bibinfo {author} {\bibfnamefont {J.~T.}\ \bibnamefont {Devreese}},\
  }\href {http://link.aps.org/doi/10.1103/PhysRevA.86.043614
  papers2://publication/doi/10.1103/PhysRevA.86.043614} {\bibfield  {journal}
  {\bibinfo  {journal} {Phys. Rev. A}\ }\textbf {\bibinfo {volume} {86}},\
  \bibinfo {pages} {043614} (\bibinfo {year} {2012})}\BibitemShut {NoStop}%
\bibitem [{\citenamefont {Casteels}\ \emph
  {et~al.}(2011{\natexlab{b}})\citenamefont {Casteels}, \citenamefont
  {Tempere},\ and\ \citenamefont {Devreese}}]{Casteels2011}%
  \BibitemOpen
  \bibfield  {author} {\bibinfo {author} {\bibfnamefont {W.}~\bibnamefont
  {Casteels}}, \bibinfo {author} {\bibfnamefont {J.}~\bibnamefont {Tempere}}, \
  and\ \bibinfo {author} {\bibfnamefont {J.~T.}\ \bibnamefont {Devreese}},\
  }\href {\doibase 10.1103/PhysRevA.84.063612} {\bibfield  {journal} {\bibinfo
  {journal} {Phys. Rev. A}\ }\textbf {\bibinfo {volume} {84}},\ \bibinfo
  {pages} {063612} (\bibinfo {year} {2011}{\natexlab{b}})}\BibitemShut
  {NoStop}%
\bibitem [{\citenamefont {Tempere}\ \emph {et~al.}(2009)\citenamefont
  {Tempere}, \citenamefont {Casteels}, \citenamefont {Oberthaler},
  \citenamefont {Knoop}, \citenamefont {Timmermans},\ and\ \citenamefont
  {Devreese}}]{Tempere2009}%
  \BibitemOpen
  \bibfield  {author} {\bibinfo {author} {\bibfnamefont {J.}~\bibnamefont
  {Tempere}}, \bibinfo {author} {\bibfnamefont {W.}~\bibnamefont {Casteels}},
  \bibinfo {author} {\bibfnamefont {M.~K.}\ \bibnamefont {Oberthaler}},
  \bibinfo {author} {\bibfnamefont {S.}~\bibnamefont {Knoop}}, \bibinfo
  {author} {\bibfnamefont {E.}~\bibnamefont {Timmermans}}, \ and\ \bibinfo
  {author} {\bibfnamefont {J.~T.}\ \bibnamefont {Devreese}},\ }\href
  {http://link.aps.org/doi/10.1103/PhysRevB.80.184504
  papers2://publication/doi/10.1103/PhysRevB.80.184504} {\bibfield  {journal}
  {\bibinfo  {journal} {Phys. Rev. B}\ }\textbf {\bibinfo {volume} {80}},\
  \bibinfo {pages} {184504} (\bibinfo {year} {2009})}\BibitemShut {NoStop}%
\bibitem [{\citenamefont {Blinova}\ \emph {et~al.}(2013)\citenamefont
  {Blinova}, \citenamefont {Boshier},\ and\ \citenamefont
  {Timmermans}}]{blinova2013single}%
  \BibitemOpen
  \bibfield  {author} {\bibinfo {author} {\bibfnamefont {A.~A.}\ \bibnamefont
  {Blinova}}, \bibinfo {author} {\bibfnamefont {M.~G.}\ \bibnamefont
  {Boshier}}, \ and\ \bibinfo {author} {\bibfnamefont {E.}~\bibnamefont
  {Timmermans}},\ }\href@noop {} {\bibfield  {journal} {\bibinfo  {journal}
  {Phys. Rev. A}\ }\textbf {\bibinfo {volume} {88}},\ \bibinfo {pages} {053610}
  (\bibinfo {year} {2013})}\BibitemShut {NoStop}%
\bibitem [{\citenamefont {Rath}\ and\ \citenamefont
  {Schmidt}(2013)}]{rath2013polaron}%
  \BibitemOpen
  \bibfield  {author} {\bibinfo {author} {\bibfnamefont {S.~P.}\ \bibnamefont
  {Rath}}\ and\ \bibinfo {author} {\bibfnamefont {R.}~\bibnamefont {Schmidt}},\
  }\href@noop {} {\bibfield  {journal} {\bibinfo  {journal} {Phys. Rev. A}\
  }\textbf {\bibinfo {volume} {88}},\ \bibinfo {pages} {053632} (\bibinfo
  {year} {2013})}\BibitemShut {NoStop}%
\bibitem [{\citenamefont {Gupta}\ \emph {et~al.}(2003)\citenamefont {Gupta},
  \citenamefont {Hadzibabic}, \citenamefont {Zwierlein}, \citenamefont {Stan},
  \citenamefont {Dieckmann}, \citenamefont {Schunck}, \citenamefont {{Van
  Kempen}}, \citenamefont {Verhaar},\ and\ \citenamefont
  {Ketterle}}]{Gupta2003}%
  \BibitemOpen
  \bibfield  {author} {\bibinfo {author} {\bibfnamefont {S.}~\bibnamefont
  {Gupta}}, \bibinfo {author} {\bibfnamefont {Z.}~\bibnamefont {Hadzibabic}},
  \bibinfo {author} {\bibfnamefont {M.~W.}\ \bibnamefont {Zwierlein}}, \bibinfo
  {author} {\bibfnamefont {C.~a.}\ \bibnamefont {Stan}}, \bibinfo {author}
  {\bibfnamefont {K.}~\bibnamefont {Dieckmann}}, \bibinfo {author}
  {\bibfnamefont {C.~H.}\ \bibnamefont {Schunck}}, \bibinfo {author}
  {\bibfnamefont {E.~G.~M.}\ \bibnamefont {{Van Kempen}}}, \bibinfo {author}
  {\bibfnamefont {B.~J.}\ \bibnamefont {Verhaar}}, \ and\ \bibinfo {author}
  {\bibfnamefont {W.}~\bibnamefont {Ketterle}},\ }\href {\doibase
  10.1126/science.1085335} {\bibfield  {journal} {\bibinfo  {journal}
  {Science}\ }\textbf {\bibinfo {volume} {300}},\ \bibinfo {pages} {1723}
  (\bibinfo {year} {2003})}\BibitemShut {NoStop}%
\bibitem [{\citenamefont {Shin}\ \emph {et~al.}(2004)\citenamefont {Shin},
  \citenamefont {Saba}, \citenamefont {Pasquini}, \citenamefont {Ketterle},
  \citenamefont {Pritchard},\ and\ \citenamefont {Leanhardt}}]{shin2004atom}%
  \BibitemOpen
  \bibfield  {author} {\bibinfo {author} {\bibfnamefont {Y.}~\bibnamefont
  {Shin}}, \bibinfo {author} {\bibfnamefont {M.}~\bibnamefont {Saba}}, \bibinfo
  {author} {\bibfnamefont {T.~A.}\ \bibnamefont {Pasquini}}, \bibinfo {author}
  {\bibfnamefont {W.}~\bibnamefont {Ketterle}}, \bibinfo {author}
  {\bibfnamefont {D.~E.}\ \bibnamefont {Pritchard}}, \ and\ \bibinfo {author}
  {\bibfnamefont {A.~E.}\ \bibnamefont {Leanhardt}},\ }\href@noop {} {\bibfield
   {journal} {\bibinfo  {journal} {Phys. Rev. Lett.}\ }\textbf {\bibinfo
  {volume} {92}},\ \bibinfo {pages} {050405} (\bibinfo {year}
  {2004})}\BibitemShut {NoStop}%
\bibitem [{\citenamefont {Chin}\ \emph {et~al.}(2010)\citenamefont {Chin},
  \citenamefont {Grimm}, \citenamefont {Julienne},\ and\ \citenamefont
  {Tiesinga}}]{Chin.paper}%
  \BibitemOpen
  \bibfield  {author} {\bibinfo {author} {\bibfnamefont {C.}~\bibnamefont
  {Chin}}, \bibinfo {author} {\bibfnamefont {R.}~\bibnamefont {Grimm}},
  \bibinfo {author} {\bibfnamefont {P.}~\bibnamefont {Julienne}}, \ and\
  \bibinfo {author} {\bibfnamefont {E.}~\bibnamefont {Tiesinga}},\ }\href@noop
  {} {\bibfield  {journal} {\bibinfo  {journal} {Rev. Mod. Phys.}\ }\textbf
  {\bibinfo {volume} {82}},\ \bibinfo {pages} {1225} (\bibinfo {year}
  {2010})}\BibitemShut {NoStop}%
\bibitem [{\citenamefont {Bloch}\ \emph {et~al.}(2008)\citenamefont {Bloch},
  \citenamefont {Dalibard},\ and\ \citenamefont {Zwerger}}]{bloch2008many}%
  \BibitemOpen
  \bibfield  {author} {\bibinfo {author} {\bibfnamefont {I.}~\bibnamefont
  {Bloch}}, \bibinfo {author} {\bibfnamefont {J.}~\bibnamefont {Dalibard}}, \
  and\ \bibinfo {author} {\bibfnamefont {W.}~\bibnamefont {Zwerger}},\
  }\href@noop {} {\bibfield  {journal} {\bibinfo  {journal} {Rev. Mod. Phys.}\
  }\textbf {\bibinfo {volume} {80}},\ \bibinfo {pages} {885} (\bibinfo {year}
  {2008})}\BibitemShut {NoStop}%
\bibitem [{\citenamefont {Zwierlein}\ \emph {et~al.}(2003)\citenamefont
  {Zwierlein}, \citenamefont {Hadzibabic}, \citenamefont {Gupta},\ and\
  \citenamefont {Ketterle}}]{Zwierlein2003}%
  \BibitemOpen
  \bibfield  {author} {\bibinfo {author} {\bibfnamefont {M.~W.}\ \bibnamefont
  {Zwierlein}}, \bibinfo {author} {\bibfnamefont {Z.}~\bibnamefont
  {Hadzibabic}}, \bibinfo {author} {\bibfnamefont {S.}~\bibnamefont {Gupta}}, \
  and\ \bibinfo {author} {\bibfnamefont {W.}~\bibnamefont {Ketterle}},\ }\href
  {\doibase 10.1103/PhysRevLett.91.250404} {\bibfield  {journal} {\bibinfo
  {journal} {Phys. Rev. Lett.}\ }\textbf {\bibinfo {volume} {91}},\ \bibinfo
  {pages} {250404} (\bibinfo {year} {2003})}\BibitemShut {NoStop}%
\bibitem [{\citenamefont {Chin}\ \emph {et~al.}(2004)\citenamefont {Chin},
  \citenamefont {Bartenstein}, \citenamefont {Altmeyer}, \citenamefont {Riedl},
  \citenamefont {Jochim}, \citenamefont {Denschlag},\ and\ \citenamefont
  {Grimm}}]{Chin2004}%
  \BibitemOpen
  \bibfield  {author} {\bibinfo {author} {\bibfnamefont {C.}~\bibnamefont
  {Chin}}, \bibinfo {author} {\bibfnamefont {M.}~\bibnamefont {Bartenstein}},
  \bibinfo {author} {\bibfnamefont {a.}~\bibnamefont {Altmeyer}}, \bibinfo
  {author} {\bibfnamefont {S.}~\bibnamefont {Riedl}}, \bibinfo {author}
  {\bibfnamefont {S.}~\bibnamefont {Jochim}}, \bibinfo {author} {\bibfnamefont
  {J.~H.}\ \bibnamefont {Denschlag}}, \ and\ \bibinfo {author} {\bibfnamefont
  {R.}~\bibnamefont {Grimm}},\ }\href {\doibase 10.1126/science.1100818}
  {\bibfield  {journal} {\bibinfo  {journal} {Science (New York, N.Y.)}\
  }\textbf {\bibinfo {volume} {305}},\ \bibinfo {pages} {1128} (\bibinfo {year}
  {2004})}\BibitemShut {NoStop}%
\bibitem [{\citenamefont {Bartenstein}\ \emph {et~al.}(2005)\citenamefont
  {Bartenstein}, \citenamefont {Altmeyer}, \citenamefont {Riedl}, \citenamefont
  {Geursen}, \citenamefont {Jochim}, \citenamefont {Chin}, \citenamefont
  {Denschlag}, \citenamefont {Grimm}, \citenamefont {Simoni}, \citenamefont
  {Tiesinga}, \citenamefont {Williams},\ and\ \citenamefont
  {Julienne}}]{Bartenstein2005}%
  \BibitemOpen
  \bibfield  {author} {\bibinfo {author} {\bibfnamefont {M.}~\bibnamefont
  {Bartenstein}}, \bibinfo {author} {\bibfnamefont {A.}~\bibnamefont
  {Altmeyer}}, \bibinfo {author} {\bibfnamefont {S.}~\bibnamefont {Riedl}},
  \bibinfo {author} {\bibfnamefont {R.}~\bibnamefont {Geursen}}, \bibinfo
  {author} {\bibfnamefont {S.}~\bibnamefont {Jochim}}, \bibinfo {author}
  {\bibfnamefont {C.}~\bibnamefont {Chin}}, \bibinfo {author} {\bibfnamefont
  {J.~H.}\ \bibnamefont {Denschlag}}, \bibinfo {author} {\bibfnamefont
  {R.}~\bibnamefont {Grimm}}, \bibinfo {author} {\bibfnamefont
  {A.}~\bibnamefont {Simoni}}, \bibinfo {author} {\bibfnamefont
  {E.}~\bibnamefont {Tiesinga}}, \bibinfo {author} {\bibfnamefont {C.~J.}\
  \bibnamefont {Williams}}, \ and\ \bibinfo {author} {\bibfnamefont {P.~S.}\
  \bibnamefont {Julienne}},\ }\href {\doibase 10.1103/PhysRevLett.94.103201}
  {\bibfield  {journal} {\bibinfo  {journal} {Phys. Rev. Lett.}\ }\textbf
  {\bibinfo {volume} {94}},\ \bibinfo {pages} {103201} (\bibinfo {year}
  {2005})}\BibitemShut {NoStop}%
\bibitem [{\citenamefont {Shin}\ \emph {et~al.}(2007)\citenamefont {Shin},
  \citenamefont {Schunck}, \citenamefont {Schirotzek},\ and\ \citenamefont
  {Ketterle}}]{Shin2007}%
  \BibitemOpen
  \bibfield  {author} {\bibinfo {author} {\bibfnamefont {Y.-I.}\ \bibnamefont
  {Shin}}, \bibinfo {author} {\bibfnamefont {C.~H.}\ \bibnamefont {Schunck}},
  \bibinfo {author} {\bibfnamefont {A.}~\bibnamefont {Schirotzek}}, \ and\
  \bibinfo {author} {\bibfnamefont {W.}~\bibnamefont {Ketterle}},\ }\href
  {\doibase 10.1103/PhysRevLett.99.090403} {\bibfield  {journal} {\bibinfo
  {journal} {Phys. Rev. Lett.}\ }\textbf {\bibinfo {volume} {99}},\ \bibinfo
  {pages} {090403} (\bibinfo {year} {2007})}\BibitemShut {NoStop}%
\bibitem [{\citenamefont {Sommer}\ \emph {et~al.}(2012)\citenamefont {Sommer},
  \citenamefont {Cheuk}, \citenamefont {Ku}, \citenamefont {Bakr},\ and\
  \citenamefont {Zwierlein}}]{Sommer2012}%
  \BibitemOpen
  \bibfield  {author} {\bibinfo {author} {\bibfnamefont {A.~T.}\ \bibnamefont
  {Sommer}}, \bibinfo {author} {\bibfnamefont {L.~W.}\ \bibnamefont {Cheuk}},
  \bibinfo {author} {\bibfnamefont {M.~J.~H.}\ \bibnamefont {Ku}}, \bibinfo
  {author} {\bibfnamefont {W.~S.}\ \bibnamefont {Bakr}}, \ and\ \bibinfo
  {author} {\bibfnamefont {M.~W.}\ \bibnamefont {Zwierlein}},\ }\href {\doibase
  10.1103/PhysRevLett.108.045302} {\bibfield  {journal} {\bibinfo  {journal}
  {Phys. Rev. Lett.}\ }\textbf {\bibinfo {volume} {108}},\ \bibinfo {pages}
  {045302} (\bibinfo {year} {2012})}\BibitemShut {NoStop}%
\bibitem [{\citenamefont {Stewart}\ \emph {et~al.}(2008)\citenamefont
  {Stewart}, \citenamefont {Gaebler},\ and\ \citenamefont {Jin}}]{Stewart2008}%
  \BibitemOpen
  \bibfield  {author} {\bibinfo {author} {\bibfnamefont {J.~T.}\ \bibnamefont
  {Stewart}}, \bibinfo {author} {\bibfnamefont {J.~P.}\ \bibnamefont
  {Gaebler}}, \ and\ \bibinfo {author} {\bibfnamefont {D.~S.}\ \bibnamefont
  {Jin}},\ }\href {\doibase 10.1038/nature07172} {\bibfield  {journal}
  {\bibinfo  {journal} {Nature}\ }\textbf {\bibinfo {volume} {454}},\ \bibinfo
  {pages} {744} (\bibinfo {year} {2008})}\BibitemShut {NoStop}%
\bibitem [{\citenamefont {Feld}\ \emph {et~al.}(2011)\citenamefont {Feld},
  \citenamefont {Fr\"{o}hlich}, \citenamefont {Vogt}, \citenamefont
  {Koschorreck},\ and\ \citenamefont {K\"{o}hl}}]{Feld2011}%
  \BibitemOpen
  \bibfield  {author} {\bibinfo {author} {\bibfnamefont {M.}~\bibnamefont
  {Feld}}, \bibinfo {author} {\bibfnamefont {B.}~\bibnamefont {Fr\"{o}hlich}},
  \bibinfo {author} {\bibfnamefont {E.}~\bibnamefont {Vogt}}, \bibinfo {author}
  {\bibfnamefont {M.}~\bibnamefont {Koschorreck}}, \ and\ \bibinfo {author}
  {\bibfnamefont {M.}~\bibnamefont {K\"{o}hl}},\ }\href {\doibase
  10.1038/nature10627} {\bibfield  {journal} {\bibinfo  {journal} {Nature}\
  }\textbf {\bibinfo {volume} {480}},\ \bibinfo {pages} {75} (\bibinfo {year}
  {2011})}\BibitemShut {NoStop}%
\bibitem [{\citenamefont {Schirotzek}\ \emph {et~al.}(2008)\citenamefont
  {Schirotzek}, \citenamefont {Shin}, \citenamefont {Schunck},\ and\
  \citenamefont {Ketterle}}]{Schirotzek2008}%
  \BibitemOpen
  \bibfield  {author} {\bibinfo {author} {\bibfnamefont {A.}~\bibnamefont
  {Schirotzek}}, \bibinfo {author} {\bibfnamefont {Y.-I.}\ \bibnamefont
  {Shin}}, \bibinfo {author} {\bibfnamefont {C.~H.}\ \bibnamefont {Schunck}}, \
  and\ \bibinfo {author} {\bibfnamefont {W.}~\bibnamefont {Ketterle}},\ }\href
  {\doibase 10.1103/PhysRevLett.101.140403} {\bibfield  {journal} {\bibinfo
  {journal} {Phys. Rev. Lett.}\ }\textbf {\bibinfo {volume} {101}},\ \bibinfo
  {pages} {140403} (\bibinfo {year} {2008})}\BibitemShut {NoStop}%
\bibitem [{\citenamefont {Schunck}\ \emph {et~al.}(2008)\citenamefont
  {Schunck}, \citenamefont {Shin}, \citenamefont {Schirotzek}, \citenamefont
  {Zwierlein},\ and\ \citenamefont
  {Ketterle}}]{C.H.SchunckY.ShinA.SchirotzekM.W.Zwierlein2008}%
  \BibitemOpen
  \bibfield  {author} {\bibinfo {author} {\bibfnamefont {C.~H.}\ \bibnamefont
  {Schunck}}, \bibinfo {author} {\bibfnamefont {Y.}~\bibnamefont {Shin}},
  \bibinfo {author} {\bibfnamefont {A.}~\bibnamefont {Schirotzek}}, \bibinfo
  {author} {\bibfnamefont {M.~W.}\ \bibnamefont {Zwierlein}}, \ and\ \bibinfo
  {author} {\bibfnamefont {W.}~\bibnamefont {Ketterle}},\ }\href@noop {}
  {\bibfield  {journal} {\bibinfo  {journal} {Science}\ }\textbf {\bibinfo
  {volume} {316}},\ \bibinfo {pages} {867} (\bibinfo {year}
  {2008})}\BibitemShut {NoStop}%
\bibitem [{\citenamefont {Veillette}\ \emph {et~al.}(2008)\citenamefont
  {Veillette}, \citenamefont {Moon}, \citenamefont {Lamacraft}, \citenamefont
  {Radzihovsky}, \citenamefont {Sachdev},\ and\ \citenamefont
  {Sheehy}}]{veillette2008radio}%
  \BibitemOpen
  \bibfield  {author} {\bibinfo {author} {\bibfnamefont {M.}~\bibnamefont
  {Veillette}}, \bibinfo {author} {\bibfnamefont {E.~G.}\ \bibnamefont {Moon}},
  \bibinfo {author} {\bibfnamefont {A.}~\bibnamefont {Lamacraft}}, \bibinfo
  {author} {\bibfnamefont {L.}~\bibnamefont {Radzihovsky}}, \bibinfo {author}
  {\bibfnamefont {S.}~\bibnamefont {Sachdev}}, \ and\ \bibinfo {author}
  {\bibfnamefont {D.~E.}\ \bibnamefont {Sheehy}},\ }\href@noop {} {\bibfield
  {journal} {\bibinfo  {journal} {Phys. Rev. A}\ }\textbf {\bibinfo {volume}
  {78}},\ \bibinfo {pages} {033614} (\bibinfo {year} {2008})}\BibitemShut
  {NoStop}%
\bibitem [{\citenamefont {Knap}\ \emph {et~al.}(2012)\citenamefont {Knap},
  \citenamefont {Shashi}, \citenamefont {Nishida}, \citenamefont {Imambekov},
  \citenamefont {Abanin},\ and\ \citenamefont {Demler}}]{Knap2012}%
  \BibitemOpen
  \bibfield  {author} {\bibinfo {author} {\bibfnamefont {M.}~\bibnamefont
  {Knap}}, \bibinfo {author} {\bibfnamefont {A.}~\bibnamefont {Shashi}},
  \bibinfo {author} {\bibfnamefont {Y.}~\bibnamefont {Nishida}}, \bibinfo
  {author} {\bibfnamefont {A.}~\bibnamefont {Imambekov}}, \bibinfo {author}
  {\bibfnamefont {D.~A.}\ \bibnamefont {Abanin}}, \ and\ \bibinfo {author}
  {\bibfnamefont {E.}~\bibnamefont {Demler}},\ }\href {\doibase
  10.1103/PhysRevX.2.041020} {\bibfield  {journal} {\bibinfo  {journal} {Phys.
  Rev. X}\ }\textbf {\bibinfo {volume} {2}},\ \bibinfo {pages} {041020}
  (\bibinfo {year} {2012})}\BibitemShut {NoStop}%
\bibitem [{\citenamefont {Pitaevskii}\ and\ \citenamefont
  {Stringari}(2003)}]{stringarii.book}%
  \BibitemOpen
  \bibfield  {author} {\bibinfo {author} {\bibfnamefont {L.~P.}\ \bibnamefont
  {Pitaevskii}}\ and\ \bibinfo {author} {\bibfnamefont {S.}~\bibnamefont
  {Stringari}},\ }\href@noop {} {\emph {\bibinfo {title} {Bose-einstein
  condensation}}},\ \bibinfo {number} {116}\ (\bibinfo  {publisher} {Oxford
  University Press},\ \bibinfo {year} {2003})\BibitemShut {NoStop}%
\bibitem [{\citenamefont {Tan}(2008{\natexlab{a}})}]{tan2008large}%
  \BibitemOpen
  \bibfield  {author} {\bibinfo {author} {\bibfnamefont {S.}~\bibnamefont
  {Tan}},\ }\href@noop {} {\bibfield  {journal} {\bibinfo  {journal} {Ann. of
  Phys.}\ }\textbf {\bibinfo {volume} {323}},\ \bibinfo {pages} {2971}
  (\bibinfo {year} {2008}{\natexlab{a}})}\BibitemShut {NoStop}%
\bibitem [{\citenamefont {Tan}(2008{\natexlab{b}})}]{tan2008energetics}%
  \BibitemOpen
  \bibfield  {author} {\bibinfo {author} {\bibfnamefont {S.}~\bibnamefont
  {Tan}},\ }\href@noop {} {\bibfield  {journal} {\bibinfo  {journal} {Ann. of
  Phys.}\ }\textbf {\bibinfo {volume} {323}},\ \bibinfo {pages} {2952}
  (\bibinfo {year} {2008}{\natexlab{b}})}\BibitemShut {NoStop}%
\bibitem [{\citenamefont {Punk}\ and\ \citenamefont
  {Zwerger}(2007)}]{punk2007theory}%
  \BibitemOpen
  \bibfield  {author} {\bibinfo {author} {\bibfnamefont {M.}~\bibnamefont
  {Punk}}\ and\ \bibinfo {author} {\bibfnamefont {W.}~\bibnamefont {Zwerger}},\
  }\href@noop {} {\bibfield  {journal} {\bibinfo  {journal} {Phys. Rev. Lett.}\
  }\textbf {\bibinfo {volume} {99}},\ \bibinfo {pages} {170404} (\bibinfo
  {year} {2007})}\BibitemShut {NoStop}%
\bibitem [{\citenamefont {Haussmann}\ \emph {et~al.}(2009)\citenamefont
  {Haussmann}, \citenamefont {Punk},\ and\ \citenamefont
  {Zwerger}}]{haussmann2009spectral}%
  \BibitemOpen
  \bibfield  {author} {\bibinfo {author} {\bibfnamefont {R.}~\bibnamefont
  {Haussmann}}, \bibinfo {author} {\bibfnamefont {M.}~\bibnamefont {Punk}}, \
  and\ \bibinfo {author} {\bibfnamefont {W.}~\bibnamefont {Zwerger}},\
  }\href@noop {} {\bibfield  {journal} {\bibinfo  {journal} {Phys. Rev. A}\
  }\textbf {\bibinfo {volume} {80}},\ \bibinfo {pages} {063612} (\bibinfo
  {year} {2009})}\BibitemShut {NoStop}%
\bibitem [{\citenamefont {Schneider}\ and\ \citenamefont
  {Randeria}(2010)}]{schneider2010universal}%
  \BibitemOpen
  \bibfield  {author} {\bibinfo {author} {\bibfnamefont {W.}~\bibnamefont
  {Schneider}}\ and\ \bibinfo {author} {\bibfnamefont {M.}~\bibnamefont
  {Randeria}},\ }\href@noop {} {\bibfield  {journal} {\bibinfo  {journal}
  {Phys. Rev. A}\ }\textbf {\bibinfo {volume} {81}},\ \bibinfo {pages} {021601}
  (\bibinfo {year} {2010})}\BibitemShut {NoStop}%
\bibitem [{\citenamefont {Braaten}\ \emph {et~al.}(2011)\citenamefont
  {Braaten}, \citenamefont {Kang},\ and\ \citenamefont
  {Platter}}]{braaten2011universal}%
  \BibitemOpen
  \bibfield  {author} {\bibinfo {author} {\bibfnamefont {E.}~\bibnamefont
  {Braaten}}, \bibinfo {author} {\bibfnamefont {D.}~\bibnamefont {Kang}}, \
  and\ \bibinfo {author} {\bibfnamefont {L.}~\bibnamefont {Platter}},\
  }\href@noop {} {\bibfield  {journal} {\bibinfo  {journal} {Phys. Rev. Lett.}\
  }\textbf {\bibinfo {volume} {106}},\ \bibinfo {pages} {153005} (\bibinfo
  {year} {2011})}\BibitemShut {NoStop}%
\bibitem [{\citenamefont {Langmack}\ \emph {et~al.}(2012)\citenamefont
  {Langmack}, \citenamefont {Barth}, \citenamefont {Zwerger},\ and\
  \citenamefont {Braaten}}]{langmack2012clock}%
  \BibitemOpen
  \bibfield  {author} {\bibinfo {author} {\bibfnamefont {C.}~\bibnamefont
  {Langmack}}, \bibinfo {author} {\bibfnamefont {M.}~\bibnamefont {Barth}},
  \bibinfo {author} {\bibfnamefont {W.}~\bibnamefont {Zwerger}}, \ and\
  \bibinfo {author} {\bibfnamefont {E.}~\bibnamefont {Braaten}},\ }\href@noop
  {} {\bibfield  {journal} {\bibinfo  {journal} {Phys. Rev. Lett.}\ }\textbf
  {\bibinfo {volume} {108}},\ \bibinfo {pages} {060402} (\bibinfo {year}
  {2012})}\BibitemShut {NoStop}%
\bibitem [{\citenamefont {Huang}(1987)}]{Huang.book}%
  \BibitemOpen
  \bibfield  {author} {\bibinfo {author} {\bibfnamefont {K.}~\bibnamefont
  {Huang}},\ }\href@noop {} {\enquote {\bibinfo {title} {Statistical
  mechanics},}\ } (\bibinfo {year} {1987})\BibitemShut {NoStop}%
\bibitem [{\citenamefont {Mahan}(2000)}]{mahan2000many}%
  \BibitemOpen
  \bibfield  {author} {\bibinfo {author} {\bibfnamefont {G.~D.}\ \bibnamefont
  {Mahan}},\ }\href@noop {} {\emph {\bibinfo {title} {Many particle physics}}}\
  (\bibinfo  {publisher} {Springer},\ \bibinfo {year} {2000})\BibitemShut
  {NoStop}%
\bibitem [{\citenamefont {Silva}(2008)}]{Silva2008}%
  \BibitemOpen
  \bibfield  {author} {\bibinfo {author} {\bibfnamefont {A.}~\bibnamefont
  {Silva}},\ }\href {http://link.aps.org/doi/10.1103/PhysRevLett.101.120603
  papers2://publication/doi/10.1103/PhysRevLett.101.120603} {\bibfield
  {journal} {\bibinfo  {journal} {Phys. Rev. Lett.}\ }\textbf {\bibinfo
  {volume} {101}},\ \bibinfo {pages} {120603} (\bibinfo {year}
  {2008})}\BibitemShut {NoStop}%
\bibitem [{\citenamefont {Chevy}(2006)}]{Chevy2006}%
  \BibitemOpen
  \bibfield  {author} {\bibinfo {author} {\bibfnamefont {F.}~\bibnamefont
  {Chevy}},\ }\href {http://link.aps.org/doi/10.1103/PhysRevA.74.063628
  papers2://publication/doi/10.1103/PhysRevA.74.063628} {\bibfield  {journal}
  {\bibinfo  {journal} {Phys. Rev. A}\ }\textbf {\bibinfo {volume} {74}},\
  \bibinfo {pages} {063628} (\bibinfo {year} {2006})}\BibitemShut {NoStop}%
\bibitem [{\citenamefont {Combescot}\ \emph {et~al.}(2007)\citenamefont
  {Combescot}, \citenamefont {Recati}, \citenamefont {Lobo},\ and\
  \citenamefont {Chevy}}]{Combescot2007}%
  \BibitemOpen
  \bibfield  {author} {\bibinfo {author} {\bibfnamefont {R.}~\bibnamefont
  {Combescot}}, \bibinfo {author} {\bibfnamefont {A.}~\bibnamefont {Recati}},
  \bibinfo {author} {\bibfnamefont {C.}~\bibnamefont {Lobo}}, \ and\ \bibinfo
  {author} {\bibfnamefont {F.}~\bibnamefont {Chevy}},\ }\href
  {http://link.aps.org/doi/10.1103/PhysRevLett.98.180402
  papers2://publication/doi/10.1103/PhysRevLett.98.180402} {\bibfield
  {journal} {\bibinfo  {journal} {Phys. Rev. Lett.}\ }\textbf {\bibinfo
  {volume} {98}},\ \bibinfo {pages} {180402} (\bibinfo {year}
  {2007})}\BibitemShut {NoStop}%
\bibitem [{\citenamefont {Anderson}(1967)}]{Anderson.OC}%
  \BibitemOpen
  \bibfield  {author} {\bibinfo {author} {\bibfnamefont {P.}~\bibnamefont
  {Anderson}},\ }\href@noop {} {\bibfield  {journal} {\bibinfo  {journal}
  {Phys. Rev. Lett.}\ }\textbf {\bibinfo {volume} {18}},\ \bibinfo {pages}
  {1049} (\bibinfo {year} {1967})}\BibitemShut {NoStop}%
\bibitem [{\citenamefont {Alexandrov}\ and\ \citenamefont
  {Mott}(1995)}]{alexandrov1995polarons}%
  \BibitemOpen
  \bibfield  {author} {\bibinfo {author} {\bibfnamefont {A.~S.}\ \bibnamefont
  {Alexandrov}}\ and\ \bibinfo {author} {\bibfnamefont {N.~F.}\ \bibnamefont
  {Mott}},\ }\href@noop {} {\emph {\bibinfo {title} {Polarons \& bipolarons}}}\
  (\bibinfo  {publisher} {World Scientific Singapore},\ \bibinfo {year}
  {1995})\BibitemShut {NoStop}%
\bibitem [{\citenamefont {Bei-Bing}\ and\ \citenamefont
  {Shao-Long}(2009)}]{bei2009polaron}%
  \BibitemOpen
  \bibfield  {author} {\bibinfo {author} {\bibfnamefont {H.}~\bibnamefont
  {Bei-Bing}}\ and\ \bibinfo {author} {\bibfnamefont {W.}~\bibnamefont
  {Shao-Long}},\ }\href@noop {} {\bibfield  {journal} {\bibinfo  {journal}
  {Chin. Phys. Lett.}\ }\textbf {\bibinfo {volume} {26}},\ \bibinfo {pages}
  {080302} (\bibinfo {year} {2009})}\BibitemShut {NoStop}%
\bibitem [{\citenamefont {Lee}\ \emph {et~al.}(1953)\citenamefont {Lee},
  \citenamefont {Low},\ and\ \citenamefont {Pines}}]{lee1953motion}%
  \BibitemOpen
  \bibfield  {author} {\bibinfo {author} {\bibfnamefont {T.}~\bibnamefont
  {Lee}}, \bibinfo {author} {\bibfnamefont {F.}~\bibnamefont {Low}}, \ and\
  \bibinfo {author} {\bibfnamefont {D.}~\bibnamefont {Pines}},\ }\href@noop {}
  {\bibfield  {journal} {\bibinfo  {journal} {Phys. Rev.}\ }\textbf {\bibinfo
  {volume} {90}},\ \bibinfo {pages} {297} (\bibinfo {year} {1953})}\BibitemShut
  {NoStop}%
\bibitem [{\citenamefont {Gerlach}\ and\ \citenamefont
  {L{\"o}wen}(1987)}]{gerlach1987proof}%
  \BibitemOpen
  \bibfield  {author} {\bibinfo {author} {\bibfnamefont {B.}~\bibnamefont
  {Gerlach}}\ and\ \bibinfo {author} {\bibfnamefont {H.}~\bibnamefont
  {L{\"o}wen}},\ }\href@noop {} {\bibfield  {journal} {\bibinfo  {journal}
  {Phys. Rev. B}\ }\textbf {\bibinfo {volume} {35}},\ \bibinfo {pages} {4291}
  (\bibinfo {year} {1987})}\BibitemShut {NoStop}%
\bibitem [{\citenamefont {Gerlach}\ and\ \citenamefont
  {L{\"o}wen}(1991)}]{gerlach1991analytical}%
  \BibitemOpen
  \bibfield  {author} {\bibinfo {author} {\bibfnamefont {B.}~\bibnamefont
  {Gerlach}}\ and\ \bibinfo {author} {\bibfnamefont {H.}~\bibnamefont
  {L{\"o}wen}},\ }\href@noop {} {\bibfield  {journal} {\bibinfo  {journal}
  {Rev. Mod. Phys.}\ }\textbf {\bibinfo {volume} {63}},\ \bibinfo {pages} {63}
  (\bibinfo {year} {1991})}\BibitemShut {NoStop}%
\bibitem [{\citenamefont {Spohn}(1986)}]{spohn1986roughening}%
  \BibitemOpen
  \bibfield  {author} {\bibinfo {author} {\bibfnamefont {H.}~\bibnamefont
  {Spohn}},\ }\href@noop {} {\bibfield  {journal} {\bibinfo  {journal} {J.
  Phys. A}\ }\textbf {\bibinfo {volume} {19}},\ \bibinfo {pages} {533}
  (\bibinfo {year} {1986})}\BibitemShut {NoStop}%
\bibitem [{\citenamefont {Glauber}(2007)}]{Glauber.book}%
  \BibitemOpen
  \bibfield  {author} {\bibinfo {author} {\bibfnamefont {R.~J.}\ \bibnamefont
  {Glauber}},\ }\href@noop {} {\emph {\bibinfo {title} {Quantum theory of
  optical coherence}}}\ (\bibinfo  {publisher} {Wiley. com},\ \bibinfo {year}
  {2007})\BibitemShut {NoStop}%
\bibitem [{\citenamefont {Zinn-Justin}(2007)}]{Zinn-Justin.book}%
  \BibitemOpen
  \bibfield  {author} {\bibinfo {author} {\bibfnamefont {J.}~\bibnamefont
  {Zinn-Justin}},\ }\href@noop {} {\emph {\bibinfo {title} {Phase transitions
  and renormalization group}}}\ (\bibinfo  {publisher} {Oxford University
  Press},\ \bibinfo {year} {2007})\BibitemShut {NoStop}%
\bibitem [{\citenamefont {Abrikosov}\ \emph {et~al.}(1965)\citenamefont
  {Abrikosov}, \citenamefont {Gorkov},\ and\ \citenamefont
  {Dzyaloshinskii}}]{abrikosov1965quantum}%
  \BibitemOpen
  \bibfield  {author} {\bibinfo {author} {\bibfnamefont {A.~A.}\ \bibnamefont
  {Abrikosov}}, \bibinfo {author} {\bibfnamefont {L.~P.}\ \bibnamefont
  {Gorkov}}, \ and\ \bibinfo {author} {\bibnamefont {Dzyaloshinskii}},\
  }\href@noop {} {\emph {\bibinfo {title} {Quantum field theoretical methods in
  statistical physics}}},\ Vol.~\bibinfo {volume} {4}\ (\bibinfo  {publisher}
  {Pergamon},\ \bibinfo {year} {1965})\BibitemShut {NoStop}%
\bibitem [{\citenamefont {Gogolin}\ \emph {et~al.}(2004)\citenamefont
  {Gogolin}, \citenamefont {Nersesyan},\ and\ \citenamefont
  {Tsvelik}}]{Tsvelik.book}%
  \BibitemOpen
  \bibfield  {author} {\bibinfo {author} {\bibfnamefont {A.~O.}\ \bibnamefont
  {Gogolin}}, \bibinfo {author} {\bibfnamefont {A.~A.}\ \bibnamefont
  {Nersesyan}}, \ and\ \bibinfo {author} {\bibfnamefont {A.~M.}\ \bibnamefont
  {Tsvelik}},\ }\href@noop {} {\emph {\bibinfo {title} {Bosonization and
  strongly correlated systems}}}\ (\bibinfo  {publisher} {Cambridge University
  Press},\ \bibinfo {year} {2004})\BibitemShut {NoStop}%
\bibitem [{\citenamefont {Combescot}\ \emph {et~al.}(2009)\citenamefont
  {Combescot}, \citenamefont {Alzetto},\ and\ \citenamefont
  {Leyronas}}]{combescot2009particle}%
  \BibitemOpen
  \bibfield  {author} {\bibinfo {author} {\bibfnamefont {R.}~\bibnamefont
  {Combescot}}, \bibinfo {author} {\bibfnamefont {F.}~\bibnamefont {Alzetto}},
  \ and\ \bibinfo {author} {\bibfnamefont {X.}~\bibnamefont {Leyronas}},\
  }\href@noop {} {\bibfield  {journal} {\bibinfo  {journal} {Phys. Rev. A}\
  }\textbf {\bibinfo {volume} {79}},\ \bibinfo {pages} {053640} (\bibinfo
  {year} {2009})}\BibitemShut {NoStop}%
\bibitem [{\citenamefont {Braaten}\ and\ \citenamefont
  {Platter}(2008)}]{braaten2008exact}%
  \BibitemOpen
  \bibfield  {author} {\bibinfo {author} {\bibfnamefont {E.}~\bibnamefont
  {Braaten}}\ and\ \bibinfo {author} {\bibfnamefont {L.}~\bibnamefont
  {Platter}},\ }\href@noop {} {\bibfield  {journal} {\bibinfo  {journal} {Phys.
  Rev. Lett.}\ }\textbf {\bibinfo {volume} {100}},\ \bibinfo {pages} {205301}
  (\bibinfo {year} {2008})}\BibitemShut {NoStop}%
\bibitem [{\citenamefont {Altman}\ \emph {et~al.}(2005)\citenamefont {Altman},
  \citenamefont {Polkovnikov}, \citenamefont {Demler}, \citenamefont
  {Halperin},\ and\ \citenamefont {Lukin}}]{altman2005superfluid}%
  \BibitemOpen
  \bibfield  {author} {\bibinfo {author} {\bibfnamefont {E.}~\bibnamefont
  {Altman}}, \bibinfo {author} {\bibfnamefont {A.}~\bibnamefont {Polkovnikov}},
  \bibinfo {author} {\bibfnamefont {E.}~\bibnamefont {Demler}}, \bibinfo
  {author} {\bibfnamefont {B.~I.}\ \bibnamefont {Halperin}}, \ and\ \bibinfo
  {author} {\bibfnamefont {M.~D.}\ \bibnamefont {Lukin}},\ }\href@noop {}
  {\bibfield  {journal} {\bibinfo  {journal} {Phys. Rev. Lett.}\ }\textbf
  {\bibinfo {volume} {95}},\ \bibinfo {pages} {020402} (\bibinfo {year}
  {2005})}\BibitemShut {NoStop}%
\bibitem [{\citenamefont {Demler}\ and\ \citenamefont
  {Maltsev}(2011)}]{demler2011semiclassical}%
  \BibitemOpen
  \bibfield  {author} {\bibinfo {author} {\bibfnamefont {E.}~\bibnamefont
  {Demler}}\ and\ \bibinfo {author} {\bibfnamefont {A.}~\bibnamefont
  {Maltsev}},\ }\href@noop {} {\bibfield  {journal} {\bibinfo  {journal} {Ann.
  of Phys.}\ }\textbf {\bibinfo {volume} {326}},\ \bibinfo {pages} {1775}
  (\bibinfo {year} {2011})}\BibitemShut {NoStop}%
\bibitem [{\citenamefont {Rigol}\ \emph {et~al.}(2008)\citenamefont {Rigol},
  \citenamefont {Dunjko},\ and\ \citenamefont
  {Olshanii}}]{rigol2008thermalization}%
  \BibitemOpen
  \bibfield  {author} {\bibinfo {author} {\bibfnamefont {M.}~\bibnamefont
  {Rigol}}, \bibinfo {author} {\bibfnamefont {V.}~\bibnamefont {Dunjko}}, \
  and\ \bibinfo {author} {\bibfnamefont {M.}~\bibnamefont {Olshanii}},\
  }\href@noop {} {\bibfield  {journal} {\bibinfo  {journal} {Nature}\ }\textbf
  {\bibinfo {volume} {452}},\ \bibinfo {pages} {854} (\bibinfo {year}
  {2008})}\BibitemShut {NoStop}%
\bibitem [{\citenamefont {Truscott}\ \emph {et~al.}(2001)\citenamefont
  {Truscott}, \citenamefont {Strecker}, \citenamefont {McAlexander},
  \citenamefont {Partridge},\ and\ \citenamefont
  {Hulet}}]{truscott2001observation}%
  \BibitemOpen
  \bibfield  {author} {\bibinfo {author} {\bibfnamefont {A.~G.}\ \bibnamefont
  {Truscott}}, \bibinfo {author} {\bibfnamefont {K.~E.}\ \bibnamefont
  {Strecker}}, \bibinfo {author} {\bibfnamefont {W.~I.}\ \bibnamefont
  {McAlexander}}, \bibinfo {author} {\bibfnamefont {G.~B.}\ \bibnamefont
  {Partridge}}, \ and\ \bibinfo {author} {\bibfnamefont {R.~G.}\ \bibnamefont
  {Hulet}},\ }\href@noop {} {\bibfield  {journal} {\bibinfo  {journal}
  {Science}\ }\textbf {\bibinfo {volume} {291}},\ \bibinfo {pages} {2570}
  (\bibinfo {year} {2001})}\BibitemShut {NoStop}%
\bibitem [{\citenamefont {Stan}\ \emph {et~al.}(2004)\citenamefont {Stan},
  \citenamefont {Zwierlein}, \citenamefont {Schunck}, \citenamefont {Raupach},\
  and\ \citenamefont {Ketterle}}]{stan2004observation}%
  \BibitemOpen
  \bibfield  {author} {\bibinfo {author} {\bibfnamefont {C.~A.}\ \bibnamefont
  {Stan}}, \bibinfo {author} {\bibfnamefont {M.~W.}\ \bibnamefont {Zwierlein}},
  \bibinfo {author} {\bibfnamefont {C.~H.}\ \bibnamefont {Schunck}}, \bibinfo
  {author} {\bibfnamefont {S.~M.~F.}\ \bibnamefont {Raupach}}, \ and\ \bibinfo
  {author} {\bibfnamefont {W.}~\bibnamefont {Ketterle}},\ }\href@noop {}
  {\bibfield  {journal} {\bibinfo  {journal} {Phys. Rev. Lett.}\ }\textbf
  {\bibinfo {volume} {93}},\ \bibinfo {pages} {143001} (\bibinfo {year}
  {2004})}\BibitemShut {NoStop}%
\bibitem [{\citenamefont {G{\"u}nter}\ \emph {et~al.}(2006)\citenamefont
  {G{\"u}nter}, \citenamefont {St{\"o}ferle}, \citenamefont {Moritz},
  \citenamefont {K{\"o}hl},\ and\ \citenamefont {Esslinger}}]{gunter2006bose}%
  \BibitemOpen
  \bibfield  {author} {\bibinfo {author} {\bibfnamefont {K.}~\bibnamefont
  {G{\"u}nter}}, \bibinfo {author} {\bibfnamefont {T.}~\bibnamefont
  {St{\"o}ferle}}, \bibinfo {author} {\bibfnamefont {H.}~\bibnamefont
  {Moritz}}, \bibinfo {author} {\bibfnamefont {M.}~\bibnamefont {K{\"o}hl}}, \
  and\ \bibinfo {author} {\bibfnamefont {T.}~\bibnamefont {Esslinger}},\
  }\href@noop {} {\bibfield  {journal} {\bibinfo  {journal} {Phys. Rev. Lett.}\
  }\textbf {\bibinfo {volume} {96}},\ \bibinfo {pages} {180402} (\bibinfo
  {year} {2006})}\BibitemShut {NoStop}%
\bibitem [{\citenamefont {Inouye}\ \emph {et~al.}(2004)\citenamefont {Inouye},
  \citenamefont {Goldwin}, \citenamefont {Olsen}, \citenamefont {Ticknor},
  \citenamefont {Bohn},\ and\ \citenamefont {Jin}}]{inouye2004observation}%
  \BibitemOpen
  \bibfield  {author} {\bibinfo {author} {\bibfnamefont {S.}~\bibnamefont
  {Inouye}}, \bibinfo {author} {\bibfnamefont {J.}~\bibnamefont {Goldwin}},
  \bibinfo {author} {\bibfnamefont {M.~L.}\ \bibnamefont {Olsen}}, \bibinfo
  {author} {\bibfnamefont {C.}~\bibnamefont {Ticknor}}, \bibinfo {author}
  {\bibfnamefont {J.~L.}\ \bibnamefont {Bohn}}, \ and\ \bibinfo {author}
  {\bibfnamefont {D.~S.}\ \bibnamefont {Jin}},\ }\href@noop {} {\bibfield
  {journal} {\bibinfo  {journal} {Phys. Rev. Lett.}\ }\textbf {\bibinfo
  {volume} {93}},\ \bibinfo {pages} {183201} (\bibinfo {year}
  {2004})}\BibitemShut {NoStop}%
\bibitem [{\citenamefont {Fukuhara}\ \emph {et~al.}(2009)\citenamefont
  {Fukuhara}, \citenamefont {Sugawa}, \citenamefont {Takasu},\ and\
  \citenamefont {Takahashi}}]{fukuhara2009all}%
  \BibitemOpen
  \bibfield  {author} {\bibinfo {author} {\bibfnamefont {T.}~\bibnamefont
  {Fukuhara}}, \bibinfo {author} {\bibfnamefont {S.}~\bibnamefont {Sugawa}},
  \bibinfo {author} {\bibfnamefont {Y.}~\bibnamefont {Takasu}}, \ and\ \bibinfo
  {author} {\bibfnamefont {Y.}~\bibnamefont {Takahashi}},\ }\href@noop {}
  {\bibfield  {journal} {\bibinfo  {journal} {Phys. Rev. A}\ }\textbf {\bibinfo
  {volume} {79}},\ \bibinfo {pages} {021601} (\bibinfo {year}
  {2009})}\BibitemShut {NoStop}%
\bibitem [{\citenamefont {Wu}\ \emph {et~al.}(2012)\citenamefont {Wu},
  \citenamefont {Park}, \citenamefont {Ahmadi}, \citenamefont {Will},\ and\
  \citenamefont {Zwierlein}}]{wu2012ultracold}%
  \BibitemOpen
  \bibfield  {author} {\bibinfo {author} {\bibfnamefont {C.-H.}\ \bibnamefont
  {Wu}}, \bibinfo {author} {\bibfnamefont {J.~W.}\ \bibnamefont {Park}},
  \bibinfo {author} {\bibfnamefont {P.}~\bibnamefont {Ahmadi}}, \bibinfo
  {author} {\bibfnamefont {S.}~\bibnamefont {Will}}, \ and\ \bibinfo {author}
  {\bibfnamefont {M.~W.}\ \bibnamefont {Zwierlein}},\ }\href@noop {} {\bibfield
   {journal} {\bibinfo  {journal} {Phys. Rev. Lett.}\ }\textbf {\bibinfo
  {volume} {109}},\ \bibinfo {pages} {085301} (\bibinfo {year}
  {2012})}\BibitemShut {NoStop}%
\bibitem [{\citenamefont {Catani}\ \emph {et~al.}(2008)\citenamefont {Catani},
  \citenamefont {De~Sarlo}, \citenamefont {Barontini}, \citenamefont
  {Minardi},\ and\ \citenamefont {Inguscio}}]{catani2008degenerate}%
  \BibitemOpen
  \bibfield  {author} {\bibinfo {author} {\bibfnamefont {J.}~\bibnamefont
  {Catani}}, \bibinfo {author} {\bibfnamefont {L.}~\bibnamefont {De~Sarlo}},
  \bibinfo {author} {\bibfnamefont {G.}~\bibnamefont {Barontini}}, \bibinfo
  {author} {\bibfnamefont {F.}~\bibnamefont {Minardi}}, \ and\ \bibinfo
  {author} {\bibfnamefont {M.}~\bibnamefont {Inguscio}},\ }\href@noop {}
  {\bibfield  {journal} {\bibinfo  {journal} {Phys. Rev. A}\ }\textbf {\bibinfo
  {volume} {77}},\ \bibinfo {pages} {011603} (\bibinfo {year}
  {2008})}\BibitemShut {NoStop}%
\bibitem [{\citenamefont {Shin}\ \emph {et~al.}(2008)\citenamefont {Shin},
  \citenamefont {Schirotzek}, \citenamefont {Schunck},\ and\ \citenamefont
  {Ketterle}}]{Shin2008}%
  \BibitemOpen
  \bibfield  {author} {\bibinfo {author} {\bibfnamefont {Y.-I.}\ \bibnamefont
  {Shin}}, \bibinfo {author} {\bibfnamefont {A.}~\bibnamefont {Schirotzek}},
  \bibinfo {author} {\bibfnamefont {C.~H.}\ \bibnamefont {Schunck}}, \ and\
  \bibinfo {author} {\bibfnamefont {W.}~\bibnamefont {Ketterle}},\ }\href
  {\doibase 10.1103/PhysRevLett.101.070404} {\bibfield  {journal} {\bibinfo
  {journal} {Phys. Rev. Lett.}\ }\textbf {\bibinfo {volume} {101}},\ \bibinfo
  {pages} {070404} (\bibinfo {year} {2008})}\BibitemShut {NoStop}%
\bibitem [{\citenamefont {Pilch}\ \emph {et~al.}(2009)\citenamefont {Pilch},
  \citenamefont {Lange}, \citenamefont {Prantner}, \citenamefont {Kerner},
  \citenamefont {Ferlaino}, \citenamefont {N{\"a}gerl},\ and\ \citenamefont
  {Grimm}}]{pilch2009observation}%
  \BibitemOpen
  \bibfield  {author} {\bibinfo {author} {\bibfnamefont {K.}~\bibnamefont
  {Pilch}}, \bibinfo {author} {\bibfnamefont {A.~D.}\ \bibnamefont {Lange}},
  \bibinfo {author} {\bibfnamefont {A.}~\bibnamefont {Prantner}}, \bibinfo
  {author} {\bibfnamefont {G.}~\bibnamefont {Kerner}}, \bibinfo {author}
  {\bibfnamefont {F.}~\bibnamefont {Ferlaino}}, \bibinfo {author}
  {\bibfnamefont {H.-C.}\ \bibnamefont {N{\"a}gerl}}, \ and\ \bibinfo {author}
  {\bibfnamefont {R.}~\bibnamefont {Grimm}},\ }\href@noop {} {\bibfield
  {journal} {\bibinfo  {journal} {Phys. Rev. A}\ }\textbf {\bibinfo {volume}
  {79}},\ \bibinfo {pages} {042718} (\bibinfo {year} {2009})}\BibitemShut
  {NoStop}%
\bibitem [{\citenamefont {Wernsdorfer}\ \emph {et~al.}(2010)\citenamefont
  {Wernsdorfer}, \citenamefont {Snoek},\ and\ \citenamefont
  {Hofstetter}}]{wernsdorfer2010lattice}%
  \BibitemOpen
  \bibfield  {author} {\bibinfo {author} {\bibfnamefont {J.}~\bibnamefont
  {Wernsdorfer}}, \bibinfo {author} {\bibfnamefont {M.}~\bibnamefont {Snoek}},
  \ and\ \bibinfo {author} {\bibfnamefont {W.}~\bibnamefont {Hofstetter}},\
  }\href@noop {} {\bibfield  {journal} {\bibinfo  {journal} {Phys. Rev. A}\
  }\textbf {\bibinfo {volume} {81}},\ \bibinfo {pages} {043620} (\bibinfo
  {year} {2010})}\BibitemShut {NoStop}%
\bibitem [{\citenamefont {McCarron}\ \emph {et~al.}(2011)\citenamefont
  {McCarron}, \citenamefont {Cho}, \citenamefont {Jenkin}, \citenamefont
  {K{\"o}ppinger},\ and\ \citenamefont {Cornish}}]{mccarron2011dual}%
  \BibitemOpen
  \bibfield  {author} {\bibinfo {author} {\bibfnamefont {D.~J.}\ \bibnamefont
  {McCarron}}, \bibinfo {author} {\bibfnamefont {H.~W.}\ \bibnamefont {Cho}},
  \bibinfo {author} {\bibfnamefont {D.~L.}\ \bibnamefont {Jenkin}}, \bibinfo
  {author} {\bibfnamefont {M.~P.}\ \bibnamefont {K{\"o}ppinger}}, \ and\
  \bibinfo {author} {\bibfnamefont {S.~L.}\ \bibnamefont {Cornish}},\
  }\href@noop {} {\bibfield  {journal} {\bibinfo  {journal} {Phys. Rev. A}\
  }\textbf {\bibinfo {volume} {84}},\ \bibinfo {pages} {011603} (\bibinfo
  {year} {2011})}\BibitemShut {NoStop}%
\bibitem [{\citenamefont {Spethmann}\ \emph {et~al.}(2012)\citenamefont
  {Spethmann}, \citenamefont {Kindermann}, \citenamefont {John}, \citenamefont
  {Weber}, \citenamefont {Meschede},\ and\ \citenamefont
  {Widera}}]{spethmann2012dynamics}%
  \BibitemOpen
  \bibfield  {author} {\bibinfo {author} {\bibfnamefont {N.}~\bibnamefont
  {Spethmann}}, \bibinfo {author} {\bibfnamefont {F.}~\bibnamefont
  {Kindermann}}, \bibinfo {author} {\bibfnamefont {S.}~\bibnamefont {John}},
  \bibinfo {author} {\bibfnamefont {C.}~\bibnamefont {Weber}}, \bibinfo
  {author} {\bibfnamefont {D.}~\bibnamefont {Meschede}}, \ and\ \bibinfo
  {author} {\bibfnamefont {A.}~\bibnamefont {Widera}},\ }\href@noop {}
  {\bibfield  {journal} {\bibinfo  {journal} {Phys. Rev. Lett.}\ }\textbf
  {\bibinfo {volume} {109}},\ \bibinfo {pages} {235301} (\bibinfo {year}
  {2012})}\BibitemShut {NoStop}%
\bibitem [{\citenamefont {Grusdt}\ \emph {et~al.}()\citenamefont {Grusdt},
  \citenamefont {Shashi}, \citenamefont {Abanin},\ and\ \citenamefont
  {Demler}}]{fabiprep}%
  \BibitemOpen
  \bibfield  {author} {\bibinfo {author} {\bibfnamefont {F.}~\bibnamefont
  {Grusdt}}, \bibinfo {author} {\bibfnamefont {A.}~\bibnamefont {Shashi}},
  \bibinfo {author} {\bibfnamefont {D.~A.}\ \bibnamefont {Abanin}}, \ and\
  \bibinfo {author} {\bibfnamefont {E.}~\bibnamefont {Demler}},\ }\href@noop {}
  {\bibinfo  {journal} {in preparation}\ }\BibitemShut {NoStop}%
\end{thebibliography}%
\end{document}